\documentclass[a4paper,11pt]{article}
\usepackage{jheppub} 
\usepackage{lineno}
\usepackage{import}

\usepackage{setspace}
\usepackage{graphicx}
\usepackage{wrapfig} 
\usepackage{caption}
\usepackage{subcaption}
\usepackage{hyperref}
\usepackage{tikz} 
\usepackage{empheq}
\usepackage{physics}
\usepackage{amsmath}
\usepackage{mathdots}
\usepackage{yhmath}
\usepackage{cancel}
\usepackage{color}
\usepackage{siunitx}
\usepackage{array}
\usepackage{multirow}
\usepackage{amssymb}
\usepackage{gensymb}
\usepackage{tabularx}
\usepackage{extarrows}
\usepackage{booktabs}
\usepackage{mathtools}
\usepackage{simplewick}
\usepackage[obeyFinal]{todonotes}

\usetikzlibrary{fadings}
\usetikzlibrary{patterns}
\usetikzlibrary{shadows.blur}
\usetikzlibrary{shapes}
\usetikzlibrary{arrows}

\definecolor{refkey}{gray}{0.45}
\definecolor{labelkey}{RGB}{155,48,48} 
\definecolor{UI_blue}{RGB}{32, 64, 151}
\definecolor{UI_red}{RGB}{187, 62, 24}
\definecolor{UI_blue2}{RGB}{0, 84, 147}
\definecolor{UI_red2}{RGB}{159, 32, 66}
\definecolor{UI_gray}{RGB}{169, 169, 169}
\definecolor{UI_sepia}{RGB}{112, 66, 20}
\definecolor{UI_bittersweet}{RGB}{254, 111, 94}
\definecolor{UI_emerald}{RGB}{80, 200, 120}
\definecolor{UI_olivegreen}{RGB}{181, 179, 92}
\definecolor{UI_cadetblue}{RGB}{95, 158, 160}
\definecolor{UI_fuchsia}{RGB}{255, 0, 255}
\definecolor{UI_midnightblue}{RGB}{25, 25, 112}
\definecolor{UI_royalblue}{RGB}{0,35, 102}
\definecolor{UI_periwinkle}{RGB}{204, 204, 255}
\definecolor{UI_redorange}{RGB}{255, 83, 73}
\definecolor{UI_brickred}{RGB}{203,65,84}	
\definecolor{UI_forestgreen}{RGB}{34, 139, 34}
\definecolor{UI_tan}{RGB}{210,180,140}	
\definecolor{UI_burlywood}{RGB}{222,184,135}
\definecolor{UI_burlywood}{RGB}{192,64,0}
\definecolor{UI_darkorchid}{RGB}{153,50,204}

\newcommand{\half}{\frac{1}{2}}

\newcommand{\eq}[1]{eq.(\ref{#1})}
\newcommand{\brf}[1]{\left(#1\right)}

\newcommand{\brt}[1]{\left[#1\right]}

\def\beq{\begin{eqnarray}}\def\eeq{\end{eqnarray}}
\def\be{\begin{equation}}\def\ee{\end{equation}}

\def\mes[#1]{d^{3}{#1}}

\def\del{\partial}

\def\del{\partial}

\reversemarginpar

\abstract{We analyse the laws of thermodynamics governing the behaviour of cosmological horizons in de Sitter space and their map  to a holographic description at future infinity, ${\cal I}^+$. In this case, the boundary can receive signals from two cosmological horizons. We find a universal form for the first law of thermodynamics, valid in general circumstances, when  matter-energy crosses both horizons and impinges  on the boundary. This universal form  leads to a  well defined notion of entropy in the    holographic dual. It is  specified on a co-dimension  one surface of the boundary, and  can be expressed as  a function of two boundary charges,  pressure and angular momentum, both of which are derived from the    Brown -York stress tensor. Additional comments on the second law, confusing factors of $i$ which arise, and comments pertaining to JT gravity, are included towards the end.}

\title{The Thermodynamics of Cosmological Horizons and Their Holographic Description in de Sitter Space}
\author[a]{Indranil Dey,}
\author[b]{Kanhu Kishore Nanda,}
\author[a]{Akashdeep Roy,}
\author[a]{Deepak Kumar Sharma,}
\author[a]{Sandip P. Trivedi}
\affiliation[a]{\it Department of Theoretical Physics,
		Tata Institute of Fundamental Research, Colaba, Mumbai, India, 400005}
\affiliation[b]{\it Chennai Mathematical Institute, H1 SIPCOT IT Park, Siruseri, Chennai, 603103}
\emailAdd{indranil.dey@tifr.res.in}
\emailAdd{kanhukishore@cmi.ac.in}
\emailAdd{akaspace99@gmail.com}
\emailAdd{deepakksharma7201@gmail.com}
\emailAdd{sandip@theory.tifr.res.in}
\preprint{\parbox{3cm}{TIFR/TH/26-13}}

\begin{document}
	\maketitle
	\flushbottom
	
	\section{Introduction}
\label{sec.intro}
De Sitter space is a frontier in the study of quantum gravity. There are many mysteries that need to be understood. A  key question is whether de Sitter space (dS)  admits a holographic description. Two versions of holography have been studied, pertaining to global de Sitter space \cite{malda_NG, Kluson:2003xh, Witten:2001kn, Strominger:2001pn, Spradlin:2001nb, Balasubramanian:2001nb} and the static patch \cite{stat2,Susskind:2021omt, Svesko:2022txo, Franken:2023pni, Galante:2023uyf}. 
In this paper we will study the global  version where the hologram can be, roughly speaking, taken to live at  asymptotic future infinity, denoted by  ${\cal I}^+$, see fig. \ref{fig.pendsst}. Some  aspects of the holographic dictionary in this case have been   worked out, using the  analogy with the AdS case as a guide, see \cite{malda_NG},  also \cite{roy_dscft}. 

As is well known, the idea of holography is  borne out in a very precise manner in AdS space, \cite{Malda_adscft},  \cite{Witten_adscft}, see also the review \cite{Magoo_ads}, the lecture notes \cite{DHoker:2002nbb, Maldacena:2003nj, DeWolfe:2018dkl}, the books \cite{Ammon:2015wua, Nastase:2015wjb} and the references therein.
A key reason to believe  that it is more general and applicable  also to  de Sitter  and flat space, stems from the universality of the laws of black hole thermodynamics. 
These laws suggest quite  generally that  the number of degrees of freedom in a gravitational system,  in some  given region,  scale like its area and not its  volume; thereby suggesting  that a  holographic description \cite{tHooft:1993dmi, Susskind:1994vu} should apply in these cases as well. 

In AdS space the laws of black hole thermodynamics map to the usual laws of thermodynamics in the non-gravitational holographic dual.  The  explicit  nature of this mapping is one of the reasons to    trust the holographic dual description and it sometimes also allows us to test the holographic correspondence in a precise manner \cite{Malda_adscft}, \cite{Magoo_ads} and references therein, and also \cite{hanada_ads, hanada_num}.   In this paper we will attempt to map the laws of black hole thermodynamics in  de Sitter space to the putative  hologram at ${\cal I}^+$.
Although holography in de Sitter space has already been a subject of considerable study, surprisingly, this question has not received the attention it deserves. 

Our  main result is that after carrying out this map we find a universal form   for  the first law of thermodynamics in the boundary theory. The boundary theory in this case is  Euclidean and we define a pressure, $P$,  in it,  which is related to an  appropriate component of the Brown York stress tensor. By studying cosmological horizons in the presence of black  holes we can also define an entropy, $S$,  in the boundary theory, which is a function of the pressure. This leads to the general form of the first law, which is then valid in general dynamical situations as well,  where propagating matter, and  gravity waves, 
can cross the cosmological horizons and impinge on the holographic screen. We show that the first law can be generalised for rotating situations as well,  by  including the effects of angular momentum. 

In contrast, we do not find a universal version of the second law. This is  not surprising,  since  the boundary theory, as mentioned above, is Euclidean and not Lorentzian, with  no analogue of the arrow of time. General considerations in the bulk do strongly suggest that the maximum entropy one can associate with de Sitter space is related to the area of the cosmological horizon of pure dS. E.g., once we include  quantum effects, black holes eventually  evaporate away, leaving empty de Sitter with this maximal entropy. In the boundary theory we find that this asymptotic statement  does have   an analogue. The boundary entropy   in the asymptotic regions of ${\cal I}^+ $  would generally  take  this maximum value (these are the regions which lie  towards the far left and far right in  ${\cal I}^+$   in   the Penrose diagram, fig. \ref{fig.pendsst}, see also fig. \ref{fig.evap}).

An important lesson, from studying general dynamical situations, is that the boundary entropy needs to  be defined on a co-dimension $1$ slice of the holographic theory. This is of course different from the bulk entropy, which is  defined on a co-dimension $1$ slice of the bulk. This lesson   ties in well with the expectations from holography. 
And it leads to the natural  suggestion  that the  boundary entropy has an interpretation as the number of degrees of freedom in the hologram. 
In the concluding comments we discuss how this suggestion is indeed borne out in the simple example of JT gravity in dS space. 

This paper is structured as follows. In section \ref{basicsa} and \ref{Sec.Holo}  we give an introduction to some aspects of dS space and to the version of holography we will be exploring. 
The laws of thermodynamics, are reviewed in section \ref{sec.first}, keeping especially cosmological horizons in mind. The first law is mapped to the boundary hologram in section \ref{Sec.hololaw} along with some additional discussion pertaining to the second law. We end with concluding comments in section \ref{Sec.con}. Appendices \ref{app.by}--\ref{app.jt} contain important supplementary material.

\section{ Some Background}
\label{basicsa}
The metric of 4 dimensional deSitter space  in global coordinates is given by 
\be
\label{metricaa}
ds^2=-d\tau^2+\cosh^2(\tau) d\Omega_3^2
\ee
A change of coordinates locally gives the metric in the form 
\be
\label{metrica}
ds^2=-{dr^2\over r^2-1} +(r^2-1) dt^2 + r^2 d\Omega_2^2
\ee
Note that we have set the Hubble constant  of dS space, $H=1$.

\begin{figure}[h]
	\centering

	\tikzset{every picture/.style={line width=0.75pt}} 
	
	\begin{tikzpicture}[x=0.75pt,y=0.75pt,yscale=-1,xscale=1]
		
		\draw   (120,86) -- (286,86) -- (286,252) -- (120,252) -- cycle ;
		\draw    (120,86) -- (286,252) ;
		\draw    (286,86) -- (120,252) ;
		\draw  [dash pattern={on 4.5pt off 4.5pt}]  (249,99) -- (336.67,99) ;
		\draw [shift={(338.67,99)}, rotate = 180] [color={rgb, 255:red, 0; green, 0; blue, 0 }  ][line width=0.75]    (10.93,-3.29) .. controls (6.95,-1.4) and (3.31,-0.3) .. (0,0) .. controls (3.31,0.3) and (6.95,1.4) .. (10.93,3.29)   ;
		\draw  [dash pattern={on 4.5pt off 4.5pt}]  (224.67,148.6) -- (339.67,148.6) ;
		\draw [shift={(341.67,148.6)}, rotate = 180] [color={rgb, 255:red, 0; green, 0; blue, 0 }  ][line width=0.75]    (10.93,-3.29) .. controls (6.95,-1.4) and (3.31,-0.3) .. (0,0) .. controls (3.31,0.3) and (6.95,1.4) .. (10.93,3.29)   ;
		\draw  [dash pattern={on 4.5pt off 4.5pt}]  (270,175.6) -- (342,175.6) ;
		\draw [shift={(344,175.6)}, rotate = 180] [color={rgb, 255:red, 0; green, 0; blue, 0 }  ][line width=0.75]    (10.93,-3.29) .. controls (6.95,-1.4) and (3.31,-0.3) .. (0,0) .. controls (3.31,0.3) and (6.95,1.4) .. (10.93,3.29)   ;
		\draw  [dash pattern={on 4.5pt off 4.5pt}]  (148,203.6) -- (342,203.6) ;
		\draw [shift={(344,203.6)}, rotate = 180] [color={rgb, 255:red, 0; green, 0; blue, 0 }  ][line width=0.75]    (10.93,-3.29) .. controls (6.95,-1.4) and (3.31,-0.3) .. (0,0) .. controls (3.31,0.3) and (6.95,1.4) .. (10.93,3.29)   ;
		\draw  [dash pattern={on 4.5pt off 4.5pt}]  (215,233.6) -- (340,233.6) ;
		\draw [shift={(342,233.6)}, rotate = 180] [color={rgb, 255:red, 0; green, 0; blue, 0 }  ][line width=0.75]    (10.93,-3.29) .. controls (6.95,-1.4) and (3.31,-0.3) .. (0,0) .. controls (3.31,0.3) and (6.95,1.4) .. (10.93,3.29)   ;
		\draw  [dash pattern={on 4.5pt off 4.5pt}]  (156.67,123.6) -- (337.67,123.6) ;
		\draw [shift={(339.67,123.6)}, rotate = 180] [color={rgb, 255:red, 0; green, 0; blue, 0 }  ][line width=0.75]    (10.93,-3.29) .. controls (6.95,-1.4) and (3.31,-0.3) .. (0,0) .. controls (3.31,0.3) and (6.95,1.4) .. (10.93,3.29)   ;
		
		\draw (149,163) node [anchor=north west][inner sep=0.75pt]   [align=left] {L};
		\draw (249,161) node [anchor=north west][inner sep=0.75pt]   [align=left] {R};
		\draw (199,109) node [anchor=north west][inner sep=0.75pt]   [align=left] {F};
		\draw (198,212) node [anchor=north west][inner sep=0.75pt]   [align=left] {B};
		\draw (198,63) node [anchor=north west][inner sep=0.75pt]   [align=left] {$\displaystyle \mathcal{I}^{+}$};
		\draw (198,253) node [anchor=north west][inner sep=0.75pt]   [align=left] {$\displaystyle \mathcal{I}^{-}$};
		\draw (205.46,141.67) node [anchor=north west][inner sep=0.75pt]  [rotate=-316.19] [align=left] {r=1};
		\draw (161.17,129.91) node [anchor=north west][inner sep=0.75pt]  [rotate=-47.54] [align=left] {r=1};
		\draw (349,115) node [anchor=north west][inner sep=0.75pt]   [align=left] {Left Cosmological Horizon};
		\draw (349,168) node [anchor=north west][inner sep=0.75pt]   [align=left] {Right Static patch};
		\draw (349,196) node [anchor=north west][inner sep=0.75pt]   [align=left] {Left Static patch};
		\draw (349,91) node [anchor=north west][inner sep=0.75pt]   [align=left] {Forward Milne patch};
		\draw (349,226) node [anchor=north west][inner sep=0.75pt]   [align=left] {Backward Milne patch};
		\draw (349,140) node [anchor=north west][inner sep=0.75pt]   [align=left] {Right Cosmological Horizon};

	\end{tikzpicture}
	
	\caption{Penrose Diagram of de Sitter Space.}
	\label{fig.pendsst}
\end{figure}
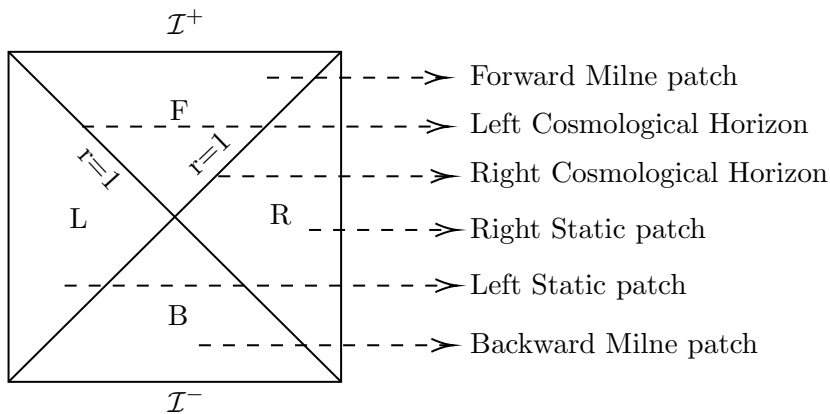

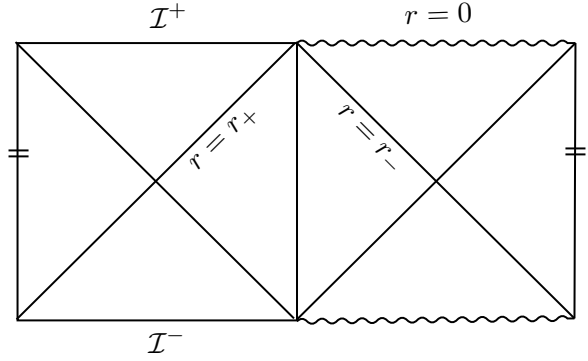
\begin{figure}[h]
	\centering
	
	\tikzset{every picture/.style={line width=0.75pt}} 
	
	\begin{tikzpicture}[x=0.75pt,y=0.75pt,yscale=-1,xscale=1]
		
		\draw   (151.5,50.5) -- (291,50.5) -- (291,190) -- (151.5,190) -- cycle ;
		\draw    (150.5,50) -- (289.5,189) ;
		\draw    (290,50.5) -- (151,189.5) ;
		\draw    (290.5,50) -- (429.5,189) ;
		\draw    (430,50.5) -- (291,189.5) ;
		\draw    (430,50.5) -- (429.5,189) ;
		\draw    (291,50.5) .. controls (292.67,48.83) and (294.33,48.83) .. (296,50.5) .. controls (297.67,52.17) and (299.33,52.17) .. (301,50.5) .. controls (302.67,48.83) and (304.33,48.83) .. (306,50.5) .. controls (307.67,52.17) and (309.33,52.17) .. (311,50.5) .. controls (312.67,48.83) and (314.33,48.83) .. (316,50.5) .. controls (317.67,52.17) and (319.33,52.17) .. (321,50.5) .. controls (322.67,48.83) and (324.33,48.83) .. (326,50.5) .. controls (327.67,52.17) and (329.33,52.17) .. (331,50.5) .. controls (332.67,48.83) and (334.33,48.83) .. (336,50.5) .. controls (337.67,52.17) and (339.33,52.17) .. (341,50.5) .. controls (342.67,48.83) and (344.33,48.83) .. (346,50.5) .. controls (347.67,52.17) and (349.33,52.17) .. (351,50.5) .. controls (352.67,48.83) and (354.33,48.83) .. (356,50.5) .. controls (357.67,52.17) and (359.33,52.17) .. (361,50.5) .. controls (362.67,48.83) and (364.33,48.83) .. (366,50.5) .. controls (367.67,52.17) and (369.33,52.17) .. (371,50.5) .. controls (372.67,48.83) and (374.33,48.83) .. (376,50.5) .. controls (377.67,52.17) and (379.33,52.17) .. (381,50.5) .. controls (382.67,48.83) and (384.33,48.83) .. (386,50.5) .. controls (387.67,52.17) and (389.33,52.17) .. (391,50.5) .. controls (392.67,48.83) and (394.33,48.83) .. (396,50.5) .. controls (397.67,52.17) and (399.33,52.17) .. (401,50.5) .. controls (402.67,48.83) and (404.33,48.83) .. (406,50.5) .. controls (407.67,52.17) and (409.33,52.17) .. (411,50.5) .. controls (412.67,48.83) and (414.33,48.83) .. (416,50.5) .. controls (417.67,52.17) and (419.33,52.17) .. (421,50.5) .. controls (422.67,48.83) and (424.33,48.83) .. (426,50.5) -- (430,50.5) -- (430,50.5) ;
		\draw    (291,190.5) .. controls (292.65,188.82) and (294.32,188.8) .. (296,190.45) .. controls (297.69,192.1) and (299.35,192.08) .. (301,190.39) .. controls (302.65,188.71) and (304.32,188.69) .. (306,190.34) .. controls (307.69,191.99) and (309.35,191.97) .. (311,190.28) .. controls (312.65,188.6) and (314.32,188.58) .. (316,190.23) .. controls (317.68,191.88) and (319.35,191.86) .. (321,190.18) .. controls (322.65,188.49) and (324.31,188.47) .. (326,190.12) .. controls (327.68,191.77) and (329.35,191.75) .. (331,190.07) .. controls (332.65,188.38) and (334.31,188.36) .. (336,190.01) .. controls (337.68,191.66) and (339.35,191.64) .. (341,189.96) .. controls (342.65,188.27) and (344.31,188.25) .. (346,189.9) .. controls (347.68,191.55) and (349.35,191.53) .. (351,189.85) .. controls (352.65,188.17) and (354.32,188.15) .. (356,189.8) .. controls (357.69,191.45) and (359.35,191.43) .. (361,189.74) .. controls (362.65,188.06) and (364.32,188.04) .. (366,189.69) .. controls (367.69,191.34) and (369.35,191.32) .. (371,189.63) .. controls (372.65,187.95) and (374.32,187.93) .. (376,189.58) .. controls (377.68,191.23) and (379.34,191.21) .. (380.99,189.53) .. controls (382.64,187.84) and (384.3,187.82) .. (385.99,189.47) .. controls (387.67,191.12) and (389.34,191.1) .. (390.99,189.42) .. controls (392.64,187.73) and (394.3,187.71) .. (395.99,189.36) .. controls (397.67,191.01) and (399.34,190.99) .. (400.99,189.31) .. controls (402.64,187.62) and (404.3,187.6) .. (405.99,189.25) .. controls (407.67,190.9) and (409.34,190.88) .. (410.99,189.2) .. controls (412.64,187.52) and (414.31,187.5) .. (415.99,189.15) .. controls (417.68,190.8) and (419.34,190.78) .. (420.99,189.09) .. controls (422.64,187.41) and (424.31,187.39) .. (425.99,189.04) -- (429.5,189) -- (429.5,189) ;
		\draw    (147,104.5) -- (157,104.5)(147,107.5) -- (157,107.5) ;
		\draw    (425,103.5) -- (435,103.5)(425,106.5) -- (435,106.5) ;
		
		\draw (216.36,28.84) node [anchor=north west][inner sep=0.75pt]   [align=left] {$\displaystyle \mathcal{I}^{+}$};
		\draw (215.36,192.36) node [anchor=north west][inner sep=0.75pt]   [align=left] {$\displaystyle \mathcal{I}^{-}$};
		\draw (343.01,29.98) node [anchor=north west][inner sep=0.75pt]  [rotate=-358.18] [align=left] {$\displaystyle r=0$};
		\draw (234.65,110.44) node [anchor=north west][inner sep=0.75pt]  [rotate=-315.03] [align=left] {$\displaystyle r=r_{+}$};
		\draw (315.28,78.73) node [anchor=north west][inner sep=0.75pt]  [rotate=-45.27] [align=left] {$\displaystyle r=r_{-}$};

	\end{tikzpicture}
	\caption{Penrose Diagram of SdS after identifications.}
	\label{fig.idpensds}
\end{figure}

\begin{figure}
	\centering

	
	\tikzset{
		pattern size/.store in=\mcSize, 
		pattern size = 5pt,
		pattern thickness/.store in=\mcThickness, 
		pattern thickness = 0.3pt,
		pattern radius/.store in=\mcRadius, 
		pattern radius = 1pt}
	\makeatletter
	\pgfutil@ifundefined{pgf@pattern@name@_0d9lu7b7z}{
		\pgfdeclarepatternformonly[\mcThickness,\mcSize]{_0d9lu7b7z}
		{\pgfqpoint{0pt}{0pt}}
		{\pgfpoint{\mcSize+\mcThickness}{\mcSize+\mcThickness}}
		{\pgfpoint{\mcSize}{\mcSize}}
		{
			\pgfsetcolor{\tikz@pattern@color}
			\pgfsetlinewidth{\mcThickness}
			\pgfpathmoveto{\pgfqpoint{0pt}{0pt}}
			\pgfpathlineto{\pgfpoint{\mcSize+\mcThickness}{\mcSize+\mcThickness}}
			\pgfusepath{stroke}
	}}
	\makeatother
	
	
	\tikzset{
		pattern size/.store in=\mcSize, 
		pattern size = 5pt,
		pattern thickness/.store in=\mcThickness, 
		pattern thickness = 0.3pt,
		pattern radius/.store in=\mcRadius, 
		pattern radius = 1pt}
	\makeatletter
	\pgfutil@ifundefined{pgf@pattern@name@_7n6ugrd8w}{
		\pgfdeclarepatternformonly[\mcThickness,\mcSize]{_7n6ugrd8w}
		{\pgfqpoint{0pt}{-\mcThickness}}
		{\pgfpoint{\mcSize}{\mcSize}}
		{\pgfpoint{\mcSize}{\mcSize}}
		{
			\pgfsetcolor{\tikz@pattern@color}
			\pgfsetlinewidth{\mcThickness}
			\pgfpathmoveto{\pgfqpoint{0pt}{\mcSize}}
			\pgfpathlineto{\pgfpoint{\mcSize+\mcThickness}{-\mcThickness}}
			\pgfusepath{stroke}
	}}
	\makeatother
	
	
	\tikzset {_yzcbkyun6/.code = {\pgfsetadditionalshadetransform{ \pgftransformshift{\pgfpoint{0 bp } { 0 bp }  }  \pgftransformrotate{0 }  \pgftransformscale{2 }  }}}
	\pgfdeclarehorizontalshading{_0j0bf6t17}{150bp}{rgb(0bp)=(1,1,0);
		rgb(37.5bp)=(1,1,0);
		rgb(62.5bp)=(0,0.5,0.5);
		rgb(100bp)=(0,0.5,0.5)}
	
	
	\tikzset {_stqvn1v9z/.code = {\pgfsetadditionalshadetransform{ \pgftransformshift{\pgfpoint{0 bp } { 0 bp }  }  \pgftransformrotate{0 }  \pgftransformscale{2 }  }}}
	\pgfdeclarehorizontalshading{_qhctrs67m}{150bp}{rgb(0bp)=(1,1,0);
		rgb(37.5bp)=(1,1,0);
		rgb(62.5bp)=(0,0.5,0.5);
		rgb(100bp)=(0,0.5,0.5)}
	\tikzset{every picture/.style={line width=0.75pt}} 
	
	\begin{tikzpicture}[x=0.75pt,y=0.75pt,yscale=-1,xscale=1]
		
		\draw   (261.5,50.5) -- (401,50.5) -- (401,190) -- (261.5,190) -- cycle ;
		\draw    (260.5,50) -- (399.5,189) ;
		\draw    (400,51.5) -- (261,190.5) ;
		\draw    (400.5,50) -- (539.5,189) ;
		\draw    (540,50.5) -- (401,189.5) ;
		\draw    (401,50.5) .. controls (402.67,48.83) and (404.33,48.83) .. (406,50.5) .. controls (407.67,52.17) and (409.33,52.17) .. (411,50.5) .. controls (412.67,48.83) and (414.33,48.83) .. (416,50.5) .. controls (417.67,52.17) and (419.33,52.17) .. (421,50.5) .. controls (422.67,48.83) and (424.33,48.83) .. (426,50.5) .. controls (427.67,52.17) and (429.33,52.17) .. (431,50.5) .. controls (432.67,48.83) and (434.33,48.83) .. (436,50.5) .. controls (437.67,52.17) and (439.33,52.17) .. (441,50.5) .. controls (442.67,48.83) and (444.33,48.83) .. (446,50.5) .. controls (447.67,52.17) and (449.33,52.17) .. (451,50.5) .. controls (452.67,48.83) and (454.33,48.83) .. (456,50.5) .. controls (457.67,52.17) and (459.33,52.17) .. (461,50.5) .. controls (462.67,48.83) and (464.33,48.83) .. (466,50.5) .. controls (467.67,52.17) and (469.33,52.17) .. (471,50.5) .. controls (472.67,48.83) and (474.33,48.83) .. (476,50.5) .. controls (477.67,52.17) and (479.33,52.17) .. (481,50.5) .. controls (482.67,48.83) and (484.33,48.83) .. (486,50.5) .. controls (487.67,52.17) and (489.33,52.17) .. (491,50.5) .. controls (492.67,48.83) and (494.33,48.83) .. (496,50.5) .. controls (497.67,52.17) and (499.33,52.17) .. (501,50.5) .. controls (502.67,48.83) and (504.33,48.83) .. (506,50.5) .. controls (507.67,52.17) and (509.33,52.17) .. (511,50.5) .. controls (512.67,48.83) and (514.33,48.83) .. (516,50.5) .. controls (517.67,52.17) and (519.33,52.17) .. (521,50.5) .. controls (522.67,48.83) and (524.33,48.83) .. (526,50.5) .. controls (527.67,52.17) and (529.33,52.17) .. (531,50.5) .. controls (532.67,48.83) and (534.33,48.83) .. (536,50.5) -- (540,50.5) -- (540,50.5) ;
		\draw    (401,190.5) .. controls (402.65,188.82) and (404.32,188.8) .. (406,190.45) .. controls (407.69,192.1) and (409.35,192.08) .. (411,190.39) .. controls (412.65,188.71) and (414.32,188.69) .. (416,190.34) .. controls (417.69,191.99) and (419.35,191.97) .. (421,190.28) .. controls (422.65,188.6) and (424.32,188.58) .. (426,190.23) .. controls (427.68,191.88) and (429.35,191.86) .. (431,190.18) .. controls (432.65,188.49) and (434.31,188.47) .. (436,190.12) .. controls (437.68,191.77) and (439.35,191.75) .. (441,190.07) .. controls (442.65,188.38) and (444.31,188.36) .. (446,190.01) .. controls (447.68,191.66) and (449.35,191.64) .. (451,189.96) .. controls (452.65,188.27) and (454.31,188.25) .. (456,189.9) .. controls (457.68,191.55) and (459.35,191.53) .. (461,189.85) .. controls (462.65,188.17) and (464.32,188.15) .. (466,189.8) .. controls (467.69,191.45) and (469.35,191.43) .. (471,189.74) .. controls (472.65,188.06) and (474.32,188.04) .. (476,189.69) .. controls (477.69,191.34) and (479.35,191.32) .. (481,189.63) .. controls (482.65,187.95) and (484.32,187.93) .. (486,189.58) .. controls (487.68,191.23) and (489.34,191.21) .. (490.99,189.53) .. controls (492.64,187.84) and (494.3,187.82) .. (495.99,189.47) .. controls (497.67,191.12) and (499.34,191.1) .. (500.99,189.42) .. controls (502.64,187.73) and (504.3,187.71) .. (505.99,189.36) .. controls (507.67,191.01) and (509.34,190.99) .. (510.99,189.31) .. controls (512.64,187.62) and (514.3,187.6) .. (515.99,189.25) .. controls (517.67,190.9) and (519.34,190.88) .. (520.99,189.2) .. controls (522.64,187.52) and (524.31,187.5) .. (525.99,189.15) .. controls (527.68,190.8) and (529.34,190.78) .. (530.99,189.09) .. controls (532.64,187.41) and (534.31,187.39) .. (535.99,189.04) -- (539.5,189) -- (539.5,189) ;
		\draw [pattern=_0d9lu7b7z,pattern size=6pt,pattern thickness=0.75pt,pattern radius=0pt, pattern color={rgb, 255:red, 65; green, 117; blue, 5}]   (192.25,120.75) -- (261.5,190) ;
		\draw [pattern=_7n6ugrd8w,pattern size=6pt,pattern thickness=0.75pt,pattern radius=0pt, pattern color={rgb, 255:red, 0; green, 0; blue, 0}]   (261,50.5) -- (190.9,120.6) ;
		\path  [shading=_0j0bf6t17,_yzcbkyun6] (188,120) .. controls (188,117.79) and (189.79,116) .. (192,116) .. controls (194.21,116) and (196,117.79) .. (196,120) .. controls (196,122.21) and (194.21,124) .. (192,124) .. controls (189.79,124) and (188,122.21) .. (188,120) -- cycle ; 
		\draw   (188,120) .. controls (188,117.79) and (189.79,116) .. (192,116) .. controls (194.21,116) and (196,117.79) .. (196,120) .. controls (196,122.21) and (194.21,124) .. (192,124) .. controls (189.79,124) and (188,122.21) .. (188,120) -- cycle ; 
		
		\path  [shading=_qhctrs67m,_stqvn1v9z] (467,120) .. controls (467,117.79) and (468.79,116) .. (471,116) .. controls (473.21,116) and (475,117.79) .. (475,120) .. controls (475,122.21) and (473.21,124) .. (471,124) .. controls (468.79,124) and (467,122.21) .. (467,120) -- cycle ; 
		\draw   (467,120) .. controls (467,117.79) and (468.79,116) .. (471,116) .. controls (473.21,116) and (475,117.79) .. (475,120) .. controls (475,122.21) and (473.21,124) .. (471,124) .. controls (468.79,124) and (467,122.21) .. (467,120) -- cycle ; 

		\draw (326.36,28.84) node [anchor=north west][inner sep=0.75pt]   [align=left] {$\displaystyle \mathcal{I}^{+}$};
		\draw (325.36,192.36) node [anchor=north west][inner sep=0.75pt]   [align=left] {$\displaystyle \mathcal{I}^{-}$};
		\draw (453.01,29.98) node [anchor=north west][inner sep=0.75pt]  [rotate=-358.18] [align=left] {$\displaystyle r=0$};
		\draw (460,67) node [anchor=north west][inner sep=0.75pt]   [align=left] {BH};
		\draw (461,160) node [anchor=north west][inner sep=0.75pt]   [align=left] {WH};
		\draw (464,125) node [anchor=north west][inner sep=0.75pt]   [align=left] {O};
		\draw (279,85) node [anchor=north west][inner sep=0.75pt]  [font=\small] [align=left] {$\displaystyle H_{1}$};
		\draw (364,90) node [anchor=north west][inner sep=0.75pt]  [font=\small] [align=left] {$\displaystyle H_{2}$};
		\draw (421,83) node [anchor=north west][inner sep=0.75pt]  [font=\small] [align=left] {$\displaystyle B_{2}$};
		\draw (369,140) node [anchor=north west][inner sep=0.75pt]  [font=\small] [align=left] {$\displaystyle \tilde{H}_{2}$};
		\draw (329,69) node [anchor=north west][inner sep=0.75pt]   [align=left] {F};
		\draw (276,116) node [anchor=north west][inner sep=0.75pt]   [align=left] {L};
		\draw (435,116) node [anchor=north west][inner sep=0.75pt]   [align=left] {R};
		\draw (436,154) node [anchor=north west][inner sep=0.75pt]  [font=\small] [align=left] {$\displaystyle \tilde{B}_{2}$};
		\draw (183,124) node [anchor=north west][inner sep=0.75pt]   [align=left] {O};
		\draw (329,159) node [anchor=north west][inner sep=0.75pt]   [align=left] {B};

	\end{tikzpicture}

	\caption{Alternate representation of Penrose Diagram of SdS after identifications.}
	\label{fig.altpensds}
\end{figure}
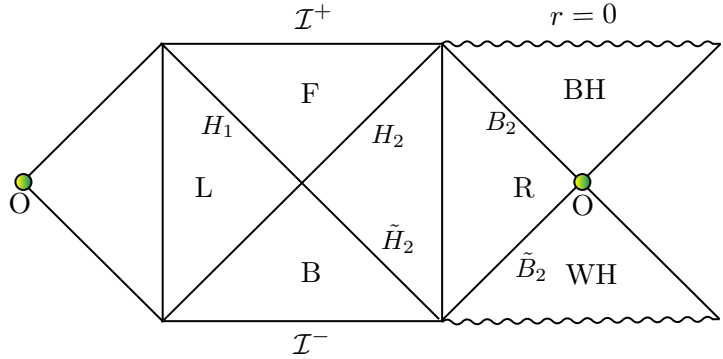

The region $r>1$ in eq.(\ref{metrica}) covers the forward Milne patch, $F$, as shown in fig. \ref{fig.pendsst}, with a boundary ${\cal I^+}$ at $r\rightarrow \infty$.
Continuing to $r<1$ gives the right static patch region $R$. Similarly analytic continuations cover the left Static patch $L$ and the backward Milne patch, $B$.  
It is manifestly clear from eq.(\ref{metrica}) that $\partial_t$ is a Killing vector, which is spacelike in the $F$ and $B$ regions and timelike in the $L,R$ static patches. 
The horizons, which we refer to as Cosmological horizons below, at $r=1$, separate the Milne  regions from the static patches. These are Killing horizons where $\partial_t$ becomes null.\\~\\
de Sitter space arises as a solution of the action
\be
\label{acta}
S={1\over 16 \pi G}\left[\int d^4x \sqrt{-g} ( R -2\Lambda )-2 \int_B d^3x \sqrt{h} K \right]
\ee
We will set $\Lambda=3$ in this paper. A non-rotating black hole geometry in asymptotic de Sitter space is given by 
\be
\label{nrmet}
ds^2=-{dr^2\over r^2-1+{2GM\over r}}+(r^2-1+{2GM\over r}) dt^2 +r^2 d\Omega_2^2
\ee
The Killing horizons are located at 
\be
\label{khloc}
r^3-r+2GM=0
\ee
We denote the larger
and smaller real roots   as $r_+$ and $r_-$ respectively. 
In the subsequent discussion we will sometimes refer to the parameter $M$, loosely, as the ``Mass" of the black hole. 
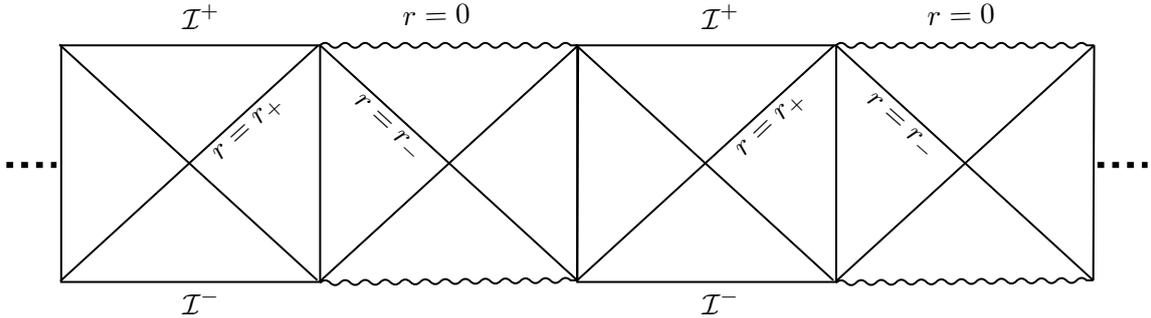
\begin{figure}[h]
	\centering
	
	\tikzset{every picture/.style={line width=0.75pt}} 
	
	\begin{tikzpicture}[x=0.75pt,y=0.75pt,yscale=-1,xscale=1]
		
		\draw   (96.33,160.95) -- (225.58,160.95) -- (225.58,279.97) -- (96.33,279.97) -- cycle ;
		\draw    (95.41,160.52) -- (224.19,279.12) ;
		\draw    (224.66,160.95) -- (95.87,279.55) ;
		\draw    (225.12,160.52) -- (353.91,279.12) ;
		\draw    (354.37,160.95) -- (225.58,279.55) ;
		\draw    (354.37,160.95) -- (353.91,279.12) ;
		\draw    (225.58,160.95) .. controls (227.25,159.28) and (228.91,159.28) .. (230.58,160.95) .. controls (232.25,162.62) and (233.91,162.62) .. (235.58,160.95) .. controls (237.25,159.28) and (238.91,159.28) .. (240.58,160.95) .. controls (242.25,162.62) and (243.91,162.62) .. (245.58,160.95) .. controls (247.25,159.28) and (248.91,159.28) .. (250.58,160.95) .. controls (252.25,162.62) and (253.91,162.62) .. (255.58,160.95) .. controls (257.25,159.28) and (258.91,159.28) .. (260.58,160.95) .. controls (262.25,162.62) and (263.91,162.62) .. (265.58,160.95) .. controls (267.25,159.28) and (268.91,159.28) .. (270.58,160.95) .. controls (272.25,162.62) and (273.91,162.62) .. (275.58,160.95) .. controls (277.25,159.28) and (278.91,159.28) .. (280.58,160.95) .. controls (282.25,162.62) and (283.91,162.62) .. (285.58,160.95) .. controls (287.25,159.28) and (288.91,159.28) .. (290.58,160.95) .. controls (292.25,162.62) and (293.91,162.62) .. (295.58,160.95) .. controls (297.25,159.28) and (298.91,159.28) .. (300.58,160.95) .. controls (302.25,162.62) and (303.91,162.62) .. (305.58,160.95) .. controls (307.25,159.28) and (308.91,159.28) .. (310.58,160.95) .. controls (312.25,162.62) and (313.91,162.62) .. (315.58,160.95) .. controls (317.25,159.28) and (318.91,159.28) .. (320.58,160.95) .. controls (322.25,162.62) and (323.91,162.62) .. (325.58,160.95) .. controls (327.25,159.28) and (328.91,159.28) .. (330.58,160.95) .. controls (332.25,162.62) and (333.91,162.62) .. (335.58,160.95) .. controls (337.25,159.28) and (338.91,159.28) .. (340.58,160.95) .. controls (342.25,162.62) and (343.91,162.62) .. (345.58,160.95) .. controls (347.25,159.28) and (348.91,159.28) .. (350.58,160.95) -- (354.37,160.95) -- (354.37,160.95) ;
		\draw    (225.58,280.4) .. controls (227.23,278.72) and (228.9,278.7) .. (230.58,280.35) .. controls (232.26,282) and (233.93,281.98) .. (235.58,280.3) .. controls (237.23,278.62) and (238.9,278.6) .. (240.58,280.25) .. controls (242.26,281.9) and (243.93,281.88) .. (245.58,280.2) .. controls (247.23,278.52) and (248.9,278.5) .. (250.58,280.15) .. controls (252.26,281.8) and (253.93,281.78) .. (255.58,280.1) .. controls (257.23,278.42) and (258.9,278.4) .. (260.58,280.05) .. controls (262.26,281.7) and (263.93,281.68) .. (265.58,280) .. controls (267.23,278.32) and (268.9,278.3) .. (270.58,279.95) .. controls (272.26,281.6) and (273.93,281.58) .. (275.58,279.9) .. controls (277.23,278.22) and (278.9,278.2) .. (280.58,279.85) .. controls (282.26,281.5) and (283.93,281.48) .. (285.58,279.8) .. controls (287.23,278.12) and (288.9,278.1) .. (290.58,279.75) .. controls (292.26,281.4) and (293.93,281.38) .. (295.58,279.7) .. controls (297.23,278.02) and (298.9,278) .. (300.58,279.65) .. controls (302.26,281.3) and (303.93,281.28) .. (305.58,279.6) .. controls (307.23,277.92) and (308.9,277.9) .. (310.58,279.55) .. controls (312.26,281.2) and (313.93,281.18) .. (315.58,279.5) .. controls (317.23,277.82) and (318.9,277.8) .. (320.58,279.45) .. controls (322.26,281.1) and (323.93,281.08) .. (325.58,279.4) .. controls (327.23,277.72) and (328.9,277.7) .. (330.58,279.35) .. controls (332.26,281) and (333.93,280.98) .. (335.58,279.3) .. controls (337.23,277.62) and (338.9,277.6) .. (340.58,279.25) .. controls (342.26,280.9) and (343.93,280.88) .. (345.58,279.2) .. controls (347.23,277.52) and (348.9,277.5) .. (350.58,279.15) -- (353.91,279.12) -- (353.91,279.12) ;
		\draw   (353.91,160.95) -- (483.16,160.95) -- (483.16,279.97) -- (353.91,279.97) -- cycle ;
		\draw    (352.98,160.52) -- (481.77,279.12) ;
		\draw    (482.24,160.95) -- (353.45,279.55) ;
		\draw    (482.7,160.52) -- (611.49,279.12) ;
		\draw    (611.95,160.95) -- (483.16,279.55) ;
		\draw    (483.16,160.95) .. controls (484.83,159.28) and (486.49,159.28) .. (488.16,160.95) .. controls (489.83,162.62) and (491.49,162.62) .. (493.16,160.95) .. controls (494.83,159.28) and (496.49,159.28) .. (498.16,160.95) .. controls (499.83,162.62) and (501.49,162.62) .. (503.16,160.95) .. controls (504.83,159.28) and (506.49,159.28) .. (508.16,160.95) .. controls (509.83,162.62) and (511.49,162.62) .. (513.16,160.95) .. controls (514.83,159.28) and (516.49,159.28) .. (518.16,160.95) .. controls (519.83,162.62) and (521.49,162.62) .. (523.16,160.95) .. controls (524.83,159.28) and (526.49,159.28) .. (528.16,160.95) .. controls (529.83,162.62) and (531.49,162.62) .. (533.16,160.95) .. controls (534.83,159.28) and (536.49,159.28) .. (538.16,160.95) .. controls (539.83,162.62) and (541.49,162.62) .. (543.16,160.95) .. controls (544.83,159.28) and (546.49,159.28) .. (548.16,160.95) .. controls (549.83,162.62) and (551.49,162.62) .. (553.16,160.95) .. controls (554.83,159.28) and (556.49,159.28) .. (558.16,160.95) .. controls (559.83,162.62) and (561.49,162.62) .. (563.16,160.95) .. controls (564.83,159.28) and (566.49,159.28) .. (568.16,160.95) .. controls (569.83,162.62) and (571.49,162.62) .. (573.16,160.95) .. controls (574.83,159.28) and (576.49,159.28) .. (578.16,160.95) .. controls (579.83,162.62) and (581.49,162.62) .. (583.16,160.95) .. controls (584.83,159.28) and (586.49,159.28) .. (588.16,160.95) .. controls (589.83,162.62) and (591.49,162.62) .. (593.16,160.95) .. controls (594.83,159.28) and (596.49,159.28) .. (598.16,160.95) .. controls (599.83,162.62) and (601.49,162.62) .. (603.16,160.95) .. controls (604.83,159.28) and (606.49,159.28) .. (608.16,160.95) -- (611.95,160.95) -- (611.95,160.95) ;
		\draw    (483.16,280.4) .. controls (484.81,278.72) and (486.48,278.7) .. (488.16,280.35) .. controls (489.84,282) and (491.51,281.98) .. (493.16,280.3) .. controls (494.81,278.62) and (496.48,278.6) .. (498.16,280.25) .. controls (499.84,281.9) and (501.51,281.88) .. (503.16,280.2) .. controls (504.81,278.52) and (506.48,278.5) .. (508.16,280.15) .. controls (509.84,281.8) and (511.51,281.78) .. (513.16,280.1) .. controls (514.81,278.42) and (516.48,278.4) .. (518.16,280.05) .. controls (519.84,281.7) and (521.51,281.68) .. (523.16,280) .. controls (524.81,278.32) and (526.48,278.3) .. (528.16,279.95) .. controls (529.84,281.6) and (531.51,281.58) .. (533.16,279.9) .. controls (534.81,278.22) and (536.48,278.2) .. (538.16,279.85) .. controls (539.84,281.5) and (541.51,281.48) .. (543.16,279.8) .. controls (544.81,278.12) and (546.48,278.1) .. (548.16,279.75) .. controls (549.84,281.4) and (551.51,281.38) .. (553.16,279.7) .. controls (554.81,278.02) and (556.48,278) .. (558.16,279.65) .. controls (559.84,281.3) and (561.51,281.28) .. (563.16,279.6) .. controls (564.81,277.92) and (566.48,277.9) .. (568.16,279.55) .. controls (569.84,281.2) and (571.51,281.18) .. (573.16,279.5) .. controls (574.81,277.82) and (576.48,277.8) .. (578.16,279.45) .. controls (579.84,281.1) and (581.51,281.08) .. (583.16,279.4) .. controls (584.81,277.72) and (586.48,277.7) .. (588.16,279.35) .. controls (589.84,281) and (591.51,280.98) .. (593.16,279.3) .. controls (594.81,277.62) and (596.48,277.6) .. (598.16,279.25) .. controls (599.84,280.9) and (601.51,280.88) .. (603.16,279.2) .. controls (604.81,277.52) and (606.48,277.5) .. (608.16,279.15) -- (611.49,279.12) -- (611.49,279.12) ;
		\draw    (611.95,160.95) -- (611.83,190.9) -- (611.49,279.12) ;
		\draw [line width=2.25]  [dash pattern={on 2.53pt off 3.02pt}]  (69,221.1) -- (94.02,221.1) ;
		\draw [line width=2.25]  [dash pattern={on 2.53pt off 3.02pt}]  (613.58,221.1) -- (638.6,221.1) ;
		
		\draw (155.36,138.84) node [anchor=north west][inner sep=0.75pt]   [align=left] {$\displaystyle \mathcal{I}^{+}$};
		\draw (414.79,138.84) node [anchor=north west][inner sep=0.75pt]   [align=left] {$\displaystyle \mathcal{I}^{+}$};
		\draw (155.36,282.36) node [anchor=north west][inner sep=0.75pt]   [align=left] {$\displaystyle \mathcal{I}^{-}$};
		\draw (414.79,282.36) node [anchor=north west][inner sep=0.75pt]   [align=left] {$\displaystyle \mathcal{I}^{-}$};
		\draw (169.25,214.15) node [anchor=north west][inner sep=0.75pt]  [rotate=-317.49] [align=left] {$\displaystyle r=r_{+}$};
		\draw (246.33,182.69) node [anchor=north west][inner sep=0.75pt]  [rotate=-42.81] [align=left] {$\displaystyle r=r_{-}$};
		\draw (430.46,212.67) node [anchor=north west][inner sep=0.75pt]  [rotate=-316.19] [align=left] {$\displaystyle r=r_{+}$};
		\draw (501.87,181.92) node [anchor=north west][inner sep=0.75pt]  [rotate=-41.59] [align=left] {$\displaystyle r=r_{-}$};
		\draw (265.01,139.98) node [anchor=north west][inner sep=0.75pt]  [rotate=-358.18] [align=left] {$\displaystyle r=0$};
		\draw (527.01,139.98) node [anchor=north west][inner sep=0.75pt]  [rotate=-358.18] [align=left] {$\displaystyle r=0$};

	\end{tikzpicture}
	
	\caption{Extended Penrose Diagram of Schwarzschild Black Hole in de Sitter space (SdS).}
	\label{fig.pensdsext}
\end{figure}
After analytically continuing the coordinates in eq.(\ref{nrmet}) one finds  a Penrose diagram consisting of an infinite sequence of space-like infinities, where $r\rightarrow \infty$, and singularities where $r\rightarrow 0$, fig. \ref{fig.pensdsext}. The solution has cosmological horizons, at $r=r_+$,  which separate Milne patch regions from static patches, and black hole horizons,
at $r=r_-$,  which separate black hole regions, with trapped or anti trapped surfaces, from static patch regions. After appropriate identifications the infinite sequence truncates to a spacetime with one future and past  Milne region, where asymptotically $r\rightarrow \infty$,  and one black hole  and one white  hole region, containing singularities at $r=0$, as shown in fig. \ref{fig.idpensds}. Equivalently this spacetime can be depicted as shown in fig. \ref{fig.altpensds}. 


For our purposes we will mostly be interested in the description of a future Milne patch $F$, in fig. \ref{fig.altpensds} bounded by two future cosmological horizons, denoted by $H_1, H_2$  and sometimes in a static patch region, e.g., $R$,  bounded, in the future,  by a   Cosmological  and black hole horizon, $H_2, B_2$, and  in the past by the Cosmological and Black Holes horizons,  ${\tilde H}_2, {\tilde B}_2$. 

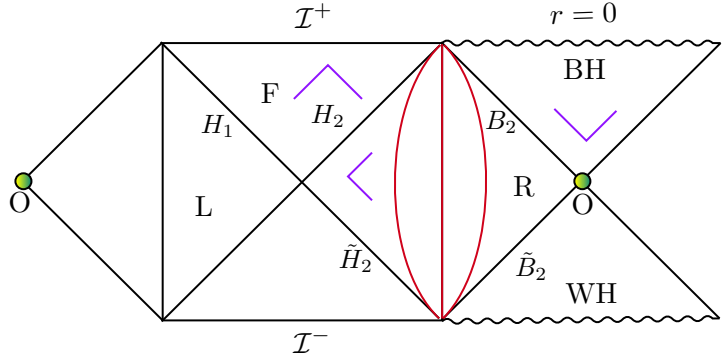
\begin{figure}[h]
	\centering

	
	\tikzset{
		pattern size/.store in=\mcSize, 
		pattern size = 5pt,
		pattern thickness/.store in=\mcThickness, 
		pattern thickness = 0.3pt,
		pattern radius/.store in=\mcRadius, 
		pattern radius = 1pt}
	\makeatletter
	\pgfutil@ifundefined{pgf@pattern@name@_npvi0fc5x}{
		\pgfdeclarepatternformonly[\mcThickness,\mcSize]{_npvi0fc5x}
		{\pgfqpoint{0pt}{0pt}}
		{\pgfpoint{\mcSize+\mcThickness}{\mcSize+\mcThickness}}
		{\pgfpoint{\mcSize}{\mcSize}}
		{
			\pgfsetcolor{\tikz@pattern@color}
			\pgfsetlinewidth{\mcThickness}
			\pgfpathmoveto{\pgfqpoint{0pt}{0pt}}
			\pgfpathlineto{\pgfpoint{\mcSize+\mcThickness}{\mcSize+\mcThickness}}
			\pgfusepath{stroke}
	}}
	\makeatother
	
	
	\tikzset{
		pattern size/.store in=\mcSize, 
		pattern size = 5pt,
		pattern thickness/.store in=\mcThickness, 
		pattern thickness = 0.3pt,
		pattern radius/.store in=\mcRadius, 
		pattern radius = 1pt}
	\makeatletter
	\pgfutil@ifundefined{pgf@pattern@name@_4zds0niq5}{
		\pgfdeclarepatternformonly[\mcThickness,\mcSize]{_4zds0niq5}
		{\pgfqpoint{0pt}{-\mcThickness}}
		{\pgfpoint{\mcSize}{\mcSize}}
		{\pgfpoint{\mcSize}{\mcSize}}
		{
			\pgfsetcolor{\tikz@pattern@color}
			\pgfsetlinewidth{\mcThickness}
			\pgfpathmoveto{\pgfqpoint{0pt}{\mcSize}}
			\pgfpathlineto{\pgfpoint{\mcSize+\mcThickness}{-\mcThickness}}
			\pgfusepath{stroke}
	}}
	\makeatother
	
	
	\tikzset {_a7bwzw0ch/.code = {\pgfsetadditionalshadetransform{ \pgftransformshift{\pgfpoint{0 bp } { 0 bp }  }  \pgftransformrotate{0 }  \pgftransformscale{2 }  }}}
	\pgfdeclarehorizontalshading{_a1hv0y66z}{150bp}{rgb(0bp)=(1,1,0);
		rgb(37.5bp)=(1,1,0);
		rgb(62.5bp)=(0,0.5,0.5);
		rgb(100bp)=(0,0.5,0.5)}
	
	
	\tikzset {_agynccy4j/.code = {\pgfsetadditionalshadetransform{ \pgftransformshift{\pgfpoint{0 bp } { 0 bp }  }  \pgftransformrotate{0 }  \pgftransformscale{2 }  }}}
	\pgfdeclarehorizontalshading{_yon8pgjw6}{150bp}{rgb(0bp)=(1,1,0);
		rgb(37.5bp)=(1,1,0);
		rgb(62.5bp)=(0,0.5,0.5);
		rgb(100bp)=(0,0.5,0.5)}
	\tikzset{every picture/.style={line width=0.75pt}} 
	
	\begin{tikzpicture}[x=0.75pt,y=0.75pt,yscale=-1,xscale=1]
		
		\draw   (261.5,50.5) -- (401,50.5) -- (401,190) -- (261.5,190) -- cycle ;
		\draw    (260.5,50) -- (399.5,189) ;
		\draw    (400,51.5) -- (261,190.5) ;
		\draw    (400.5,50) -- (539.5,189) ;
		\draw    (540,50.5) -- (401,189.5) ;
		\draw    (401,50.5) .. controls (402.67,48.83) and (404.33,48.83) .. (406,50.5) .. controls (407.67,52.17) and (409.33,52.17) .. (411,50.5) .. controls (412.67,48.83) and (414.33,48.83) .. (416,50.5) .. controls (417.67,52.17) and (419.33,52.17) .. (421,50.5) .. controls (422.67,48.83) and (424.33,48.83) .. (426,50.5) .. controls (427.67,52.17) and (429.33,52.17) .. (431,50.5) .. controls (432.67,48.83) and (434.33,48.83) .. (436,50.5) .. controls (437.67,52.17) and (439.33,52.17) .. (441,50.5) .. controls (442.67,48.83) and (444.33,48.83) .. (446,50.5) .. controls (447.67,52.17) and (449.33,52.17) .. (451,50.5) .. controls (452.67,48.83) and (454.33,48.83) .. (456,50.5) .. controls (457.67,52.17) and (459.33,52.17) .. (461,50.5) .. controls (462.67,48.83) and (464.33,48.83) .. (466,50.5) .. controls (467.67,52.17) and (469.33,52.17) .. (471,50.5) .. controls (472.67,48.83) and (474.33,48.83) .. (476,50.5) .. controls (477.67,52.17) and (479.33,52.17) .. (481,50.5) .. controls (482.67,48.83) and (484.33,48.83) .. (486,50.5) .. controls (487.67,52.17) and (489.33,52.17) .. (491,50.5) .. controls (492.67,48.83) and (494.33,48.83) .. (496,50.5) .. controls (497.67,52.17) and (499.33,52.17) .. (501,50.5) .. controls (502.67,48.83) and (504.33,48.83) .. (506,50.5) .. controls (507.67,52.17) and (509.33,52.17) .. (511,50.5) .. controls (512.67,48.83) and (514.33,48.83) .. (516,50.5) .. controls (517.67,52.17) and (519.33,52.17) .. (521,50.5) .. controls (522.67,48.83) and (524.33,48.83) .. (526,50.5) .. controls (527.67,52.17) and (529.33,52.17) .. (531,50.5) .. controls (532.67,48.83) and (534.33,48.83) .. (536,50.5) -- (540,50.5) -- (540,50.5) ;
		\draw    (401,190.5) .. controls (402.65,188.82) and (404.32,188.8) .. (406,190.45) .. controls (407.69,192.1) and (409.35,192.08) .. (411,190.39) .. controls (412.65,188.71) and (414.32,188.69) .. (416,190.34) .. controls (417.69,191.99) and (419.35,191.97) .. (421,190.28) .. controls (422.65,188.6) and (424.32,188.58) .. (426,190.23) .. controls (427.68,191.88) and (429.35,191.86) .. (431,190.18) .. controls (432.65,188.49) and (434.31,188.47) .. (436,190.12) .. controls (437.68,191.77) and (439.35,191.75) .. (441,190.07) .. controls (442.65,188.38) and (444.31,188.36) .. (446,190.01) .. controls (447.68,191.66) and (449.35,191.64) .. (451,189.96) .. controls (452.65,188.27) and (454.31,188.25) .. (456,189.9) .. controls (457.68,191.55) and (459.35,191.53) .. (461,189.85) .. controls (462.65,188.17) and (464.32,188.15) .. (466,189.8) .. controls (467.69,191.45) and (469.35,191.43) .. (471,189.74) .. controls (472.65,188.06) and (474.32,188.04) .. (476,189.69) .. controls (477.69,191.34) and (479.35,191.32) .. (481,189.63) .. controls (482.65,187.95) and (484.32,187.93) .. (486,189.58) .. controls (487.68,191.23) and (489.34,191.21) .. (490.99,189.53) .. controls (492.64,187.84) and (494.3,187.82) .. (495.99,189.47) .. controls (497.67,191.12) and (499.34,191.1) .. (500.99,189.42) .. controls (502.64,187.73) and (504.3,187.71) .. (505.99,189.36) .. controls (507.67,191.01) and (509.34,190.99) .. (510.99,189.31) .. controls (512.64,187.62) and (514.3,187.6) .. (515.99,189.25) .. controls (517.67,190.9) and (519.34,190.88) .. (520.99,189.2) .. controls (522.64,187.52) and (524.31,187.5) .. (525.99,189.15) .. controls (527.68,190.8) and (529.34,190.78) .. (530.99,189.09) .. controls (532.64,187.41) and (534.31,187.39) .. (535.99,189.04) -- (539.5,189) -- (539.5,189) ;
		\draw [pattern=_npvi0fc5x,pattern size=6pt,pattern thickness=0.75pt,pattern radius=0pt, pattern color={rgb, 255:red, 65; green, 117; blue, 5}]   (192.25,120.75) -- (261.5,190) ;
		\draw [pattern=_4zds0niq5,pattern size=6pt,pattern thickness=0.75pt,pattern radius=0pt, pattern color={rgb, 255:red, 0; green, 0; blue, 0}]   (261,50.5) -- (190.9,120.6) ;
		\draw [color={rgb, 255:red, 208; green, 2; blue, 27 }  ,draw opacity=1 ]   (401,51) -- (401,189) ;
		\draw [color={rgb, 255:red, 208; green, 2; blue, 27 }  ,draw opacity=1 ]   (400,51.5) .. controls (370,81.6) and (370,160.6) .. (399.5,189) ;
		\draw [color={rgb, 255:red, 208; green, 2; blue, 27 }  ,draw opacity=1 ]   (402,51.5) .. controls (430,80.6) and (430,159.6) .. (401.5,189) ;
		\draw [color={rgb, 255:red, 144; green, 19; blue, 254 }  ,draw opacity=1 ]   (327,78.6) -- (344,61.6) -- (361,78.6) ;
		\draw [color={rgb, 255:red, 144; green, 19; blue, 254 }  ,draw opacity=1 ]   (366,105.6) -- (354,117.6) -- (366,129.6) ;
		\draw [color={rgb, 255:red, 144; green, 19; blue, 254 }  ,draw opacity=1 ]   (457,83.6) -- (473,99.6) -- (488,84.6) ;
		\path  [shading=_a1hv0y66z,_a7bwzw0ch] (188,120) .. controls (188,117.79) and (189.79,116) .. (192,116) .. controls (194.21,116) and (196,117.79) .. (196,120) .. controls (196,122.21) and (194.21,124) .. (192,124) .. controls (189.79,124) and (188,122.21) .. (188,120) -- cycle ; 
		\draw   (188,120) .. controls (188,117.79) and (189.79,116) .. (192,116) .. controls (194.21,116) and (196,117.79) .. (196,120) .. controls (196,122.21) and (194.21,124) .. (192,124) .. controls (189.79,124) and (188,122.21) .. (188,120) -- cycle ; 
		
		\path  [shading=_yon8pgjw6,_agynccy4j] (467,120) .. controls (467,117.79) and (468.79,116) .. (471,116) .. controls (473.21,116) and (475,117.79) .. (475,120) .. controls (475,122.21) and (473.21,124) .. (471,124) .. controls (468.79,124) and (467,122.21) .. (467,120) -- cycle ; 
		\draw   (467,120) .. controls (467,117.79) and (468.79,116) .. (471,116) .. controls (473.21,116) and (475,117.79) .. (475,120) .. controls (475,122.21) and (473.21,124) .. (471,124) .. controls (468.79,124) and (467,122.21) .. (467,120) -- cycle ; 

		\draw (326.36,28.84) node [anchor=north west][inner sep=0.75pt]   [align=left] {$\displaystyle \mathcal{I}^{+}$};
		\draw (325.36,192.36) node [anchor=north west][inner sep=0.75pt]   [align=left] {$\displaystyle \mathcal{I}^{-}$};
		\draw (453.01,29.98) node [anchor=north west][inner sep=0.75pt]  [rotate=-358.18] [align=left] {$\displaystyle r=0$};
		\draw (460,57) node [anchor=north west][inner sep=0.75pt]   [align=left] {BH};
		\draw (461,170) node [anchor=north west][inner sep=0.75pt]   [align=left] {WH};
		\draw (464,125) node [anchor=north west][inner sep=0.75pt]   [align=left] {O};
		\draw (279,85) node [anchor=north west][inner sep=0.75pt]  [font=\small] [align=left] {$\displaystyle H_{1}$};
		\draw (334,80) node [anchor=north west][inner sep=0.75pt]  [font=\small] [align=left] {$\displaystyle H_{2}$};
		\draw (421,83) node [anchor=north west][inner sep=0.75pt]  [font=\small] [align=left] {$\displaystyle B_{2}$};
		\draw (348,150) node [anchor=north west][inner sep=0.75pt]  [font=\small] [align=left] {$\displaystyle \tilde{H}_{2}$};
		\draw (309,69) node [anchor=north west][inner sep=0.75pt]   [align=left] {F};
		\draw (276,126) node [anchor=north west][inner sep=0.75pt]   [align=left] {L};
		\draw (435,116) node [anchor=north west][inner sep=0.75pt]   [align=left] {R};
		\draw (436,154) node [anchor=north west][inner sep=0.75pt]  [font=\small] [align=left] {$\displaystyle \tilde{B}_{2}$};
		\draw (183,124) node [anchor=north west][inner sep=0.75pt]   [align=left] {O};

	\end{tikzpicture}

	\caption{Worldlines and Light-sheets in SdS geometry. Red lines denote worldlines of observers at constant $r$. Violet lines denote light-sheets.}
	\label{fig.lshsds}
\end{figure}

As mentioned   the solutions considered above have $\partial_t$ as   a Killing vector and we will sometimes  refer to  them  as ``stationary"  solutions below\footnote{We hope this is not confusing, the space-times we are considering are   of course evolving in time and have no globally defined time-like Killing vector.  The Killing vector $\partial_t$ is only time-like in the static patches.}  
We will also be interested in more general situations where the isometry generated by $\partial_t$ is broken, which we  will  refer to  as ``non-stationary" solutions.
One can  define  cosmological and black hole horizons  in this more general context as follows. One considers a time like observer $\gamma$; more precisely, an  inextendible time like curve; and then considers the chronological past of $\gamma$, $I(\gamma)$.
The boundary of $I(\gamma)$  then defines a future event horizon. It is the boundary of  the region of spacetime from which the observer  can receive a signal.
This definition can be extended to a family of observers.  Past event horizons can be similarly defined by taking the boundary of the chronological future of an observer or a family of observers.  In the example above of a black hole in dS, consider the family of observers moving along constant $r$ in the static patch R,  with $r_-<r<r_+$, as shown in fig. \ref{fig.lshsds}. The cosmological and black hole horizons are event horizons for these observers, with $H_2, B_2$ being the future and ${\tilde H}_2,{\tilde B}_2$ being the past horizons. 

Another way to think about horizons is by considering light sheets in spacetime \cite{Bousso1999}. 
A light sheet is a surface, with non-increasing area,  generated  by  null rays which emanate from  a codimension-2 surface in spacetime. In  the  static patch $R$  of the stationary solution, fig. \ref{fig.lshsds}, let us  consider light sheets emanating from spherical symmetric surfaces, at constant $t$ and $r$. It is easy to see that there is one light sheet of this type which is forward directed in time and one which is backward directed. However, in the forward Milne patch, $F$, fig. \ref{fig.lshsds}, both light sheets associated with a spherically symmetric surface are easily seen to be  backward directed, since the accelerated expansion of the universe makes the  forward directed null congruences  to have an expanding area.
The future cosmological horizon $H_2$ can then be defined to be  the boundary between these two regions. Similarly, in the black hole region `BH', fig. \ref{fig.lshsds}, both light sheets emanating from a spherically symmetric surface must be future directed. And the future black hole event horizon $B_2$  can be defined to be  the boundary between the static patch, with one forward and one backward directed light sheet,  and the black hole region, with two future directed light sheets. Past cosmological or black hole horizons can also be defined in a similar fashion. In non-stationary cases, which eventually ``settle down" to a stationary solution, one can use a similar criterion to distinguish, for example, the future cosmological horizon from the future black hole horizon, by considering light sheets in the far future. 


The above discussion can be extended to rotating and charged black holes. The metric, in the presence of rotation represented by parameter $a$, takes the form in Boyer-Lindquist coordinates as, 

\begin{align}
	ds^2=&-\left({\Delta_r-\Delta_\theta a^2\sin^2\theta\over \sigma}\right)dt^2-{2a\sin^2\theta\over\sigma\zeta}(\Delta_\theta(r^2+a^2)-\Delta_r)dtd\phi\nonumber\\&+{\sin^2\theta\over \sigma\zeta^2}(\Delta_\theta(r^2+a^2)^2-\Delta_ra^2\sin^2\theta)d\phi^2+{\sigma\over \Delta_r}dr^2+{\sigma\over\Delta_\theta}d\theta^2  \label{rotmet}
\end{align}
where 
\begin{equation}
	\label{not1}
	\Delta_r=r^2+a^2-2GMr-{r^2}(r^2+a^2)~~,~~\Delta_\theta=1+{ a^2}\cos^2\theta~~,~~\sigma=r^2+a^2\cos^2\theta~~,~~\zeta=1+{a^2}
\end{equation}
To note, the parameter $a$ is related to $J$, referred loosely as the ``angular momentum" of the black hole, as
\begin{equation}
	\label{J}
	J=aM/\zeta^2
\end{equation}

Now there are three types of horizons,  corresponding to the three values of $r$, where $\Delta_r$ vanishes, satisfying
\begin{equation}
	\label{kerhoreq}
	r^4+(a^2-1)r^2+2GMr-a^2=0
\end{equation}
The solutions of \eq{kerhoreq} are given by $r_c, r_+, r_-$, which denote the locations of cosmological, black hole's outer and inner horizons respectively. Going to the region $r>r_c$ one reaches asymptotic space-like infinity ${\cal I}^+ $ or  ${\cal I}^-$. Going to the region $r<r_-$ one hits a singularity at $r=0,~ \theta={\pi\over 2}$ . The static patch in the region $r_+<r<r_c$. 
More discussion on these solutions can be found in appendix \ref{app.kdsprop}. 

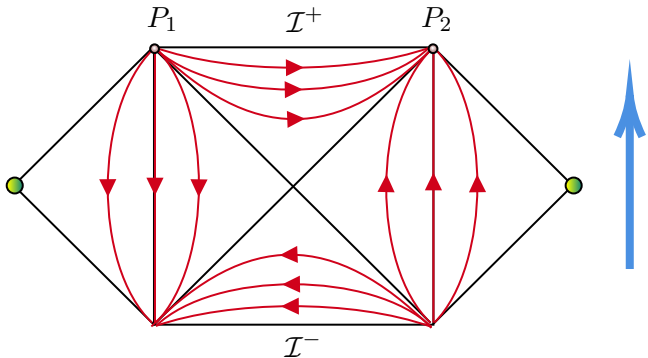
\begin{figure}[h]
	\centering

	
	\tikzset{
		pattern size/.store in=\mcSize, 
		pattern size = 5pt,
		pattern thickness/.store in=\mcThickness, 
		pattern thickness = 0.3pt,
		pattern radius/.store in=\mcRadius, 
		pattern radius = 1pt}
	\makeatletter
	\pgfutil@ifundefined{pgf@pattern@name@_vdrvvtuvg}{
		\pgfdeclarepatternformonly[\mcThickness,\mcSize]{_vdrvvtuvg}
		{\pgfqpoint{0pt}{0pt}}
		{\pgfpoint{\mcSize+\mcThickness}{\mcSize+\mcThickness}}
		{\pgfpoint{\mcSize}{\mcSize}}
		{
			\pgfsetcolor{\tikz@pattern@color}
			\pgfsetlinewidth{\mcThickness}
			\pgfpathmoveto{\pgfqpoint{0pt}{0pt}}
			\pgfpathlineto{\pgfpoint{\mcSize+\mcThickness}{\mcSize+\mcThickness}}
			\pgfusepath{stroke}
	}}
	\makeatother
	
	
	\tikzset{
		pattern size/.store in=\mcSize, 
		pattern size = 5pt,
		pattern thickness/.store in=\mcThickness, 
		pattern thickness = 0.3pt,
		pattern radius/.store in=\mcRadius, 
		pattern radius = 1pt}
	\makeatletter
	\pgfutil@ifundefined{pgf@pattern@name@_z2xwr3uji}{
		\pgfdeclarepatternformonly[\mcThickness,\mcSize]{_z2xwr3uji}
		{\pgfqpoint{0pt}{-\mcThickness}}
		{\pgfpoint{\mcSize}{\mcSize}}
		{\pgfpoint{\mcSize}{\mcSize}}
		{
			\pgfsetcolor{\tikz@pattern@color}
			\pgfsetlinewidth{\mcThickness}
			\pgfpathmoveto{\pgfqpoint{0pt}{\mcSize}}
			\pgfpathlineto{\pgfpoint{\mcSize+\mcThickness}{-\mcThickness}}
			\pgfusepath{stroke}
	}}
	\makeatother
	
	
	\tikzset {_gygoz65h0/.code = {\pgfsetadditionalshadetransform{ \pgftransformshift{\pgfpoint{0 bp } { 0 bp }  }  \pgftransformrotate{0 }  \pgftransformscale{2 }  }}}
	\pgfdeclarehorizontalshading{_8g3i2lqer}{150bp}{rgb(0bp)=(1,1,0);
		rgb(37.5bp)=(1,1,0);
		rgb(62.5bp)=(0,0.5,0.5);
		rgb(100bp)=(0,0.5,0.5)}
	
	
	\tikzset {_09sl95m34/.code = {\pgfsetadditionalshadetransform{ \pgftransformshift{\pgfpoint{0 bp } { 0 bp }  }  \pgftransformrotate{0 }  \pgftransformscale{2 }  }}}
	\pgfdeclarehorizontalshading{_jp288yq17}{150bp}{rgb(0bp)=(1,1,0);
		rgb(37.5bp)=(1,1,0);
		rgb(62.5bp)=(0,0.5,0.5);
		rgb(100bp)=(0,0.5,0.5)}
	
	
	\tikzset {_rh3rqtl5z/.code = {\pgfsetadditionalshadetransform{ \pgftransformshift{\pgfpoint{89.1 bp } { -128.7 bp }  }  \pgftransformscale{1.32 }  }}}
	\pgfdeclareradialshading{_tnpihuzf4}{\pgfpoint{-72bp}{104bp}}{rgb(0bp)=(1,1,1);
		rgb(0bp)=(1,1,1);
		rgb(25bp)=(0.48,0.15,0.15);
		rgb(400bp)=(0.48,0.15,0.15)}
	
	
	\tikzset {_g0uoxhsor/.code = {\pgfsetadditionalshadetransform{ \pgftransformshift{\pgfpoint{89.1 bp } { -128.7 bp }  }  \pgftransformscale{1.32 }  }}}
	\pgfdeclareradialshading{_i9jm55v38}{\pgfpoint{-72bp}{104bp}}{rgb(0bp)=(1,1,1);
		rgb(0bp)=(1,1,1);
		rgb(25bp)=(0.48,0.15,0.15);
		rgb(400bp)=(0.48,0.15,0.15)}
	\tikzset{every picture/.style={line width=0.75pt}} 
	
	\begin{tikzpicture}[x=0.75pt,y=0.75pt,yscale=-1,xscale=1]
		
		\draw   (261.5,50.5) -- (401,50.5) -- (401,190) -- (261.5,190) -- cycle ;
		\draw    (260.5,50) -- (399.5,189) ;
		\draw    (400,51.5) -- (261,190.5) ;
		\draw    (400.5,50) -- (471,120.5) ;
		\draw    (471,119.5) -- (401,189.5) ;
		\draw [pattern=_vdrvvtuvg,pattern size=6pt,pattern thickness=0.75pt,pattern radius=0pt, pattern color={rgb, 255:red, 65; green, 117; blue, 5}]   (192.25,120.75) -- (261.5,190) ;
		\draw [pattern=_z2xwr3uji,pattern size=6pt,pattern thickness=0.75pt,pattern radius=0pt, pattern color={rgb, 255:red, 0; green, 0; blue, 0}]   (261,50.5) -- (190.9,120.6) ;
		\draw [color={rgb, 255:red, 208; green, 2; blue, 27 }  ,draw opacity=1 ]   (401,51) -- (401,189) ;
		\draw [shift={(401,113.5)}, rotate = 90] [fill={rgb, 255:red, 208; green, 2; blue, 27 }  ,fill opacity=1 ][line width=0.08]  [draw opacity=0] (8.93,-4.29) -- (0,0) -- (8.93,4.29) -- cycle    ;
		\draw [color={rgb, 255:red, 208; green, 2; blue, 27 }  ,draw opacity=1 ]   (400,51.5) .. controls (370,81.6) and (370,160.6) .. (399.5,189) ;
		\draw [shift={(377.6,114.39)}, rotate = 91.55] [fill={rgb, 255:red, 208; green, 2; blue, 27 }  ,fill opacity=1 ][line width=0.08]  [draw opacity=0] (8.93,-4.29) -- (0,0) -- (8.93,4.29) -- cycle    ;
		\draw [color={rgb, 255:red, 208; green, 2; blue, 27 }  ,draw opacity=1 ]   (402,51.5) .. controls (430,80.6) and (430,159.6) .. (401.5,189) ;
		\draw [shift={(422.85,114.46)}, rotate = 89.09] [fill={rgb, 255:red, 208; green, 2; blue, 27 }  ,fill opacity=1 ][line width=0.08]  [draw opacity=0] (8.93,-4.29) -- (0,0) -- (8.93,4.29) -- cycle    ;
		\path  [shading=_8g3i2lqer,_gygoz65h0] (188,120) .. controls (188,117.79) and (189.79,116) .. (192,116) .. controls (194.21,116) and (196,117.79) .. (196,120) .. controls (196,122.21) and (194.21,124) .. (192,124) .. controls (189.79,124) and (188,122.21) .. (188,120) -- cycle ; 
		\draw   (188,120) .. controls (188,117.79) and (189.79,116) .. (192,116) .. controls (194.21,116) and (196,117.79) .. (196,120) .. controls (196,122.21) and (194.21,124) .. (192,124) .. controls (189.79,124) and (188,122.21) .. (188,120) -- cycle ; 
		
		\path  [shading=_jp288yq17,_09sl95m34] (467,120) .. controls (467,117.79) and (468.79,116) .. (471,116) .. controls (473.21,116) and (475,117.79) .. (475,120) .. controls (475,122.21) and (473.21,124) .. (471,124) .. controls (468.79,124) and (467,122.21) .. (467,120) -- cycle ; 
		\draw   (467,120) .. controls (467,117.79) and (468.79,116) .. (471,116) .. controls (473.21,116) and (475,117.79) .. (475,120) .. controls (475,122.21) and (473.21,124) .. (471,124) .. controls (468.79,124) and (467,122.21) .. (467,120) -- cycle ; 
		
		\draw [color={rgb, 255:red, 208; green, 2; blue, 27 }  ,draw opacity=1 ]   (262,51) -- (262,189) ;
		\draw [shift={(262,125)}, rotate = 270] [fill={rgb, 255:red, 208; green, 2; blue, 27 }  ,fill opacity=1 ][line width=0.08]  [draw opacity=0] (8.93,-4.29) -- (0,0) -- (8.93,4.29) -- cycle    ;
		\draw [color={rgb, 255:red, 208; green, 2; blue, 27 }  ,draw opacity=1 ]   (261,51.5) .. controls (231,81.6) and (231,160.6) .. (260.5,189) ;
		\draw [shift={(238.51,125.75)}, rotate = 269.5] [fill={rgb, 255:red, 208; green, 2; blue, 27 }  ,fill opacity=1 ][line width=0.08]  [draw opacity=0] (8.93,-4.29) -- (0,0) -- (8.93,4.29) -- cycle    ;
		\draw [color={rgb, 255:red, 208; green, 2; blue, 27 }  ,draw opacity=1 ]   (263,51.5) .. controls (291,80.6) and (291,159.6) .. (262.5,189) ;
		\draw [shift={(283.82,125.82)}, rotate = 271.03] [fill={rgb, 255:red, 208; green, 2; blue, 27 }  ,fill opacity=1 ][line width=0.08]  [draw opacity=0] (8.93,-4.29) -- (0,0) -- (8.93,4.29) -- cycle    ;
		\draw [color={rgb, 255:red, 208; green, 2; blue, 27 }  ,draw opacity=1 ]   (401,50.5) .. controls (353,98.6) and (310.19,98.91) .. (261.19,49.81) ;
		\draw [shift={(336.95,86.38)}, rotate = 177.68] [fill={rgb, 255:red, 208; green, 2; blue, 27 }  ,fill opacity=1 ][line width=0.08]  [draw opacity=0] (8.93,-4.29) -- (0,0) -- (8.93,4.29) -- cycle    ;
		\draw [color={rgb, 255:red, 208; green, 2; blue, 27 }  ,draw opacity=1 ]   (398.69,50.81) .. controls (369.59,78.81) and (290.59,78.81) .. (261.19,50.31) ;
		\draw [shift={(335.73,71.66)}, rotate = 179.09] [fill={rgb, 255:red, 208; green, 2; blue, 27 }  ,fill opacity=1 ][line width=0.08]  [draw opacity=0] (8.93,-4.29) -- (0,0) -- (8.93,4.29) -- cycle    ;
		\draw [color={rgb, 255:red, 208; green, 2; blue, 27 }  ,draw opacity=1 ]   (398.69,50.81) .. controls (362,63.6) and (301,64.6) .. (261.19,50.31) ;
		\draw [shift={(336.31,60.65)}, rotate = 179.39] [fill={rgb, 255:red, 208; green, 2; blue, 27 }  ,fill opacity=1 ][line width=0.08]  [draw opacity=0] (8.93,-4.29) -- (0,0) -- (8.93,4.29) -- cycle    ;
		\draw [color={rgb, 255:red, 208; green, 2; blue, 27 }  ,draw opacity=1 ]   (260.19,190.92) .. controls (308.19,142.82) and (351,142.51) .. (400,191.61) ;
		\draw [shift={(324.24,155.03)}, rotate = 357.68] [fill={rgb, 255:red, 208; green, 2; blue, 27 }  ,fill opacity=1 ][line width=0.08]  [draw opacity=0] (8.93,-4.29) -- (0,0) -- (8.93,4.29) -- cycle    ;
		\draw [color={rgb, 255:red, 208; green, 2; blue, 27 }  ,draw opacity=1 ]   (262.5,190.61) .. controls (291.6,162.61) and (370.6,162.61) .. (400,191.11) ;
		\draw [shift={(325.46,169.76)}, rotate = 359.09] [fill={rgb, 255:red, 208; green, 2; blue, 27 }  ,fill opacity=1 ][line width=0.08]  [draw opacity=0] (8.93,-4.29) -- (0,0) -- (8.93,4.29) -- cycle    ;
		\draw [color={rgb, 255:red, 208; green, 2; blue, 27 }  ,draw opacity=1 ]   (262.5,190.61) .. controls (299.19,177.82) and (360.19,176.82) .. (400,191.11) ;
		\draw [shift={(324.88,180.77)}, rotate = 359.39] [fill={rgb, 255:red, 208; green, 2; blue, 27 }  ,fill opacity=1 ][line width=0.08]  [draw opacity=0] (8.93,-4.29) -- (0,0) -- (8.93,4.29) -- cycle    ;
		\draw [color={rgb, 255:red, 74; green, 144; blue, 226 }  ,draw opacity=1 ][line width=3]    (499,162) -- (499,79.6) ;
		\draw [shift={(499,74.6)}, rotate = 90] [color={rgb, 255:red, 74; green, 144; blue, 226 }  ,draw opacity=1 ][line width=3]    (20.77,-6.25) .. controls (13.2,-2.65) and (6.28,-0.57) .. (0,0) .. controls (6.28,0.57) and (13.2,2.66) .. (20.77,6.25)   ;
		\path  [shading=_tnpihuzf4,_rh3rqtl5z] (259.6,51.2) .. controls (259.6,49.98) and (260.58,49) .. (261.8,49) .. controls (263.02,49) and (264,49.98) .. (264,51.2) .. controls (264,52.42) and (263.02,53.4) .. (261.8,53.4) .. controls (260.58,53.4) and (259.6,52.42) .. (259.6,51.2) -- cycle ; 
		\draw   (259.6,51.2) .. controls (259.6,49.98) and (260.58,49) .. (261.8,49) .. controls (263.02,49) and (264,49.98) .. (264,51.2) .. controls (264,52.42) and (263.02,53.4) .. (261.8,53.4) .. controls (260.58,53.4) and (259.6,52.42) .. (259.6,51.2) -- cycle ; 
		
		\path  [shading=_i9jm55v38,_g0uoxhsor] (398.6,51.2) .. controls (398.6,49.98) and (399.58,49) .. (400.8,49) .. controls (402.02,49) and (403,49.98) .. (403,51.2) .. controls (403,52.42) and (402.02,53.4) .. (400.8,53.4) .. controls (399.58,53.4) and (398.6,52.42) .. (398.6,51.2) -- cycle ; 
		\draw   (398.6,51.2) .. controls (398.6,49.98) and (399.58,49) .. (400.8,49) .. controls (402.02,49) and (403,49.98) .. (403,51.2) .. controls (403,52.42) and (402.02,53.4) .. (400.8,53.4) .. controls (399.58,53.4) and (398.6,52.42) .. (398.6,51.2) -- cycle ; 

		\draw (326.36,28.84) node [anchor=north west][inner sep=0.75pt]   [align=left] {$\displaystyle \mathcal{I}^{+}$};
		\draw (325.36,192.36) node [anchor=north west][inner sep=0.75pt]   [align=left] {$\displaystyle \mathcal{I}^{-}$};
		\draw (256.36,28.84) node [anchor=north west][inner sep=0.75pt]   [align=left] {$\displaystyle P_{1}$};
		\draw (393.36,28.84) node [anchor=north west][inner sep=0.75pt]   [align=left] {$\displaystyle P_{2}$};

	\end{tikzpicture}
	
	\caption{Red lines denote the direction of the Killing vector $\del_t$. Blue arrow denotes the direction of chronological time.}
	\label{fig.tsds}
\end{figure}

Let us end with discussing one more point. As mentioned above the killing vector $\partial_t$ is space like in the Future and past Milne patches in fig. \ref{fig.altpensds}. Thus $t$ denotes a spatial coordinate in these patches and also on ${\cal I}^+$, where the hologram lives- we hope this  notation will  not cause confusion for the reader. 
In our discussion of horizon thermodynamics it will be useful to consider the change of the future cosmological horizons. The black hole spacetime fig. \ref{fig.tsds} is time orientable and we can choose a globally well-defined  time coordinate and a related arrow of time which is future directed everywhere. This arrow  is denoted by the solid blue line in fig. \ref{fig.tsds}. To avoid confusion we will refer to this time coordinate as chronological time below. Similar comments also apply in the rotating case. 

\section{Aspects of Holography}
\label{Sec.Holo}
Here we review some basic aspects of holography for the  hologram living at ${\cal I}^+$. 

In the static coordinate system, eq.(\ref{nrmet}) the boundary, ${\cal I}^+$, fig. \ref{fig.pendsst} and \ref{fig.altpensds},  is at $r\rightarrow \infty$. In carrying out calculations, in practice, one first  considers a boundary at a finite location, at  $r=r_B$ with $r_B\gg1$,   and then takes the $r_B\rightarrow \infty$ limit, after taking into account divergences that might occur. 

The basic dictionary, \cite{malda_NG}, for holography at ${\cal I}^+$ is that the wave function of the bulk state is equal to the partition function of the boundary theory in the presence of sources. 
These sources are determined  by boundary values that the bulk fields take. 
Some more details are as follows. In $dS_{d+1}$, 
the asymptotic behaviour for a scalar of mass $m$ is given by 
\be
\label{assb}
\phi\rightarrow r^{-\Delta_\pm}
\ee
where 
\be
\label{assc}
\Delta_\pm={d\over 2}\pm \sqrt{{d^2\over 4}-m^2}
\ee
For $m^2\le {d^2\over 4}$ the corresponding operator in the boundary has dimension
$\Delta_+={d\over 2}+ \sqrt{{d^2\over 4}-m^2}$. While for $m^2>{d^2\over 4}$ the correspondence has to be established with more care by going to the coherent state basis, and the operator has the dimension $\Delta={d\over2}+i \sqrt{m^2-{d^2\over 4}}$, see \cite{roy_dscft}. 

The stress tensor which corresponds to the metric in the bulk will be particularly important for us. 
It follows from the general dictionary that the expectation value of the stress tensor in the hologram  is related to the extrinsic curvature of the boundary hypersurface \cite{Balasubramanian:1999re,Balasubramanian:2001nb,Klemm:2001ea} by the relation
\be
\label{By}
\sqrt{\gamma} \langle T^a_b \rangle = i ({1\over 8 \pi G}) \sqrt{h} (K^a_b-h^a_b K)
\ee
Here indices $a,b$ denote indices for coordinates along the boundary, $K^a_b$ is the extrinsic curvature of the boundary  and $h_{ab}$ is the induced metric at the boundary $r=r_B$.  And $\gamma_{ab}$ is related to $h_{ab}$  
by 
\be
\label{relsrb}
h_{ab}=r_B^2 \gamma_{ab}
\ee
Note that in eq.(\ref{By}) the indices are raised and lowered on the LHS by $\gamma_{ab}$ and on the RHS by $h_{ab}$. 
We see from eq.(\ref{By}) that the boundary stress tensor is related to the extrinsic curvature, in a manner analogous to the definition of the Brown -York stress tensor in AdS/CFT. For further discussion pertaining to eq.(\ref{By}) see appendix \ref{app.by}.       \\~\\ 
An important difference with the AdS case is the factor of $i$ which appears on the RHS in eq.(\ref{By}). 
The origin of this factor can be understood as follows. In the semi-classical limit the wave function is of the WKB form and given by 
\be
\label{formpsi}
\Psi=e^{iS}. 
\ee
The general dictionary tells us that 
\be
\label{gendby}
\langle T^{ab}\rangle={2\over\sqrt{\gamma}}{\delta \log(\Psi) \over \delta \gamma_{ab}}
\ee
where $h_{ab}$, the induced metric is related to $\gamma_{ab}$ by the relation \eq{relsrb}.
The action  $S$ appearing in  eq.(\ref{formpsi}) includes the Einstein-Hilbert term along with the Gibbons-Hawking boundary term, see \eq{acta}.
The relation in eq.(\ref{gendby}) then  gives rise to eq.(\ref{By}). 
Another important difference with the AdS case has to do with divergent terms which arise in the expectation value $\langle T^a_b \rangle$, and more generally in correlation functions. In the AdS case these divergences are removed by adding suitable ``counter terms", i.e. terms in the action which have support only at the boundary.
However, in the dS space case the divergences are physical and should not be subtracted. In fact they are required for the wave function to satisfy the Wheeler de Witt equation \cite{Tuneer_dSH, KK_JText}. Nevertheless it will be useful in our discussion below of black hole thermodynamics,  to define a renormalised stress tensor by adding suitable  boundary  terms in the action and also adding their contribution on the RHS of eq.(\ref{By}) to define a renormalised stress tensor.\\

In the subsequent discussion we now specialise to $d=3$, i.e. $dS_{4}$. The required counter terms for obtaining a finite stress tensor are of the form, 
\be
\label{cta}
S_{ct}={1\over 8\pi G}\int d^3y \sqrt{h} ( \Lambda_B + c_2 R[h_{ab}])
\ee
where, as mentioned above, $h_{ab}$ denotes the $3$ dimensional induced metric on the boundary and $\Lambda_B$ and $c_2$ are constants.
$R[h_{ab}]$ is the Ricci scalar one obtains from $h_{ab}$. 
Including $S_{ct}$, \eq{cta}, yields the renormalised stress tensor
\be
\label{restra}
\sqrt{\gamma}\langle ( T^{r})^a_b\rangle={i \over (8 \pi G)} \sqrt{h} \left[(K^a_b-h^a_b K)  + h^a_b \Lambda_B - 2c_2 \left(R^a_b-{1\over 2 } h^a_b R \right)\right]
\ee
Choosing 
\begin{equation}
	\label{retrab}
	\Lambda_B = {2}~~,~~c_2 = -{1\over2}
\end{equation}
makes the renormalised stress tensor finite. See appendix \ref{app.renorm} for more details. 

In the subsequent discussion we will also find it convenient to define the pressure  density\footnote{We call $p$ the ``pressure density" because $t$ is a space-like coordinate on the boundary.} $p$  which is  related to the $t-t$ component of the stress tensor:
\be
\label{defp}
p=i \langle (T^{r})^t_t\rangle
\ee
And the pressure $P$ which is the integral over the $\theta-\phi$ two-sphere of $p$, at a constant value of $t$ :
\be
\label{press}
P=\int d^2 \Omega \ p
\ee

For a black hole with mass parameter $M$, eq.(\ref{nrmet}), we get 
\be
\label{Pvala}
P=M
\ee
$P$ is real because  of the factor of $i$ on the RHS of eq.(\ref{defp}). 


For the case of rotating black holes we will also need to consider the angular momentum, as defined in the boundary theory, and its renormalised value. 
In carrying out the calculations it is convenient to work in the  Fefferman-Graham coordinate system in which the metric components take the asymptotic form 
	\begin{equation}
		\label{holometa}
		\lim_{{\bar r}\rightarrow\infty}ds^2={\bar r}^2d{\bar t}^2-{d\bar{r}^2\over {\bar r}^2}+{\bar r}^2(d{\bar \theta}^2+\sin^2{\bar \theta}~ d\bar{\phi}^2)
	\end{equation}	
	In this coordinate system, we  define  the angular momentum density, $j$, which is related to the ${\bar t}-\bar{\phi}$ component of the stress tensor, as, 
	\begin{equation}
		j=-i\langle(T^r)^{\bar t}_{\bar \phi}\rangle
	\end{equation}
	And the angular momentum, $J$, which is the integral over ${\bar\theta}-{\bar\phi}$ two-sphere of $q$, at a constant value of ${\bar t}$, to be, 
	\begin{equation}
		\label{shear}
		J=\int d^2\Omega j
	\end{equation}
	For a black hole with mass parameter $M$ and spin parameter $a$, from eq.(\ref{rotmet}), we get the pressure $P$, \eq{press}, eq.(\ref{defp}), to be
	\begin{equation}
		\label{Pval}
		P={M\over (1+a^2)^2}
	\end{equation}
	and the angular momentum $J$, \eq{shear}, to be
	\begin{equation}
		\label{Jval}	
		J={aM\over (1+a^2)^2}	
	\end{equation}
	More discussion on these formulae can be found in appendix \ref{app.KdScharge}.

	We end this section with two comments. First, let us note  another important   convention that we adopt: we take   the Killing vector $\partial_t$, see  eq.(\ref{nrmet}), in the vicinity of the boundary ${\cal I}^+$ to point along the  direction as shown by the red  arrows  in fig. \ref{fig.tsds},  i.e. from left to right,   with $P_1$ and $P_2$ being at $t=-\infty$ and $\infty$ respectively. The direction of the flow generated by  $\partial_t$ in various regions, from continuity,  then takes the form as shown in fig. \ref{fig.tsds}. We see that while in the static patch region $R$ the flow  goes forward in chronological time, in the static patch $L$ it goes backwards. The direction  of chronological time is denoted by the blue arrow in fig. \ref{fig.tsds}.
	This point will be important when we interpret the first law of thermodynamics  from the boundary perspective in section \ref{Sec.hololaw}. 
	
	Second, as has been discussed in the literature \cite{malda_NG, roy_dscft, Strominger:2001pn}, the late time wave function, as one approaches ${\cal I}^+$, takes the form of a CFT
	partition function in the presence of sources. More precisely, for any physical state, satisfying the time and spatial reparametrisation constraints, 
	the wave function can be expanded in terms of coefficient functions \cite{Chakraborty:2025izq, Chakraborty:2023los, Tuneer_dSH}. These coefficient functions behave like the correlation functions of a CFT satisfying the Ward identities of conformal invariance.
	
	It is worth giving an example of such a Ward identity explicitly. 
	In a  theory with the metric and a scalar field, the wave function, asymptotically,  takes the form,
	\begin{align}
		\Psi[\gamma_{ij}, {\hat \phi}] =\exp&\left[{1\over 2}\int d^3x d^3y \sqrt{\hat{g}({\bf x})}\sqrt{\hat{g}({\bf y})} {\hat \phi}({\bf x}) {\hat \phi}({\bf y}) \langle O({\bf x}) O({\bf y})\rangle \right.\nonumber\\
		&\left.+ {1\over 4}\int d^3x d^3y d^3z \sqrt{\hat{g}({\bf x})}\sqrt{\hat{g}({\bf y})}\sqrt{\hat{g}({\bf z})} \gamma_{ij}({\bf z}) {\hat \phi}({\bf x}) {\hat \phi}({\bf y}) \langle T^{ij}({\bf z}) O({\bf x}) O({\bf y})\rangle + \cdots\right], \label{wfa}
	\end{align}
	where the ellipses denote additional  terms. Here, ${\hat g}_{ij}=\delta_{ij}+\gamma_{ij}$ where $\gamma_{ij}$ is related to the asymptotic value of the metric, and ${\hat \phi}$ is related to the asymptotic value of the scalar field, in case of the complementary series, or to the  coherent state basis, in case of the principal series \cite{roy_dscft}. $\langle O({\bf x}) O({\bf y})\rangle,~\langle T^{ij}({\bf z}) O({\bf x}) O({\bf y})\rangle$ are two examples of the coefficient functions referred to above. 
	The Ward identities of conformal invariance include those of translational invariance along the boundary directions. These take the following form,  for  the two coefficient functions  explicitly shown above, 
	\be
	\label{transward}
	\partial_j^{\bf z}\langle T^{ij}({\bf z})O({\bf x})O({\bf y})\rangle+\delta ({\bf x}-{\bf z}) \partial_i^{\bf x} \langle O({\bf x})O({\bf y})\rangle+\delta({\bf y}-{\bf z}) \partial_i^{\bf y} \langle O({\bf y})O({\bf x})\rangle=0
	\ee
	Note this ward identity cannot be preserved by rescaling the stress tensor, and thus fixes the normalisation of $T_{ij}$ completely. 
	This is a more general property. In general, Ward identities completely fix the normalisation of  conserved charges.

	\section{The Laws of Horizon Thermodynamics}
	\label{sec.first}
	Here we consider the behaviour of horizons in cosmological spacetimes of the kind discussed above and their close connection with the laws of thermodynamics. 
	We will be particularly interested in the behaviour of cosmological horizons here, but  our discussion will apply to  black hole horizons as well. 
	We start by considering Schwarzschild black holes, and discuss  the rotating case as we proceed.  Generalisations to include charge, i.e. the Kerr-Neumann case, are also straightforward, but we do not discuss these here. This section is  a review and included to establish notation and for completeness. Standard references include \cite{Wald_GRQFT, Poisson_Toolkit, Wald:1984rg} . 
	
	\subsection{Zeroth Law} 
	To define the surface gravity of a Horizon we consider, in the Schwarzschild case, the Killing vector 
	\be
	\label{defkv}
	\xi=\partial_t
	\ee
	This is normalised so that at ${\cal I}^+$, where $r\rightarrow \infty$, 
	\be
	\label{normkv}
	{1\over r^2} \xi^\mu\xi_\mu\rightarrow 1
	\ee
	This Killing vector field is normal to both the cosmological and black hole horizons.
	In the rotating case the Killing vector normal to the horizons is 
	\be
	\label{defkvrot}
	\xi=\partial_t + \Omega'_H \partial_\phi
	\ee
	where $\Omega'_H$ is the angular velocity of the black hole.
	The surface gravity, $\kappa$,  for the cosmological or black hole horizon, is  then defined by the relation
	\be
	\label{defsg}
	\nabla^\nu(\xi^\mu\xi_\mu)=-2 \kappa \xi^\nu
	\ee
	It is easy to see from the solution eq.(\ref{nrmet}) that $\kappa$ is a constant on both horizons for the Schwarzschild case and from eq.(\ref{rotmet}) for the killing horizons in the rotating case. 
	
	
	In fact, the constancy of the surface gravity on a Killing horizon follows from the Einstein equations in general, see \cite{Poisson_Toolkit, Wald:1984rg}, for discussion. 
	
	\subsection{First Law}
	\label{sec.1l}
\begin{figure}[h]
		\centering


\tikzset {_j11jknkyh/.code = {\pgfsetadditionalshadetransform{ \pgftransformshift{\pgfpoint{89.1 bp } { -128.7 bp }  }  \pgftransformscale{1.32 }  }}}
\pgfdeclareradialshading{_l1bf1qz6p}{\pgfpoint{-72bp}{104bp}}{rgb(0bp)=(1,1,1);
	rgb(0bp)=(1,1,1);
	rgb(25bp)=(0.48,0.15,0.15);
	rgb(400bp)=(0.48,0.15,0.15)}


\tikzset {_f1xpuxn88/.code = {\pgfsetadditionalshadetransform{ \pgftransformshift{\pgfpoint{89.1 bp } { -128.7 bp }  }  \pgftransformscale{1.32 }  }}}
\pgfdeclareradialshading{_r9al192hf}{\pgfpoint{-72bp}{104bp}}{rgb(0bp)=(1,1,1);
	rgb(0bp)=(1,1,1);
	rgb(25bp)=(0.48,0.15,0.15);
	rgb(400bp)=(0.48,0.15,0.15)}
\tikzset{every picture/.style={line width=0.75pt}} 

\begin{tikzpicture}[x=0.75pt,y=0.75pt,yscale=-1,xscale=1]
	
	\draw [color={rgb, 255:red, 208; green, 2; blue, 27 }  ,draw opacity=1 ]   (238.27,108.13) -- (284.43,152.54) ;
	\draw [color={rgb, 255:red, 208; green, 2; blue, 27 }  ,draw opacity=1 ]   (257.86,106.14) -- (296.43,140.54) ;
	\draw  [color={rgb, 255:red, 255; green, 255; blue, 255 }  ,draw opacity=1 ][fill={rgb, 255:red, 126; green, 211; blue, 33 }  ,fill opacity=1 ] (256.17,107.13) -- (239.03,107.12) -- (340.27,201.55) -- (357.41,201.56) -- cycle ;
	\draw   (333,106) -- (167,106) -- (167,272) -- (333,272) -- cycle ;
	\draw    (333,106) -- (167,272) ;
	\draw    (167,106) -- (333,272) ;
	\draw [color={rgb, 255:red, 74; green, 144; blue, 226 }  ,draw opacity=1 ][line width=3]    (523,236) -- (523,153.6) ;
	\draw [shift={(523,148.6)}, rotate = 90] [color={rgb, 255:red, 74; green, 144; blue, 226 }  ,draw opacity=1 ][line width=3]    (20.77,-6.25) .. controls (13.2,-2.65) and (6.28,-0.57) .. (0,0) .. controls (6.28,0.57) and (13.2,2.66) .. (20.77,6.25)   ;
	\draw    (333,106) .. controls (334.67,104.33) and (336.33,104.33) .. (338,106) .. controls (339.67,107.67) and (341.33,107.67) .. (343,106) .. controls (344.67,104.33) and (346.33,104.33) .. (348,106) .. controls (349.67,107.67) and (351.33,107.67) .. (353,106) .. controls (354.67,104.33) and (356.33,104.33) .. (358,106) .. controls (359.67,107.67) and (361.33,107.67) .. (363,106) .. controls (364.67,104.33) and (366.33,104.33) .. (368,106) .. controls (369.67,107.67) and (371.33,107.67) .. (373,106) .. controls (374.67,104.33) and (376.33,104.33) .. (378,106) .. controls (379.67,107.67) and (381.33,107.67) .. (383,106) .. controls (384.67,104.33) and (386.33,104.33) .. (388,106) .. controls (389.67,107.67) and (391.33,107.67) .. (393,106) .. controls (394.67,104.33) and (396.33,104.33) .. (398,106) .. controls (399.67,107.67) and (401.33,107.67) .. (403,106) .. controls (404.67,104.33) and (406.33,104.33) .. (408,106) .. controls (409.67,107.67) and (411.33,107.67) .. (413,106) .. controls (414.67,104.33) and (416.33,104.33) .. (418,106) .. controls (419.67,107.67) and (421.33,107.67) .. (423,106) .. controls (424.67,104.33) and (426.33,104.33) .. (428,106) .. controls (429.67,107.67) and (431.33,107.67) .. (433,106) .. controls (434.67,104.33) and (436.33,104.33) .. (438,106) .. controls (439.67,107.67) and (441.33,107.67) .. (443,106) .. controls (444.67,104.33) and (446.33,104.33) .. (448,106) .. controls (449.67,107.67) and (451.33,107.67) .. (453,106) .. controls (454.67,104.33) and (456.33,104.33) .. (458,106) .. controls (459.67,107.67) and (461.33,107.67) .. (463,106) .. controls (464.67,104.33) and (466.33,104.33) .. (468,106) .. controls (469.67,107.67) and (471.33,107.67) .. (473,106) .. controls (474.67,104.33) and (476.33,104.33) .. (478,106) .. controls (479.67,107.67) and (481.33,107.67) .. (483,106) .. controls (484.67,104.33) and (486.33,104.33) .. (488,106) .. controls (489.67,107.67) and (491.33,107.67) .. (493,106) .. controls (494.67,104.33) and (496.33,104.33) .. (498,106) -- (499,106) -- (499,106) ;
	\draw    (499,106) -- (499,272) ;
	\draw    (333,272) .. controls (334.67,270.33) and (336.33,270.33) .. (338,272) .. controls (339.67,273.67) and (341.33,273.67) .. (343,272) .. controls (344.67,270.33) and (346.33,270.33) .. (348,272) .. controls (349.67,273.67) and (351.33,273.67) .. (353,272) .. controls (354.67,270.33) and (356.33,270.33) .. (358,272) .. controls (359.67,273.67) and (361.33,273.67) .. (363,272) .. controls (364.67,270.33) and (366.33,270.33) .. (368,272) .. controls (369.67,273.67) and (371.33,273.67) .. (373,272) .. controls (374.67,270.33) and (376.33,270.33) .. (378,272) .. controls (379.67,273.67) and (381.33,273.67) .. (383,272) .. controls (384.67,270.33) and (386.33,270.33) .. (388,272) .. controls (389.67,273.67) and (391.33,273.67) .. (393,272) .. controls (394.67,270.33) and (396.33,270.33) .. (398,272) .. controls (399.67,273.67) and (401.33,273.67) .. (403,272) .. controls (404.67,270.33) and (406.33,270.33) .. (408,272) .. controls (409.67,273.67) and (411.33,273.67) .. (413,272) .. controls (414.67,270.33) and (416.33,270.33) .. (418,272) .. controls (419.67,273.67) and (421.33,273.67) .. (423,272) .. controls (424.67,270.33) and (426.33,270.33) .. (428,272) .. controls (429.67,273.67) and (431.33,273.67) .. (433,272) .. controls (434.67,270.33) and (436.33,270.33) .. (438,272) .. controls (439.67,273.67) and (441.33,273.67) .. (443,272) .. controls (444.67,270.33) and (446.33,270.33) .. (448,272) .. controls (449.67,273.67) and (451.33,273.67) .. (453,272) .. controls (454.67,270.33) and (456.33,270.33) .. (458,272) .. controls (459.67,273.67) and (461.33,273.67) .. (463,272) .. controls (464.67,270.33) and (466.33,270.33) .. (468,272) .. controls (469.67,273.67) and (471.33,273.67) .. (473,272) .. controls (474.67,270.33) and (476.33,270.33) .. (478,272) .. controls (479.67,273.67) and (481.33,273.67) .. (483,272) .. controls (484.67,270.33) and (486.33,270.33) .. (488,272) .. controls (489.67,273.67) and (491.33,273.67) .. (493,272) .. controls (494.67,270.33) and (496.33,270.33) .. (498,272) -- (499,272) -- (499,272) ;
	\draw    (499,106) -- (333,272) ;
	\draw    (334,106) -- (500,272) ;
	
	\draw  [draw opacity=0][shading=_l1bf1qz6p,_j11jknkyh] (253.71,106.14) .. controls (253.71,103.85) and (255.57,102) .. (257.86,102) .. controls (260.15,102) and (262,103.85) .. (262,106.14) .. controls (262,108.43) and (260.15,110.29) .. (257.86,110.29) .. controls (255.57,110.29) and (253.71,108.43) .. (253.71,106.14) -- cycle ;
	\draw  [draw opacity=0][shading=_r9al192hf,_f1xpuxn88] (233.13,107.13) .. controls (233.13,104.84) and (234.99,102.99) .. (237.27,102.99) .. controls (239.56,102.99) and (241.42,104.84) .. (241.42,107.13) .. controls (241.42,109.42) and (239.56,111.27) .. (237.27,111.27) .. controls (234.99,111.27) and (233.13,109.42) .. (233.13,107.13) -- cycle ;
	\draw [line width=1.5]    (286.43,143.54) -- (258.62,117.48) ;
	\draw [shift={(256.43,115.43)}, rotate = 43.14] [color={rgb, 255:red, 0; green, 0; blue, 0 }  ][line width=1.5]    (14.21,-4.28) .. controls (9.04,-1.82) and (4.3,-0.39) .. (0,0) .. controls (4.3,0.39) and (9.04,1.82) .. (14.21,4.28)   ;
	\draw    (284.43,152.54) -- (296.43,140.54) ;
	\draw [line width=0.75]    (296,150.97) -- (324.59,122.39) ;
	\draw [shift={(326,120.97)}, rotate = 135] [color={rgb, 255:red, 0; green, 0; blue, 0 }  ][line width=0.75]    (10.93,-3.29) .. controls (6.95,-1.4) and (3.31,-0.3) .. (0,0) .. controls (3.31,0.3) and (6.95,1.4) .. (10.93,3.29)   ;
	
	\draw (176,138) node [anchor=north east][inner sep=0.75pt]  [font=\small,xscale=-1] [align=left] {$ $};
	\draw (271,169) node [anchor=north west][inner sep=0.75pt]  [font=\small] [align=left] {$\displaystyle H_{2}$};
	\draw (232,87) node [anchor=north west][inner sep=0.75pt]  [font=\small] [align=left] {$\displaystyle L$};
	\draw (253,87.14) node [anchor=north west][inner sep=0.75pt]  [font=\small] [align=left] {$\displaystyle R$};
	\draw (193,151) node [anchor=north west][inner sep=0.75pt]  [font=\small] [align=left] {$\displaystyle H_{1}$};
	\draw (310,133) node [anchor=north west][inner sep=0.75pt]  [font=\small] [align=left] {$\displaystyle \lambda $};

\end{tikzpicture}

		\caption{Example of a physical process in the Schwarzschild dS geometry : Matter crossing through cosmological horizon $H_2$.}
		\label{fig.physsds}
	\end{figure}
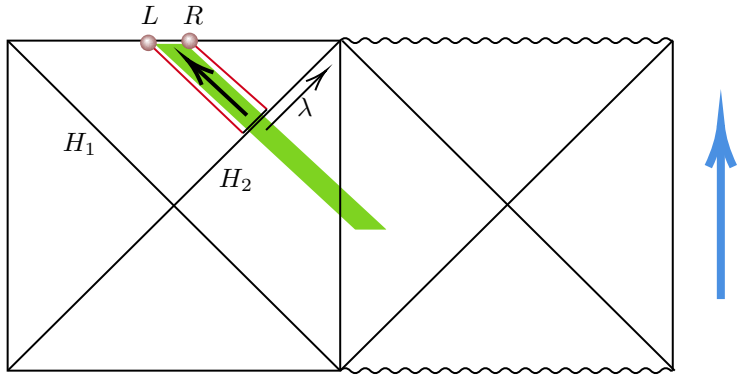

	Next,  we turn to the First Law of Thermodynamics.  Starting with a stationary solution we consider a perturbation which arises due to some matter being present in the black hole background which crosses a cosmological horizon. We take the matter stress tensor to have compact support and  make the  important  assumption that the space time sufficiently far in the future will settle down again into a stationary solution. This is a reasonable assumption to make especially for small perturbations. For concreteness, we consider matter which  crosses the cosmological horizon $H_2$, fig. \ref{fig.physsds}. For now we consider  the Schwarzschild case, with a non-rotating matter perturbation. We will incorporate the effects of rotation later on.   We  work to first order in the matter stress tensor perturbation and calculate the resulting change in area. 
	
	Note first that the current
	\be
	\label{defcurrent}
	j^\mu={\cal T}^{\mu\nu} \xi_\nu,
	\ee
	is conserved. This follows from the facts that 
	${\cal T}^{\mu\nu}$- the matter  stress tensor  perturbation- is conserved,  $\xi_\nu$ is a Killing vector in the unperturbed stationary solution we started with, and that we are working to first order in ${\cal T}^{\mu\nu}$. 
	Next, we  define the  total flux of energy  $\Delta {\cal E}$  crossing the horizon to be, 
	\be
	\label{defmass1}
	\Delta {\cal E}=\int_{0}^\infty d\lambda \sqrt{\sigma} d^2 y {\cal T}^{\mu\nu}\xi_\mu k_\nu 
	\ee
	Here $\lambda$ is the affine parameter along  null generators of the horizon, $y^1,y^2$ are additional space-like coordinates along the horizon, with 
	\be
	\label{defvolel}
	dV=d\lambda \sqrt{\sigma} d^2y
	\ee
	being the  volume element along the horizon, and   $k^\nu$ is the null generator which is parallel transported along the horizon, 
	\be
	\label{defk}
	k^\mu\nabla_\mu k^\nu=0.
	\ee
	We take the matter perturbation flux across the horizon to vanish for $\lambda<0$. This allows the lower limit in the integral in eq.(\ref{defmass1}) to be set at $\lambda=0$.  
	With these definitions we see  that the integral on the RHS  of eq.(\ref{defmass1}) computes the flux of the conserved current, eq(\ref{defcurrent}), across the horizon. 
	
	The Raychaudhuri equation for the expansion $\theta$ of the horizon then takes the form 
	\be
	\label{rc1}
	{d\theta\over d\lambda}=-(8\pi G){\cal T}_{\mu\nu}k^\mu k^\nu
	\ee
	In addition $k^\mu$ is related to $\xi^\mu$ by 
	\be
	\label{rela}
	\xi^\mu=\kappa \lambda k^\mu
	\ee
	Inserting eq.(\ref{rc1}) on the RHS of eq.(\ref{defmass1}) then gives that the change in energy $\Delta {\cal E}$ is 
	\be
	\label{chnma}
	(8 \pi G) \Delta {\cal E}=-\kappa  \int_0^\infty d\lambda \int \sqrt{\sigma} d^2y \lambda {d\theta\over d\lambda}
	\ee
	where we have taken the surface gravity out of the  integral, since it is a constant\footnote{To first order in the perturbation this follows just from the constancy of $\kappa$ in the stationary solution.}. 
	After noting that the area element $\sqrt{\sigma}$ can treated to be independent of $\lambda$, since we are working to first order in perturbation theory, 
	and  carrying out an integration by parts in $\lambda$, we get
	\be
	\label{sefr}
	(8 \pi G) \Delta {\cal E}=\kappa  \int d\lambda d^2y \sqrt{\sigma}  \theta 
	\ee
	In obtaining this result we have used the fact that in the far future, $\lambda\rightarrow \infty$, as per our assumption  the system settles down into a stationary solution again,   $\theta\rightarrow 0$.\\
	We now note that 
	the area element is related to the expansion parameter $\theta$ by 
	\be
	\label{expa}
	\theta={1\over \sqrt{ \sigma}}{d\sqrt{\sigma}\over d\lambda},
	\ee
	see e.g. section (2.4.8) of \cite{Poisson_Toolkit}. And, the area of the event horizon  along its evolution, at any value of $\lambda$, is given by 
	\be
	\label{defa}
	A=\int d^2y \sqrt{\sigma}
	\ee
	This gives, on integrating the RHS of eq.(\ref{sefr}), 
	\be
	\label{sefra}
	(8 \pi G) \Delta {\cal E}=\kappa  \Delta A,
	\ee
	where $\Delta A$ is the change in the area of the event horizon,
	\be
	\label{defdelta}
	\Delta A= A_f-A_i
	\ee
	Including the effects of rotation, with the Killing vector now being,  eq.(\ref{newkil}), yields, similarly, 
	\be
	\label{firstlaw1}
	\kappa \Delta A= (8 \pi G) [\Delta {\cal E} + \Omega_H \Delta { J}]
	\ee
	where $\Omega_H$, the angular velocity related to $\Omega'_H$ mentioned in \eq{defkvrot}, as
	\begin{equation}
		\Omega_H=\Omega'_H-a
	\end{equation}  
	and
	\be
	\label{defdelE}
	\Delta {\cal E}=\int_0^{\infty} d\lambda\int  d^2y \sqrt{\sigma} {\cal T}_{\mu\nu}{(\xi_{t})}^{\mu} k^{\nu} 
	\ee 
	\be
	\label{defdelJ}
	\Delta {J}=\int_0^{\infty} d\lambda\int  d^2y \sqrt{\sigma} {\cal T}_{\mu\nu}{(\xi_{\bar\phi})}^{\mu} k^{\nu} 
	\ee
	with the space-like killing vector associated with rotations being 
	\be
	\label{defkvrota}
	\xi_{\bar\phi}=\del_{\bar\phi}
	\ee
	where the coordinate $\bar\phi$ is defined as $\bar{\phi}\equiv\phi-at$; see appendix \ref{app.kdsprop} for more details.
	
	\subsection{Second Law}
	Finally we consider the second law of thermodynamics. We are interested in  the change of the  area of an event  horizon as it evolves into the future.  
	Our considerations here are more general and not restricted to small matter perturbations. 
	From the Raychaudhuri equation it follows, as long as the matter satisfies the null energy condition, and the horizon is hypersurface orthogonal, that 
	\be
	\label{exprel}
	{d\theta \over d \lambda}\le -{1\over 2} \theta^2
	\ee
	It can be argued in general that the    generators of the horizon are future extendible, i.e. do not end on caustics, see \cite{Wald_GRQFT,Poisson_Toolkit}. It then follows  that 
	$\theta\ge 0$ everywhere along the horizon. 
	From eq.(\ref{expa}) and eq.(\ref{defa}) it then follows that 
	\be
	\label{cons2b}
	{dA \over d\lambda}\ge 0
	\ee
	everywhere along the event horizon. This establishes the area increase theorem which is the analogue of the second law of thermodynamics. 
	
	The assumption that the generators must be future extendible, for predictable horizons, can be shown to be true, see \cite{Wald_GRQFT} for a discussion. 
	\subsection{Additional Comments}
	\label{sec.addcom}
	Note that, in the non-rotating case, the area of the cosmological horizon is given by 
	\be
	\label{areacos}
	A=4\pi r_c^2
	\ee
	where $r_c$ is the larger of the two real roots of \eq{khloc}. 
	It is easy to see that as the parameter $M$ decreases, and the black hole  spacetime goes towards being global de Sitter space, the area increases, so that 
	\be
	\label{ratea}
	{dA\over dM}<0
	\ee
	The smallest area arises in the Nariai case, where 
	\be
	\label{naria}
	A={8\pi\over 3}
	\ee
	And the largest area for global  dS space
	\be
	\label{entgds}
	A=4 \pi
	\ee
	Denoting the emblackening factor in eq.(\ref{nrmet}) by 
	\be
	\label{fdef}
	f(r)=1-r^2-{2GM \over r}
	\ee
	The surface gravity of the cosmological horizon is given by 
	\be
	\label{dvalkc}
	\kappa= {f'(r_c)\over2}
	\ee
	
	Similarly in rotating case, the area of the cosmological horizon is given by
	\begin{equation}
		\label{Acker}
		A=4\pi{r_c^2+a^2\over1+{a^2}}	
	\end{equation}
	with $r_c$ being the largest root of \eq{kerhoreq}. $r_c$ depends on $M,a$, and $a$ in turn can be taken to be a function of $J,M$, eq.(\ref{J}). Thus we see that $A$ is a function of both $M,J$. 
	The surface gravity of the cosmological horizon is given by
	\begin{equation}
		\label{valkappa}
		\kappa=\frac{a^2 \left(r_c^2+1\right)+3 r_c^4-r_c^2}{2 r_c \left(a^2+r_c^2\right)}	
	\end{equation}
	
	In the following discussion on the first law it will be convenient to define the variable  
	\begin{equation}
		\label{defMtilde}
		\tilde{M}={M\over \zeta^2}
	\end{equation}
	From eq.(\ref{Pval}) we see that 
	\be
	\label{relpval}
	P={\tilde M}.
	\ee
	Thus the area can be taken to be a function of the pressure $P$ and angular momentum $J$, $A(P,J)$. 
	See appendix \ref{app.kdsprop} for more details.

	\section{The  First Law as Viewed in the Hologram}
	\label{Sec.hololaw}
	We turn now to considering the First Law  from the perspective of the hologram at ${\cal I}^+$. 
	One important feature is that the boundary at ${\cal I}^+$ can receive signals from both cosmological horizons, $H_1,H_2$, fig. \ref{fig.physsds} and we seek a universal first law which would hold in the most general situation. 
	
	We start by first considering a situation where matter crosses one of the horizons, $H_2$, and impinges on the boundary as shown in fig. \ref{fig.physsds}.
	This yields a version of the first law, as we  see below. We  will subsequently   argue in subsection \ref{sec.hololaw1}  that this law is also valid in the more general situation when there is a flux of matter-energy from both horizons. 
	
	We take the matter crossing   $H_2$  to be  localised as shown in fig. \ref{fig.physsds}. 
	And consider the conserved current $j^\mu$ defined in eq.(\ref{defcurrent}). 
	We then get that
	\be
	\label{conscur}
	\int_\Sigma d^3y\sqrt{h} {\cal T}_{\mu\nu}\xi^\mu n^\nu =\int_{H_2} d\lambda \sqrt{\sigma} d^2y {\cal T}_{\mu \nu} \xi^\mu k^\nu=\Delta {\cal E}
	\ee
	where $\Sigma$ is the late time surface at $r=r_B$ mentioned in section \ref{Sec.Holo}, with  $r_B \rightarrow \infty$, and $h_{ab}$ is the induced boundary metric on this surface. 
	The last equality on the RHS is arrived at by using eq.(\ref{defmass1}).
	
	Next, note that the Gauss Codazzi relation  for any hypersurface takes the form, see eq.(3.42) in \cite{Poisson_Toolkit}
	\be\label{gca}
	G_{\alpha\beta}e^\alpha_a n^\beta=(K_a^b-h_a^b K)|_b
	\ee
	where $G_{\alpha\beta}=R_{\alpha\beta}-{1\over 2} g_{\alpha\beta} R$, denotes the Einstein  tensor; $(K_a^b)|_b$ denotes the covariant derivative of the extrinsic curvature tensor with respect to the induced metric on $\Sigma$, etc; and 
	\be
	\label{defea}
	e^\alpha_a={\partial x^\alpha\over \partial y^a},
	\ee
	with $y^a$ being coordinates along the hypersurface.
	Applying this relation at the surface $\Sigma$, with $r=r_B$, and 
	taking the index $a$, in eq.(\ref{gca}), to    be $t$ 
	then gives, 
	\be\label{gcb}
	G_{t\beta} n^\beta=(K_t^b-h_t^b K)|_b
	\ee
	Next using the Einstein equations,  noting that the only non-zero component of $\xi^\mu$ is $\xi^t=1$, and integrating over $\Sigma$,  we get  from eq.(\ref{gcb}) 
	\be\label{gcc}
	\int_\Sigma d^3y\sqrt{h}{\cal T}_{\alpha\beta} \xi^\alpha n^\beta=({1\over 8 \pi G})\int_\Sigma d^3y\sqrt{h} (K_t^b-h_t^b K)|_b
	\ee
	In obtaining this equation we have used the fact that  $g_{\alpha\beta}\xi^\alpha n^\beta=0$, and as a result the cosmological constant term drops out from the LHS. 
	
	Now note from eq.(\ref{By})  that the RHS is related to $\langle T_t^b \rangle$, the expectation value of the boundary stress tensor. And further, that  adding the extra terms in eq.(\ref{restra}) which relate the boundary stress tensor to its renormalised version, $\langle (T^r)_t^b \rangle$,  does not change the RHS, since the metric $h_{ab}$ and Einstein tensor, $R_{ab}-{1\over 2} h_{ab} R$ are both covariantly constant, i.e. $h_{ab}|^b=0$, etc.

	Using  eq.(\ref{conscur}) this gives,  
	\be
	\label{gcda}
	\Delta {\cal E}=(-i)\int_\Sigma d^3y \sqrt{\gamma} \langle (T^r)_t^b \rangle |_b 
	\ee
	where on the RHS $\langle (T^r)_t^b \rangle$ denotes the renormalised stress tensor of the boundary theory. We note here that the covariant derivate denoted as $|_b$ in eq.(\ref{gcc}) is with respect to $h_{ab}$, whereas we can take it to be with respect to the metric $\gamma_{ab}$ in eq.(\ref{gcda}), since the two metrics are related by a constant rescaling on the boundary, eq.(\ref{relsrb}).

	A few more steps are needed to arrive at our main result. We can write 
	\be
	\label{fm1}
	\int_\Sigma d^3y \sqrt{\gamma} \langle (T^r)_t^b \rangle|_b =\int_\Sigma d^3y\sqrt{\gamma} \langle (T^r)_a^b \rangle |_b \xi^a
	\ee
	since, as was mentioned above $\xi^t=1$ and its other components vanish.  
	This leads to 
	\be
	\label{fm2}
	\int_\Sigma d^3y \sqrt{\gamma} \langle (T^r)_t^b \rangle |_b=\int_\Sigma d^3y\sqrt{\gamma} [(\langle(T^r)_a^b\rangle \xi^a)|_b - \langle (T^r)_a^b \rangle \xi^a|_b]
	\ee
	
	The first term on the RHS is the divergence of a current $\langle (T^r)^a_b \rangle\xi^b$ and can be integrated and expressed as a difference between two surface terms. The second, would have vanished  if $\xi^a\partial_a$ was a Killing vector with respect to the induced metric $h_{ab}$ on $\Sigma$. This would be true for a stationary black hole solution, with $\Sigma$ being located  surface at $r=r_B$, but it will not be true in general for  time dependent situations. Note that the renormalised stress tensor is non-vanishing in the  stationary solution one begins with, so even working to first order in the perturbation this term cannot be immediately dropped. 
	However, as we argue in appendix \ref{app.covxi}, $\xi^a|_b$, even after allowing for the perturbation,   asymptotically goes like $1/r^3$, while $\langle (T^r)^a_b \rangle$, being the renormalised stress tensor, is finite. Thus, the second term can indeed be dropped in the limit when the boundary $\Sigma$ is located at $r_B \rightarrow \infty$, 
	This then leads from eq.(\ref{gcda}) and eq.(\ref{fm1}) to 
	\be
	\label{gce}
	\Delta {\cal E}=(-i) \brt{ \int_{S_R} d\theta d\phi \sqrt{\gamma} \langle (T^r)_t^t\rangle-\int_{S_L} d\theta d\phi \sqrt{\gamma} \langle (T^r)_t^t \rangle}
	\ee
	where $S_R, S_L$ are two spherical surfaces, bounding the region where the perturbation has support on $\Sigma$, see fig. \ref{fig.physsds}. 
	$S_R$ is at the larger value of $t$, compared to $S_L$, i.e. further to the right. 
	From eq.(\ref{press}) we see that this equation can also be expressed in terms of the pressure as 
	\be
	\label{gde}
	\Delta {\cal E}=-(P_R-P_L)=-\Delta P
	\ee
	where $P_R, P_L$ refer to the pressure at $t_R,t_L$ respectively. 
	
	Inserting this expression for the energy flux $\Delta{\cal E}$  in eq.(\ref{sefra}) now leads to 
	\be
	\label{finforma}
	\kappa \Delta A=-(8 \pi G) \Delta P
	\ee
	where 
	\be
	\label{valdeltaA}
	\Delta A=A_R-A_L
	\ee
	is the change of the horizon area as one evolves  into the future, i.e. along the direction of advancing chronological time, as is denoted in solid blue in fig. \ref{fig.tsds}, see also fig. \ref{fig.physsds}. 
	To summarise the final form of the first law we have arrived at is given by Eq.(\ref{finforma}).
	
	Now notice that the initial value of the Pressure $P_L$, to the left of the localised pulse in fig. \ref{fig.physsds} is related to the initial value of the mass parameter $M_L$ through the relation
	\be
	\label{inval}
	M_L=P_L
	\ee 
	Once the perturbation (which we have taken to be localised in space and time) dies out, it is reasonable to assume, as was mentioned earlier, that the geometry also settles down to a black hole spacetime. Its mass parameter should be related to the final value of the pressure $P_R$, to the right of the pulse, similarly as 
	\be
	\label{fival}
	M_R=P_R
	\ee
	This allows us to also state the first law eq.(\ref{finforma}) as 
	\be
	\label{finforc}
	\kappa \Delta A= -(8 \pi G) \Delta M=-(8 \pi G) (M_R-M_L)
	\ee
	The minus sign on the RHS is tied to the minus sign we found in eq.(\ref{ratea}). A time like observer in the static patch R will see that a positive flux $\Delta{\cal E}$ crossing  the   horizon leads to a decrease in the mass parameter, as the black hole becomes smaller, so that $\Delta M<0$. On the other hand the spacetime goes more towards  pure dS, so  the area grows, making $\Delta A>0$.  
	
	
	Before proceeding we note that a specific example consisting of a spherical shell of matter has been worked out in appendix \ref{app.shell}. 
	
	The derivation above, corresponds to what is usually referred to as the `physical processes" version of the first law. For a black hole in flat or AdS space there is also an ``equilibrium version", see \cite{Wald:1984rg}, obtained by comparing different black hole solutions. The version is obtained by comparing the solution corresponding to the initial stationary black hole with mass parameter $M_L$ and the final one,  valid to the right of the localised perturbation in fig. \ref{fig.physsds}, with mass parameter $M_R$. 
	We note that comparing the two  also lead to  eq.(\ref{finforc}) with $\kappa$ being given by eq.(\ref{dvalkc}). 
	
	So far our discussion has been in classical GR.  At this stage we can promote it to the semiclassical theory. The horizon $H_2$ has a Hawking temperature in the semi-classical theory given by \cite{GibHawking},  
	\be
	\label{hawkt}
	T={\kappa\over 2 \pi}
	\ee
	We can also attribute to the horizon,  as we do  black hole horizons in AdS or flat space, an entropy
	\be
	\label{defentropy}
	S={A\over 4 G}
	\ee
	This yields from eq.(\ref{finforma}) the first law in the form 
	\be
	\label{qfirstlaw}
	T \Delta S=-\Delta P
	\ee
	Eq.(\ref{qfirstlaw}) is one of the main results of the paper.

	\subsection{Both Horizons And The First Law in General} 
	\label{sec.hololaw1}
\begin{figure}[h]
		\centering

		
		\tikzset {_l138u4ewr/.code = {\pgfsetadditionalshadetransform{ \pgftransformshift{\pgfpoint{89.1 bp } { -128.7 bp }  }  \pgftransformscale{1.32 }  }}}
		\pgfdeclareradialshading{_tjwqe12qe}{\pgfpoint{-72bp}{104bp}}{rgb(0bp)=(1,1,1);
			rgb(0bp)=(1,1,1);
			rgb(25bp)=(0.48,0.15,0.15);
			rgb(400bp)=(0.48,0.15,0.15)}
		
		
		\tikzset {_ve911eqqp/.code = {\pgfsetadditionalshadetransform{ \pgftransformshift{\pgfpoint{89.1 bp } { -128.7 bp }  }  \pgftransformscale{1.32 }  }}}
		\pgfdeclareradialshading{_zzcgu11el}{\pgfpoint{-72bp}{104bp}}{rgb(0bp)=(1,1,1);
			rgb(0bp)=(1,1,1);
			rgb(25bp)=(0.48,0.15,0.15);
			rgb(400bp)=(0.48,0.15,0.15)}
		\tikzset{every picture/.style={line width=0.75pt}} 
		
		\begin{tikzpicture}[x=0.75pt,y=0.75pt,yscale=-1,xscale=1]
			
			\draw  [color={rgb, 255:red, 255; green, 255; blue, 255 }  ,draw opacity=1 ][fill={rgb, 255:red, 245; green, 166; blue, 35 }  ,fill opacity=1 ] (370.13,66.13) -- (386.25,66.12) -- (285.45,160.55) -- (269.32,160.56) -- cycle ;
			\draw   (433,66) -- (267,66) -- (267,232) -- (433,232) -- cycle ;
			\draw    (433,66) -- (267,232) ;
			\draw    (267,66) -- (433,232) ;
			\draw    (267,66) .. controls (265.33,67.67) and (263.67,67.67) .. (262,66) .. controls (260.33,64.33) and (258.67,64.33) .. (257,66) .. controls (255.33,67.67) and (253.67,67.67) .. (252,66) .. controls (250.33,64.33) and (248.67,64.33) .. (247,66) .. controls (245.33,67.67) and (243.67,67.67) .. (242,66) .. controls (240.33,64.33) and (238.67,64.33) .. (237,66) .. controls (235.33,67.67) and (233.67,67.67) .. (232,66) .. controls (230.33,64.33) and (228.67,64.33) .. (227,66) .. controls (225.33,67.67) and (223.67,67.67) .. (222,66) .. controls (220.33,64.33) and (218.67,64.33) .. (217,66) .. controls (215.33,67.67) and (213.67,67.67) .. (212,66) .. controls (210.33,64.33) and (208.67,64.33) .. (207,66) .. controls (205.33,67.67) and (203.67,67.67) .. (202,66) .. controls (200.33,64.33) and (198.67,64.33) .. (197,66) .. controls (195.33,67.67) and (193.67,67.67) .. (192,66) .. controls (190.33,64.33) and (188.67,64.33) .. (187,66) .. controls (185.33,67.67) and (183.67,67.67) .. (182,66) .. controls (180.33,64.33) and (178.67,64.33) .. (177,66) .. controls (175.33,67.67) and (173.67,67.67) .. (172,66) .. controls (170.33,64.33) and (168.67,64.33) .. (167,66) .. controls (165.33,67.67) and (163.67,67.67) .. (162,66) .. controls (160.33,64.33) and (158.67,64.33) .. (157,66) .. controls (155.33,67.67) and (153.67,67.67) .. (152,66) .. controls (150.33,64.33) and (148.67,64.33) .. (147,66) .. controls (145.33,67.67) and (143.67,67.67) .. (142,66) .. controls (140.33,64.33) and (138.67,64.33) .. (137,66) .. controls (135.33,67.67) and (133.67,67.67) .. (132,66) .. controls (130.33,64.33) and (128.67,64.33) .. (127,66) .. controls (125.33,67.67) and (123.67,67.67) .. (122,66) .. controls (120.33,64.33) and (118.67,64.33) .. (117,66) .. controls (115.33,67.67) and (113.67,67.67) .. (112,66) .. controls (110.33,64.33) and (108.67,64.33) .. (107,66) .. controls (105.33,67.67) and (103.67,67.67) .. (102,66) -- (101,66) -- (101,66) ;
			\draw    (101,66) -- (101,232) ;
			\draw    (267,232) .. controls (265.33,233.67) and (263.67,233.67) .. (262,232) .. controls (260.33,230.33) and (258.67,230.33) .. (257,232) .. controls (255.33,233.67) and (253.67,233.67) .. (252,232) .. controls (250.33,230.33) and (248.67,230.33) .. (247,232) .. controls (245.33,233.67) and (243.67,233.67) .. (242,232) .. controls (240.33,230.33) and (238.67,230.33) .. (237,232) .. controls (235.33,233.67) and (233.67,233.67) .. (232,232) .. controls (230.33,230.33) and (228.67,230.33) .. (227,232) .. controls (225.33,233.67) and (223.67,233.67) .. (222,232) .. controls (220.33,230.33) and (218.67,230.33) .. (217,232) .. controls (215.33,233.67) and (213.67,233.67) .. (212,232) .. controls (210.33,230.33) and (208.67,230.33) .. (207,232) .. controls (205.33,233.67) and (203.67,233.67) .. (202,232) .. controls (200.33,230.33) and (198.67,230.33) .. (197,232) .. controls (195.33,233.67) and (193.67,233.67) .. (192,232) .. controls (190.33,230.33) and (188.67,230.33) .. (187,232) .. controls (185.33,233.67) and (183.67,233.67) .. (182,232) .. controls (180.33,230.33) and (178.67,230.33) .. (177,232) .. controls (175.33,233.67) and (173.67,233.67) .. (172,232) .. controls (170.33,230.33) and (168.67,230.33) .. (167,232) .. controls (165.33,233.67) and (163.67,233.67) .. (162,232) .. controls (160.33,230.33) and (158.67,230.33) .. (157,232) .. controls (155.33,233.67) and (153.67,233.67) .. (152,232) .. controls (150.33,230.33) and (148.67,230.33) .. (147,232) .. controls (145.33,233.67) and (143.67,233.67) .. (142,232) .. controls (140.33,230.33) and (138.67,230.33) .. (137,232) .. controls (135.33,233.67) and (133.67,233.67) .. (132,232) .. controls (130.33,230.33) and (128.67,230.33) .. (127,232) .. controls (125.33,233.67) and (123.67,233.67) .. (122,232) .. controls (120.33,230.33) and (118.67,230.33) .. (117,232) .. controls (115.33,233.67) and (113.67,233.67) .. (112,232) .. controls (110.33,230.33) and (108.67,230.33) .. (107,232) .. controls (105.33,233.67) and (103.67,233.67) .. (102,232) -- (101,232) -- (101,232) ;
			\draw    (101,66) -- (267,232) ;
			\draw    (266,66) -- (100,232) ;
			\draw [color={rgb, 255:red, 74; green, 144; blue, 226 }  ,draw opacity=1 ][line width=3]    (463,196) -- (463,113.6) ;
			\draw [shift={(463,108.6)}, rotate = 90] [color={rgb, 255:red, 74; green, 144; blue, 226 }  ,draw opacity=1 ][line width=3]    (20.77,-6.25) .. controls (13.2,-2.65) and (6.28,-0.57) .. (0,0) .. controls (6.28,0.57) and (13.2,2.66) .. (20.77,6.25)   ;
			\draw [line width=1.5]    (327.79,113.34) -- (358.22,85.32) ;
			\draw [shift={(360.43,83.29)}, rotate = 137.36] [color={rgb, 255:red, 0; green, 0; blue, 0 }  ][line width=1.5]    (14.21,-4.28) .. controls (9.04,-1.82) and (4.3,-0.39) .. (0,0) .. controls (4.3,0.39) and (9.04,1.82) .. (14.21,4.28)   ;
			\draw [color={rgb, 255:red, 208; green, 2; blue, 27 }  ,draw opacity=1 ]   (315,114.6) -- (368.13,65.13) ;
			\draw [color={rgb, 255:red, 208; green, 2; blue, 27 }  ,draw opacity=1 ]   (327,124.6) -- (389.13,66.13) ;
			\draw  [draw opacity=0][shading=_tjwqe12qe,_l138u4ewr] (384.71,66.14) .. controls (384.71,63.85) and (386.57,62) .. (388.86,62) .. controls (391.15,62) and (393,63.85) .. (393,66.14) .. controls (393,68.43) and (391.15,70.29) .. (388.86,70.29) .. controls (386.57,70.29) and (384.71,68.43) .. (384.71,66.14) -- cycle ;
			\draw  [draw opacity=0][shading=_zzcgu11el,_ve911eqqp] (362.13,65.13) .. controls (362.13,62.84) and (363.99,60.99) .. (366.27,60.99) .. controls (368.56,60.99) and (370.42,62.84) .. (370.42,65.13) .. controls (370.42,67.42) and (368.56,69.27) .. (366.27,69.27) .. controls (363.99,69.27) and (362.13,67.42) .. (362.13,65.13) -- cycle ;
			\draw [color={rgb, 255:red, 65; green, 117; blue, 5 }  ,draw opacity=1 ]   (316,113.6) -- (327,124.6) ;
			
			\draw (276,98) node [anchor=north east][inner sep=0.75pt]  [font=\small,xscale=-1] [align=left] {$ $};
			\draw (277,94) node [anchor=north west][inner sep=0.75pt]  [font=\small] [align=left] {$\displaystyle H_{1}$};
			\draw (363,44) node [anchor=north west][inner sep=0.75pt]  [font=\small] [align=left] {$\displaystyle L$};
			\draw (386,44.14) node [anchor=north west][inner sep=0.75pt]  [font=\small] [align=left] {$\displaystyle R$};
			\draw (375,128) node [anchor=north west][inner sep=0.75pt]  [font=\small] [align=left] {$\displaystyle H_{2}$};

		\end{tikzpicture}

		\caption{Example of a physical process in the SdS geometry : Matter crossing through cosmological horizon $H_1$.}
		\label{fig.physleft}
	\end{figure}
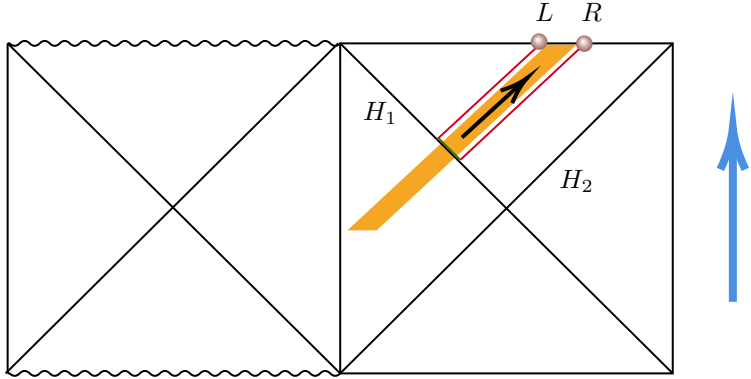

	We  turn next to extending our discussion of the first law by including  the more general situation where there is a flux of matter-energy through both cosmological horizons.
	First consider the case where there is only a perturbation crossing the left horizon $H_1$ in fig. \ref{fig.physleft}, with no perturbation cross $H_2$. 
	It is easy to see that the discussion above can be applied to this situation by simply exchanging the $L$ and $R$ labels above. As a result 
	\begin{eqnarray}
		\label{flipsigns}
		\Delta A & \rightarrow  & -\Delta A \\
		\Delta P & \rightarrow &  -\Delta P
	\end{eqnarray}
	keeping the form of the first law, eq.(\ref{finforma})  unchanged. 
	In more detail note that the flow of the killing vector $\partial_t$ shown in fig. \ref{fig.tsds} is oriented towards the chronological future  for $H_2$, but the past for $H_1$. 
	As a result going towards the future in chronological time the change in the area of $H_1$ is given by 
	\be
	\label{chngareas}
	\Delta A=A_L-A_R
	\ee
	see fig. \ref{fig.physleft}. 
	This accounts for the first change in sign in eq.(\ref{flipsigns}) 
	Also due to the opposite orientation of the Killing vector, the flux crossing $H_1$ is given by 
	\be
	\label{fluxh1}
	\Delta {\cal E}=-\int^\infty_{0} d\lambda \sqrt{\sigma}  d^2y {\cal T}_{\mu \nu} \xi^\nu k^\mu 
	\ee
	where $\xi=\partial_t$, $k^\mu$ is the null generator which is oriented along the chronological future and $\lambda$ is the affine parameter, which also increases towards the chronological future. Comparing with eq.(\ref{defmass1}) we see that there is an extra sign in front. This accounts for the change of sign in the second line in eq.(\ref{flipsigns}). 
	
	The bottom line is that for  perturbations crossing either of the two horizons, $H_1,H_2$ we find the same first law, relating the change in pressure on the boundary to the change of the horizon area.   Its universal form suggests that we can define the entropy, which so far has  being associated with one or the other  horizon, to be in  fact an attribute of the boundary theory itself. 
	The stationary solutions eq.(\ref{nrmet}) lead to a relation between the entropy and pressure $P$ given implicitly by the conditions
	\be
	\label{relentropya}
	S(P)={\pi \over G} r^2_c(P)
	\ee
	Here  $r_c(P)$ is determined as a function of $P$ as follows. First note that $r_c$ is the   larger real  root of \eq{khloc}, and therefore a function of $M$, $r_c(M)$. Also  note from \eq{inval}, \eqref{fival}, see also eq.\eqref{Pvala}, that $P=M$, this yields the function $r_c(P)$.  
	The relation eq.(\ref{relentropya}) between $S$ and $P$ can be thought of as the analogue of an  equation of state relating the two quantities. 
	
\begin{figure}[h]
		\centering

		
		\tikzset {_5r6w6urdz/.code = {\pgfsetadditionalshadetransform{ \pgftransformshift{\pgfpoint{89.1 bp } { -128.7 bp }  }  \pgftransformscale{1.32 }  }}}
		\pgfdeclareradialshading{_t5ns5jbtm}{\pgfpoint{-72bp}{104bp}}{rgb(0bp)=(1,1,1);
			rgb(0bp)=(1,1,1);
			rgb(25bp)=(0.48,0.15,0.15);
			rgb(400bp)=(0.48,0.15,0.15)}
		
		
		\tikzset {_d9surekcz/.code = {\pgfsetadditionalshadetransform{ \pgftransformshift{\pgfpoint{89.1 bp } { -128.7 bp }  }  \pgftransformscale{1.32 }  }}}
		\pgfdeclareradialshading{_7n0oc098d}{\pgfpoint{-72bp}{104bp}}{rgb(0bp)=(1,1,1);
			rgb(0bp)=(1,1,1);
			rgb(25bp)=(0.48,0.15,0.15);
			rgb(400bp)=(0.48,0.15,0.15)}
		
		
		\tikzset {_22724ujjm/.code = {\pgfsetadditionalshadetransform{ \pgftransformshift{\pgfpoint{89.1 bp } { -128.7 bp }  }  \pgftransformscale{1.32 }  }}}
		\pgfdeclareradialshading{_jbtb8a75y}{\pgfpoint{-72bp}{104bp}}{rgb(0bp)=(1,1,1);
			rgb(0bp)=(1,1,1);
			rgb(25bp)=(0.48,0.15,0.15);
			rgb(400bp)=(0.48,0.15,0.15)}
		
		
		\tikzset {_pkr4j3v5b/.code = {\pgfsetadditionalshadetransform{ \pgftransformshift{\pgfpoint{89.1 bp } { -128.7 bp }  }  \pgftransformscale{1.32 }  }}}
		\pgfdeclareradialshading{_osze7iv7f}{\pgfpoint{-72bp}{104bp}}{rgb(0bp)=(1,1,1);
			rgb(0bp)=(1,1,1);
			rgb(25bp)=(0.48,0.15,0.15);
			rgb(400bp)=(0.48,0.15,0.15)}
		\tikzset{every picture/.style={line width=0.75pt}} 
		
		\begin{tikzpicture}[x=0.75pt,y=0.75pt,yscale=-1,xscale=1]
			
			\draw  [color={rgb, 255:red, 255; green, 255; blue, 255 }  ,draw opacity=1 ][fill={rgb, 255:red, 245; green, 166; blue, 35 }  ,fill opacity=1 ] (370.13,66.13) -- (386.25,66.12) -- (285.45,160.55) -- (269.32,160.56) -- cycle ;
			\draw [color={rgb, 255:red, 208; green, 2; blue, 27 }  ,draw opacity=1 ]   (306.27,67.13) -- (369.43,126.54) ;
			\draw [color={rgb, 255:red, 208; green, 2; blue, 27 }  ,draw opacity=1 ]   (327.86,66.14) -- (381.43,114.54) ;
			\draw  [color={rgb, 255:red, 255; green, 255; blue, 255 }  ,draw opacity=1 ][fill={rgb, 255:red, 126; green, 211; blue, 33 }  ,fill opacity=1 ] (326.17,67.13) -- (309.03,67.12) -- (410.27,161.55) -- (427.41,161.56) -- cycle ;
			\draw   (433,66) -- (267,66) -- (267,232) -- (433,232) -- cycle ;
			\draw    (433,66) -- (267,232) ;
			\draw    (267,66) -- (433,232) ;
			\draw    (267,66) .. controls (265.33,67.67) and (263.67,67.67) .. (262,66) .. controls (260.33,64.33) and (258.67,64.33) .. (257,66) .. controls (255.33,67.67) and (253.67,67.67) .. (252,66) .. controls (250.33,64.33) and (248.67,64.33) .. (247,66) .. controls (245.33,67.67) and (243.67,67.67) .. (242,66) .. controls (240.33,64.33) and (238.67,64.33) .. (237,66) .. controls (235.33,67.67) and (233.67,67.67) .. (232,66) .. controls (230.33,64.33) and (228.67,64.33) .. (227,66) .. controls (225.33,67.67) and (223.67,67.67) .. (222,66) .. controls (220.33,64.33) and (218.67,64.33) .. (217,66) .. controls (215.33,67.67) and (213.67,67.67) .. (212,66) .. controls (210.33,64.33) and (208.67,64.33) .. (207,66) .. controls (205.33,67.67) and (203.67,67.67) .. (202,66) .. controls (200.33,64.33) and (198.67,64.33) .. (197,66) .. controls (195.33,67.67) and (193.67,67.67) .. (192,66) .. controls (190.33,64.33) and (188.67,64.33) .. (187,66) .. controls (185.33,67.67) and (183.67,67.67) .. (182,66) .. controls (180.33,64.33) and (178.67,64.33) .. (177,66) .. controls (175.33,67.67) and (173.67,67.67) .. (172,66) .. controls (170.33,64.33) and (168.67,64.33) .. (167,66) .. controls (165.33,67.67) and (163.67,67.67) .. (162,66) .. controls (160.33,64.33) and (158.67,64.33) .. (157,66) .. controls (155.33,67.67) and (153.67,67.67) .. (152,66) .. controls (150.33,64.33) and (148.67,64.33) .. (147,66) .. controls (145.33,67.67) and (143.67,67.67) .. (142,66) .. controls (140.33,64.33) and (138.67,64.33) .. (137,66) .. controls (135.33,67.67) and (133.67,67.67) .. (132,66) .. controls (130.33,64.33) and (128.67,64.33) .. (127,66) .. controls (125.33,67.67) and (123.67,67.67) .. (122,66) .. controls (120.33,64.33) and (118.67,64.33) .. (117,66) .. controls (115.33,67.67) and (113.67,67.67) .. (112,66) .. controls (110.33,64.33) and (108.67,64.33) .. (107,66) .. controls (105.33,67.67) and (103.67,67.67) .. (102,66) -- (101,66) -- (101,66) ;
			\draw    (101,66) -- (101,232) ;
			\draw    (267,232) .. controls (265.33,233.67) and (263.67,233.67) .. (262,232) .. controls (260.33,230.33) and (258.67,230.33) .. (257,232) .. controls (255.33,233.67) and (253.67,233.67) .. (252,232) .. controls (250.33,230.33) and (248.67,230.33) .. (247,232) .. controls (245.33,233.67) and (243.67,233.67) .. (242,232) .. controls (240.33,230.33) and (238.67,230.33) .. (237,232) .. controls (235.33,233.67) and (233.67,233.67) .. (232,232) .. controls (230.33,230.33) and (228.67,230.33) .. (227,232) .. controls (225.33,233.67) and (223.67,233.67) .. (222,232) .. controls (220.33,230.33) and (218.67,230.33) .. (217,232) .. controls (215.33,233.67) and (213.67,233.67) .. (212,232) .. controls (210.33,230.33) and (208.67,230.33) .. (207,232) .. controls (205.33,233.67) and (203.67,233.67) .. (202,232) .. controls (200.33,230.33) and (198.67,230.33) .. (197,232) .. controls (195.33,233.67) and (193.67,233.67) .. (192,232) .. controls (190.33,230.33) and (188.67,230.33) .. (187,232) .. controls (185.33,233.67) and (183.67,233.67) .. (182,232) .. controls (180.33,230.33) and (178.67,230.33) .. (177,232) .. controls (175.33,233.67) and (173.67,233.67) .. (172,232) .. controls (170.33,230.33) and (168.67,230.33) .. (167,232) .. controls (165.33,233.67) and (163.67,233.67) .. (162,232) .. controls (160.33,230.33) and (158.67,230.33) .. (157,232) .. controls (155.33,233.67) and (153.67,233.67) .. (152,232) .. controls (150.33,230.33) and (148.67,230.33) .. (147,232) .. controls (145.33,233.67) and (143.67,233.67) .. (142,232) .. controls (140.33,230.33) and (138.67,230.33) .. (137,232) .. controls (135.33,233.67) and (133.67,233.67) .. (132,232) .. controls (130.33,230.33) and (128.67,230.33) .. (127,232) .. controls (125.33,233.67) and (123.67,233.67) .. (122,232) .. controls (120.33,230.33) and (118.67,230.33) .. (117,232) .. controls (115.33,233.67) and (113.67,233.67) .. (112,232) .. controls (110.33,230.33) and (108.67,230.33) .. (107,232) .. controls (105.33,233.67) and (103.67,233.67) .. (102,232) -- (101,232) -- (101,232) ;
			\draw    (101,66) -- (267,232) ;
			\draw    (266,66) -- (100,232) ;
			\draw [color={rgb, 255:red, 74; green, 144; blue, 226 }  ,draw opacity=1 ][line width=3]    (623,196) -- (623,113.6) ;
			\draw [shift={(623,108.6)}, rotate = 90] [color={rgb, 255:red, 74; green, 144; blue, 226 }  ,draw opacity=1 ][line width=3]    (20.77,-6.25) .. controls (13.2,-2.65) and (6.28,-0.57) .. (0,0) .. controls (6.28,0.57) and (13.2,2.66) .. (20.77,6.25)   ;
			\draw    (433,66) .. controls (434.67,64.33) and (436.33,64.33) .. (438,66) .. controls (439.67,67.67) and (441.33,67.67) .. (443,66) .. controls (444.67,64.33) and (446.33,64.33) .. (448,66) .. controls (449.67,67.67) and (451.33,67.67) .. (453,66) .. controls (454.67,64.33) and (456.33,64.33) .. (458,66) .. controls (459.67,67.67) and (461.33,67.67) .. (463,66) .. controls (464.67,64.33) and (466.33,64.33) .. (468,66) .. controls (469.67,67.67) and (471.33,67.67) .. (473,66) .. controls (474.67,64.33) and (476.33,64.33) .. (478,66) .. controls (479.67,67.67) and (481.33,67.67) .. (483,66) .. controls (484.67,64.33) and (486.33,64.33) .. (488,66) .. controls (489.67,67.67) and (491.33,67.67) .. (493,66) .. controls (494.67,64.33) and (496.33,64.33) .. (498,66) .. controls (499.67,67.67) and (501.33,67.67) .. (503,66) .. controls (504.67,64.33) and (506.33,64.33) .. (508,66) .. controls (509.67,67.67) and (511.33,67.67) .. (513,66) .. controls (514.67,64.33) and (516.33,64.33) .. (518,66) .. controls (519.67,67.67) and (521.33,67.67) .. (523,66) .. controls (524.67,64.33) and (526.33,64.33) .. (528,66) .. controls (529.67,67.67) and (531.33,67.67) .. (533,66) .. controls (534.67,64.33) and (536.33,64.33) .. (538,66) .. controls (539.67,67.67) and (541.33,67.67) .. (543,66) .. controls (544.67,64.33) and (546.33,64.33) .. (548,66) .. controls (549.67,67.67) and (551.33,67.67) .. (553,66) .. controls (554.67,64.33) and (556.33,64.33) .. (558,66) .. controls (559.67,67.67) and (561.33,67.67) .. (563,66) .. controls (564.67,64.33) and (566.33,64.33) .. (568,66) .. controls (569.67,67.67) and (571.33,67.67) .. (573,66) .. controls (574.67,64.33) and (576.33,64.33) .. (578,66) .. controls (579.67,67.67) and (581.33,67.67) .. (583,66) .. controls (584.67,64.33) and (586.33,64.33) .. (588,66) .. controls (589.67,67.67) and (591.33,67.67) .. (593,66) .. controls (594.67,64.33) and (596.33,64.33) .. (598,66) -- (599,66) -- (599,66) ;
			\draw    (599,66) -- (599,232) ;
			\draw    (433,232) .. controls (434.67,230.33) and (436.33,230.33) .. (438,232) .. controls (439.67,233.67) and (441.33,233.67) .. (443,232) .. controls (444.67,230.33) and (446.33,230.33) .. (448,232) .. controls (449.67,233.67) and (451.33,233.67) .. (453,232) .. controls (454.67,230.33) and (456.33,230.33) .. (458,232) .. controls (459.67,233.67) and (461.33,233.67) .. (463,232) .. controls (464.67,230.33) and (466.33,230.33) .. (468,232) .. controls (469.67,233.67) and (471.33,233.67) .. (473,232) .. controls (474.67,230.33) and (476.33,230.33) .. (478,232) .. controls (479.67,233.67) and (481.33,233.67) .. (483,232) .. controls (484.67,230.33) and (486.33,230.33) .. (488,232) .. controls (489.67,233.67) and (491.33,233.67) .. (493,232) .. controls (494.67,230.33) and (496.33,230.33) .. (498,232) .. controls (499.67,233.67) and (501.33,233.67) .. (503,232) .. controls (504.67,230.33) and (506.33,230.33) .. (508,232) .. controls (509.67,233.67) and (511.33,233.67) .. (513,232) .. controls (514.67,230.33) and (516.33,230.33) .. (518,232) .. controls (519.67,233.67) and (521.33,233.67) .. (523,232) .. controls (524.67,230.33) and (526.33,230.33) .. (528,232) .. controls (529.67,233.67) and (531.33,233.67) .. (533,232) .. controls (534.67,230.33) and (536.33,230.33) .. (538,232) .. controls (539.67,233.67) and (541.33,233.67) .. (543,232) .. controls (544.67,230.33) and (546.33,230.33) .. (548,232) .. controls (549.67,233.67) and (551.33,233.67) .. (553,232) .. controls (554.67,230.33) and (556.33,230.33) .. (558,232) .. controls (559.67,233.67) and (561.33,233.67) .. (563,232) .. controls (564.67,230.33) and (566.33,230.33) .. (568,232) .. controls (569.67,233.67) and (571.33,233.67) .. (573,232) .. controls (574.67,230.33) and (576.33,230.33) .. (578,232) .. controls (579.67,233.67) and (581.33,233.67) .. (583,232) .. controls (584.67,230.33) and (586.33,230.33) .. (588,232) .. controls (589.67,233.67) and (591.33,233.67) .. (593,232) .. controls (594.67,230.33) and (596.33,230.33) .. (598,232) -- (599,232) -- (599,232) ;
			\draw    (599,66) -- (433,232) ;
			\draw    (434,66) -- (600,232) ;
			
			\draw  [draw opacity=0][shading=_t5ns5jbtm,_5r6w6urdz] (325.71,66.14) .. controls (325.71,63.85) and (327.57,62) .. (329.86,62) .. controls (332.15,62) and (334,63.85) .. (334,66.14) .. controls (334,68.43) and (332.15,70.29) .. (329.86,70.29) .. controls (327.57,70.29) and (325.71,68.43) .. (325.71,66.14) -- cycle ;
			\draw  [draw opacity=0][shading=_7n0oc098d,_d9surekcz] (302.13,66.13) .. controls (302.13,63.84) and (303.99,61.99) .. (306.27,61.99) .. controls (308.56,61.99) and (310.42,63.84) .. (310.42,66.13) .. controls (310.42,68.42) and (308.56,70.27) .. (306.27,70.27) .. controls (303.99,70.27) and (302.13,68.42) .. (302.13,66.13) -- cycle ;
			\draw [line width=1.5]    (374.43,119.54) -- (357.65,104.39) ;
			\draw [shift={(355.43,102.37)}, rotate = 42.1] [color={rgb, 255:red, 0; green, 0; blue, 0 }  ][line width=1.5]    (14.21,-4.28) .. controls (9.04,-1.82) and (4.3,-0.39) .. (0,0) .. controls (4.3,0.39) and (9.04,1.82) .. (14.21,4.28)   ;
			\draw    (369.43,126.54) -- (381.43,114.54) ;
			\draw [line width=1.5]    (322,120.4) -- (339.23,104.42) ;
			\draw [shift={(341.43,102.37)}, rotate = 137.15] [color={rgb, 255:red, 0; green, 0; blue, 0 }  ][line width=1.5]    (14.21,-4.28) .. controls (9.04,-1.82) and (4.3,-0.39) .. (0,0) .. controls (4.3,0.39) and (9.04,1.82) .. (14.21,4.28)   ;
			\draw [color={rgb, 255:red, 208; green, 2; blue, 27 }  ,draw opacity=1 ]   (315,114.6) -- (368.13,65.13) ;
			\draw [color={rgb, 255:red, 208; green, 2; blue, 27 }  ,draw opacity=1 ]   (327,124.6) -- (389.13,66.13) ;
			\draw  [draw opacity=0][shading=_jbtb8a75y,_22724ujjm] (384.71,66.14) .. controls (384.71,63.85) and (386.57,62) .. (388.86,62) .. controls (391.15,62) and (393,63.85) .. (393,66.14) .. controls (393,68.43) and (391.15,70.29) .. (388.86,70.29) .. controls (386.57,70.29) and (384.71,68.43) .. (384.71,66.14) -- cycle ;
			\draw  [draw opacity=0][shading=_osze7iv7f,_pkr4j3v5b] (362.13,65.13) .. controls (362.13,62.84) and (363.99,60.99) .. (366.27,60.99) .. controls (368.56,60.99) and (370.42,62.84) .. (370.42,65.13) .. controls (370.42,67.42) and (368.56,69.27) .. (366.27,69.27) .. controls (363.99,69.27) and (362.13,67.42) .. (362.13,65.13) -- cycle ;
			\draw [color={rgb, 255:red, 65; green, 117; blue, 5 }  ,draw opacity=1 ]   (316,113.6) -- (327,124.6) ;
			
			\draw (276,98) node [anchor=north east][inner sep=0.75pt]  [font=\small,xscale=-1] [align=left] {$ $};
			\draw (277,94) node [anchor=north west][inner sep=0.75pt]  [font=\small] [align=left] {$\displaystyle H_{1}$};
			\draw (405,92) node [anchor=north west][inner sep=0.75pt]  [font=\small] [align=left] {$\displaystyle H_{2}$};
			\draw (346,67) node [anchor=north west][inner sep=0.75pt]  [font=\small] [align=left] {$\displaystyle I$};

		\end{tikzpicture}
		
		\caption{Example of a physical process in the Schwarzschild de Sitter geometry : Matter crossing through both the cosmological horizons $H_1$ and $H_2$.}
		\label{fig.physLR}
	\end{figure}
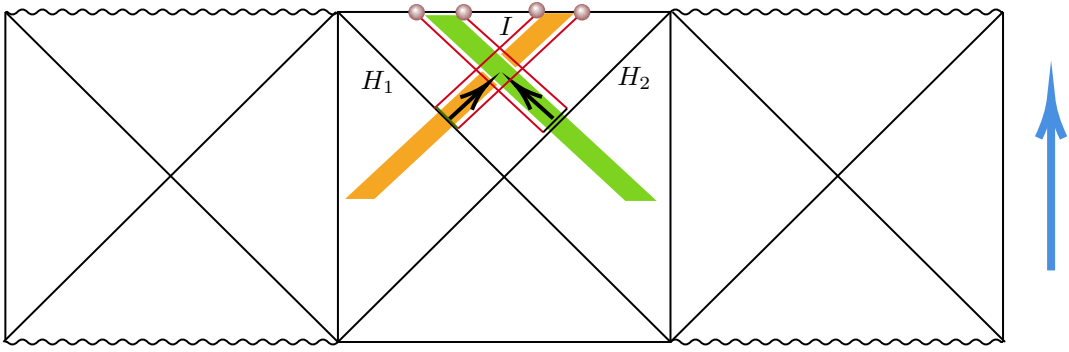

	Our discussion  of the first law suggests that the same  entropy, as a function of $P$,  can be attributed as a property of   the hologram.
	More specifically, suppose during the evolution of the spacetime, there is a quiescent region in between, along the boundary,  denoted by $I$ in fig. \ref{fig.physLR}, 
	where the perturbations have died down and no flux from either horizon is impinging on the boundary. Then in this region one could compute the Pressure using the Brown York stress tensor, \eq{By} and assign an  entropy for this region to be $S(P)$ given by eq.(\ref{relentropya}). 
	The resulting change in entropy would be  related to the change in pressure as given by  the first law eq.(\ref{qfirstlaw}) in this more general case as well. 
	An example of this kind is worked out in considerable detail in appendix \ref{app.jt} for the JT theory in $2$ dimensions.

	Note that in these more general situations the entropy in the in -between region will not agree with the   entropy of  either horizon, $H_1$ or $H_2$,  during the course of their evolution. 
	
	These considerations are quite general and extend to generic situations, where there are multiple sets of fluxes crossing both horizons, or a  flux which impinges on the boundary simultaneously from both Horizons. 
	
	
	In concluding this section, let us also note  that the entropy as we have defined it, intrinsic to the boundary,  is specified  on a co-dimension $1$ surface of the boundary, i.e. at a specific instance that the spatial coordinate $t$ takes. This is different from what one might have naively thought, since  entropy in the bulk is defined 
	on an entire hypersurface of the bulk,  i.e. a specific instance of time in the bulk.

	\subsection{Including Rotation}
	\label{sec.1lkerr}
	Rotation can be easily included in the discussion above. 
	The Killing vector which becomes null at the horizon in this case is given by, \eq{newkil},  
	\be
	\label{kvhor}
	\xi=\partial_t+\Omega_H\partial_{\bar \phi}
	\ee
	The Raychaudhuri equation yields eq.(\ref{firstlaw1}), where $\Delta {\cal E}$ is given in \eq{defdelE} and $\Delta {J}$ in \eq{defdelJ}. 
	
	In turn the Gauss Codazzi relation eq.(\ref{gcc}) becomes 
	\be
	\label{gccrot}
	\int_{\Sigma} d^3y \sqrt{h} {\cal T}_{\alpha \beta}\xi^\alpha n^\beta=\brf{1\over 8 \pi G} \left[\int_{\Sigma}d^3y \sqrt{h} (K^b_t-h^b_tK)|_b +\Omega_H \int_{\Sigma}d^3y \sqrt{h} (K^b_{\bar \phi}-h^b_{\bar \phi}K)|_b\right]
	\ee
	From \eq{angcharge} of the appendix \ref{app.KdScharge} we see that 
	\be
	\label{deferot}
	\brf{1\over 8 \pi G} \int_{\Sigma}d^3y \sqrt{h} (K^b_t-h^b_tK)|_b =-\Delta P
	\ee
	and 
	\be
	\label{defjrot}
	\brf{1\over 8 \pi G} \int_{\Sigma}d^3y \sqrt{h} (K^b_{\bar \phi}-h^b_{\bar \phi}K)|_b =\Delta { J}
	\ee
	It then follows that the first law takes the form 
	\be
	\label{flrot}
	\kappa \Delta A=-(8 \pi G) (\Delta P -\Omega_H \Delta {J})
	\ee
	And after making the identifications eq.(\ref{hawkt}), eq.(\ref{defentropy})
	\be
	\label{flrota}
	T \Delta S= - (\Delta P -\Omega_H \Delta { J})
	\ee
	
	Note that in the rotating case the boundary  entropy is defined as  a function of both $P,J$. This function is obtained as follows. For black holes the area of the cosmological horizon  is a function of $P$ and $J$, as discussed in section \ref{sec.addcom}, see {\eq{Acker}, eq.(\ref{defMtilde}), eq.(\ref{relpval}) and \eq{Jval}.
	This gives the entropy of the cosmological horizon in a black hole solution as a function of $P,J$, to be,
	\be
	\label{entdef}
	S(P,J)=\brf{1\over 4 G} A(P,J)
	\ee
	This same functional form now provides a definition of the entropy in the hologram as a function of the boundary charges, $P, J$. 
	Note, that like  the non-rotating case, it follows  that in general dynamical situations,  with a flux of energy across both cosmological horizons, the entropy defined in this manner must satisfy the first law, eq.(\ref{flrota}). 
	
	Eq.(\ref{flrota}) is therefore the  general universal form of the first law in the hologram and  a central result of our paper.

	\subsection{Evaporation and the Second Law}
	\label{sec.evap}
	Most of the discussion in this section has been about the Holographic description of the first law. But before concluding let us also make some comments about a  second law as seen on the hologram. 
	The boundary is Euclidean, so one  does not expect the entropy to  always increase along any of its directions. Nor could we find any such direction and therefore a version of the second law in our investigation while considering classical processes. For example, going from left to right, i.e. towards increasing values of the spatial $t$ coordinate in fig. \ref{fig.tsds}, is as good as going from right to left, i.e. towards decreasing values of $t$, in general. In the instance considered in fig. \ref{fig.physsds} where the flux emanated from $H_2$ the entropy increased going from left to right, whereas the opposite is true in fig. \ref{fig.physleft}, when the flux was emanating from $H_1$. 
	
	One comment about entropy increase can be made, once black hole evaporation is included, though. Due to the fact that the surface gravity of the black holes horizons, $B_1, B_2$, is  higher than the cosmological horizons, $H_2,H_1$, their temperature is also higher:
	\be
	\label{temcom}
	T_{BH}>T_{cos}
	\ee
	As a result if one waits sufficiently long the black hole will completely evaporate away. For a de Sitter space of radius $R_{ds}={1\over H}$, and a black hole of mass $M$ with $GM\sim R_{ds}$ this will happen on a time scale (as seen in the static patches) of order 
	${t\over R_{ds}} \sim {M^2\over M_{Pl}^2}$, which is long when $R_{ds}$ and $M$ are big in Planck units. 
	
	\begin{figure}[h]
		\centering

		
		\tikzset {_pn7bdc9yt/.code = {\pgfsetadditionalshadetransform{ \pgftransformshift{\pgfpoint{0 bp } { 0 bp }  }  \pgftransformrotate{0 }  \pgftransformscale{2 }  }}}
		\pgfdeclarehorizontalshading{_z6q61iigw}{150bp}{rgb(0bp)=(1,1,0);
			rgb(37.5bp)=(1,1,0);
			rgb(62.5bp)=(0,0.5,0.5);
			rgb(100bp)=(0,0.5,0.5)}
		
		
		\tikzset {_qyqakppwi/.code = {\pgfsetadditionalshadetransform{ \pgftransformshift{\pgfpoint{0 bp } { 0 bp }  }  \pgftransformrotate{0 }  \pgftransformscale{2 }  }}}
		\pgfdeclarehorizontalshading{_2z1wrukgx}{150bp}{rgb(0bp)=(1,1,0);
			rgb(37.5bp)=(1,1,0);
			rgb(62.5bp)=(0,0.5,0.5);
			rgb(100bp)=(0,0.5,0.5)}
		\tikzset{every picture/.style={line width=0.75pt}} 
		
		\begin{tikzpicture}[x=0.75pt,y=0.75pt,yscale=-1,xscale=1]
			
			\draw  [color={rgb, 255:red, 0; green, 0; blue, 0 }  ,draw opacity=1 ] (270.21,212) -- (340.43,283.92) -- (200,283.92) -- cycle ;
			\draw  [color={rgb, 255:red, 0; green, 0; blue, 0 }  ,draw opacity=1 ] (410.21,212) -- (480.43,283.92) -- (340,283.92) -- cycle ;
			\draw [color={rgb, 255:red, 208; green, 2; blue, 27 }  ,draw opacity=1 ]   (270,154.92) -- (270,212) ;
			\draw [color={rgb, 255:red, 10; green, 96; blue, 196 }  ,draw opacity=1 ]   (270,154.92) -- (410,154.92) ;
			\draw [color={rgb, 255:red, 208; green, 2; blue, 27 }  ,draw opacity=1 ]   (410,154.92) -- (410,212) ;
			\draw [color={rgb, 255:red, 144; green, 19; blue, 254 }  ,draw opacity=1 ]   (298.43,223.92) .. controls (298.4,221.56) and (299.57,220.37) .. (301.93,220.35) .. controls (304.29,220.32) and (305.45,219.13) .. (305.42,216.77) .. controls (305.39,214.41) and (306.56,213.22) .. (308.92,213.2) .. controls (311.28,213.18) and (312.45,211.99) .. (312.42,209.63) .. controls (312.39,207.27) and (313.56,206.08) .. (315.92,206.05) .. controls (318.27,206.02) and (319.44,204.83) .. (319.41,202.48) .. controls (319.38,200.12) and (320.55,198.93) .. (322.91,198.91) .. controls (325.27,198.88) and (326.44,197.69) .. (326.41,195.33) .. controls (326.38,192.98) and (327.55,191.79) .. (329.9,191.76) .. controls (332.26,191.74) and (333.43,190.55) .. (333.4,188.19) .. controls (333.37,185.83) and (334.54,184.64) .. (336.9,184.61) .. controls (339.26,184.59) and (340.43,183.4) .. (340.4,181.04) .. controls (340.37,178.69) and (341.54,177.5) .. (343.89,177.47) .. controls (346.25,177.44) and (347.42,176.25) .. (347.39,173.89) .. controls (347.36,171.53) and (348.53,170.34) .. (350.89,170.32) -- (353.43,167.72) -- (359.03,162) ;
			\draw [shift={(360.43,160.57)}, rotate = 134.38] [color={rgb, 255:red, 144; green, 19; blue, 254 }  ,draw opacity=1 ][line width=0.75]    (10.93,-3.29) .. controls (6.95,-1.4) and (3.31,-0.3) .. (0,0) .. controls (3.31,0.3) and (6.95,1.4) .. (10.93,3.29)   ;
			\draw [color={rgb, 255:red, 144; green, 19; blue, 254 }  ,draw opacity=1 ]   (304.43,237.57) .. controls (304.43,235.22) and (305.61,234.04) .. (307.96,234.04) .. controls (310.32,234.04) and (311.5,232.86) .. (311.5,230.5) .. controls (311.5,228.14) and (312.68,226.96) .. (315.04,226.96) .. controls (317.39,226.96) and (318.57,225.78) .. (318.57,223.43) .. controls (318.57,221.07) and (319.75,219.89) .. (322.11,219.89) .. controls (324.46,219.89) and (325.64,218.71) .. (325.64,216.36) .. controls (325.64,214) and (326.82,212.82) .. (329.18,212.82) .. controls (331.53,212.82) and (332.71,211.64) .. (332.71,209.29) .. controls (332.71,206.93) and (333.89,205.75) .. (336.25,205.75) .. controls (338.6,205.75) and (339.78,204.57) .. (339.78,202.22) .. controls (339.78,199.86) and (340.96,198.68) .. (343.32,198.68) .. controls (345.67,198.68) and (346.85,197.5) .. (346.85,195.15) .. controls (346.85,192.79) and (348.03,191.61) .. (350.39,191.61) .. controls (352.75,191.61) and (353.93,190.43) .. (353.93,188.07) .. controls (353.93,185.72) and (355.11,184.54) .. (357.46,184.54) .. controls (359.82,184.54) and (361,183.36) .. (361,181) .. controls (361,178.65) and (362.18,177.47) .. (364.53,177.47) .. controls (366.89,177.47) and (368.07,176.29) .. (368.07,173.93) .. controls (368.07,171.58) and (369.25,170.4) .. (371.6,170.4) -- (372.01,169.99) -- (377.67,164.33) ;
			\draw [shift={(379.08,162.92)}, rotate = 135] [color={rgb, 255:red, 144; green, 19; blue, 254 }  ,draw opacity=1 ][line width=0.75]    (10.93,-3.29) .. controls (6.95,-1.4) and (3.31,-0.3) .. (0,0) .. controls (3.31,0.3) and (6.95,1.4) .. (10.93,3.29)   ;
			\draw [color={rgb, 255:red, 144; green, 19; blue, 254 }  ,draw opacity=1 ]   (369,231) .. controls (366.64,231) and (365.46,229.82) .. (365.46,227.46) .. controls (365.46,225.11) and (364.28,223.93) .. (361.93,223.93) .. controls (359.57,223.93) and (358.39,222.75) .. (358.39,220.39) .. controls (358.39,218.04) and (357.21,216.86) .. (354.86,216.86) .. controls (352.5,216.86) and (351.32,215.68) .. (351.32,213.32) .. controls (351.32,210.97) and (350.14,209.79) .. (347.79,209.79) .. controls (345.43,209.79) and (344.25,208.61) .. (344.25,206.25) .. controls (344.25,203.9) and (343.07,202.72) .. (340.72,202.72) .. controls (338.36,202.72) and (337.18,201.54) .. (337.18,199.18) .. controls (337.18,196.82) and (336,195.64) .. (333.64,195.64) .. controls (331.29,195.64) and (330.11,194.46) .. (330.11,192.11) .. controls (330.11,189.75) and (328.93,188.57) .. (326.57,188.57) .. controls (324.22,188.57) and (323.04,187.39) .. (323.04,185.04) .. controls (323.04,182.68) and (321.86,181.5) .. (319.5,181.5) .. controls (317.15,181.5) and (315.97,180.32) .. (315.97,177.97) .. controls (315.97,175.61) and (314.79,174.43) .. (312.43,174.43) .. controls (310.08,174.43) and (308.9,173.25) .. (308.9,170.9) .. controls (308.9,168.54) and (307.72,167.36) .. (305.36,167.36) -- (304.99,166.99) -- (299.33,161.33) ;
			\draw [shift={(297.92,159.92)}, rotate = 45] [color={rgb, 255:red, 144; green, 19; blue, 254 }  ,draw opacity=1 ][line width=0.75]    (10.93,-3.29) .. controls (6.95,-1.4) and (3.31,-0.3) .. (0,0) .. controls (3.31,0.3) and (6.95,1.4) .. (10.93,3.29)   ;
			\draw [color={rgb, 255:red, 144; green, 19; blue, 254 }  ,draw opacity=1 ]   (379.43,220.43) .. controls (377.07,220.43) and (375.89,219.25) .. (375.89,216.89) .. controls (375.89,214.54) and (374.71,213.36) .. (372.36,213.36) .. controls (370,213.36) and (368.82,212.18) .. (368.82,209.82) .. controls (368.82,207.47) and (367.64,206.29) .. (365.29,206.29) .. controls (362.93,206.29) and (361.75,205.11) .. (361.75,202.75) .. controls (361.75,200.4) and (360.57,199.22) .. (358.22,199.22) .. controls (355.86,199.22) and (354.68,198.04) .. (354.68,195.68) .. controls (354.68,193.32) and (353.5,192.14) .. (351.14,192.14) .. controls (348.79,192.14) and (347.61,190.96) .. (347.61,188.61) .. controls (347.61,186.25) and (346.43,185.07) .. (344.07,185.07) .. controls (341.72,185.07) and (340.54,183.89) .. (340.54,181.54) .. controls (340.54,179.18) and (339.36,178) .. (337,178) .. controls (334.65,178) and (333.47,176.82) .. (333.47,174.47) .. controls (333.47,172.11) and (332.29,170.93) .. (329.93,170.93) .. controls (327.58,170.93) and (326.4,169.75) .. (326.4,167.4) -- (324.99,165.99) -- (319.33,160.33) ;
			\draw [shift={(317.92,158.92)}, rotate = 45] [color={rgb, 255:red, 144; green, 19; blue, 254 }  ,draw opacity=1 ][line width=0.75]    (10.93,-3.29) .. controls (6.95,-1.4) and (3.31,-0.3) .. (0,0) .. controls (3.31,0.3) and (6.95,1.4) .. (10.93,3.29)   ;
			\draw  [fill={rgb, 255:red, 126; green, 211; blue, 33 }  ,fill opacity=1 ] (418.63,129.53) -- (435.25,131.6) -- (437.31,148.22) -- (432.64,143.55) -- (421.72,154.47) -- (412.38,145.12) -- (423.3,134.21) -- cycle ;
			\draw  [fill={rgb, 255:red, 126; green, 211; blue, 33 }  ,fill opacity=1 ] (241.99,148.22) -- (244.05,131.6) -- (260.68,129.53) -- (256,134.21) -- (266.92,145.12) -- (257.58,154.47) -- (246.66,143.55) -- cycle ;
			\path  [shading=_z6q61iigw,_pn7bdc9yt] (477,283) .. controls (477,280.79) and (478.79,279) .. (481,279) .. controls (483.21,279) and (485,280.79) .. (485,283) .. controls (485,285.21) and (483.21,287) .. (481,287) .. controls (478.79,287) and (477,285.21) .. (477,283) -- cycle ; 
			\draw   (477,283) .. controls (477,280.79) and (478.79,279) .. (481,279) .. controls (483.21,279) and (485,280.79) .. (485,283) .. controls (485,285.21) and (483.21,287) .. (481,287) .. controls (478.79,287) and (477,285.21) .. (477,283) -- cycle ; 
			
			\path  [shading=_2z1wrukgx,_qyqakppwi] (197,283) .. controls (197,280.79) and (198.79,279) .. (201,279) .. controls (203.21,279) and (205,280.79) .. (205,283) .. controls (205,285.21) and (203.21,287) .. (201,287) .. controls (198.79,287) and (197,285.21) .. (197,283) -- cycle ; 
			\draw   (197,283) .. controls (197,280.79) and (198.79,279) .. (201,279) .. controls (203.21,279) and (205,280.79) .. (205,283) .. controls (205,285.21) and (203.21,287) .. (201,287) .. controls (198.79,287) and (197,285.21) .. (197,283) -- cycle ; 

			\draw (355,285.49) node [anchor=north west][inner sep=0.75pt]  [font=\small] [align=left] {$\displaystyle \Sigma_c $};
			\draw (190,104) node [anchor=north west][inner sep=0.75pt]   [align=left] {Far Left};
			\draw (440,104) node [anchor=north west][inner sep=0.75pt]   [align=left] {Far Right};
			\draw (186,121.32) node [anchor=north west][inner sep=0.75pt]    {$t=-\infty $};
			\draw (446,119.32) node [anchor=north west][inner sep=0.75pt]    {$t=+\infty $};

		\end{tikzpicture}
		
		\caption{A sketch of the space-time diagram including Hawking evaporation in de Sitter space.}
		\label{fig.evap}
	\end{figure}
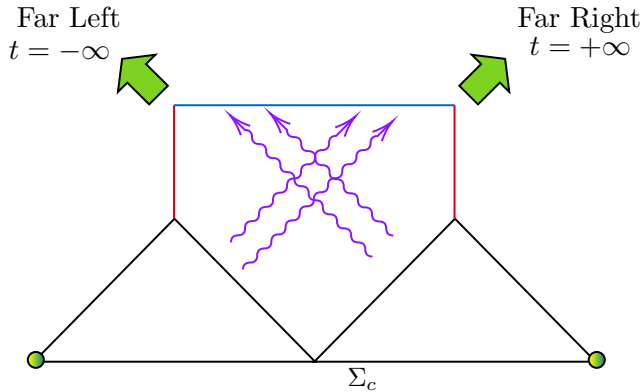

	However, after waiting for such a long time, the black holes in the left and right static patches would have eventually evaporated away leaving empty de Sitter space, which is the configuration with largest entropy, as mentioned in section \ref{sec.intro}. A sketch of the resulting space-time diagram is given in fig. \ref{fig.evap}. 
	At space-like infinity ${\cal I}^+$ in the resulting  spacetime this means  that the  entropy we would associate on the far left and far right (denoted by $t_{-\infty}, t_{\infty}$ in fig. \ref{fig.evap}), after all the Hawking radiation has impinged on the boundary, would take the value $S_{ds}={\pi \over G} R_{ds}^2$. 
	This is the maximum value for the entropy.  Thus, once evaporation is included we can  quite generically conclude that the  entropy in the hologram at the two end points (extreme left and right in fig. \ref{fig.evap}) always becomes the  de Sitter entropy and moreover this is the maximum entropy possible. This is perhaps the closest one can come to a second law in the boundary. Note that this conclusion is invariant under carrying out a left-right exchange of the boundary, i.e. under reflection, $t\rightarrow -t$ on the boundary \footnote{Time reversal symmetry, or more correctly CRT,  would exchange ${\cal I}^+$ and ${\cal I}^-$ in an antipodal manner. States which are invariant under this symmetry would also have maximum dS entropy at the two end points at ${\cal I}^+$. }. 
	
	\section{Concluding Comments}
	\label{Sec.con}
	We saw above that the first law takes a universal form on the future boundary of  de Sitter space, ${\cal I}^+$,  eq.(\ref{qfirstlaw}), eq.(\ref{flrota}). This is the main result of the paper. 
	As a consequence, we can associate a notion of  entropy with the boundary. In the non-rotating case this entropy  is determined as a function of the pressure by the functional form, $S(P)$, \eq{relentropya}.  While this functional form  can be obtained by studying the thermodynamics of the Schwarzschild black hole, it  is more general and also applies to dynamical situations where matter, with energy-momentum  flux, crosses the cosmological horizon and ultimately impinges on the boundary. In such dynamical situations the pressure changes on the boundary and the entropy also then changes in a way given by the functional form, $S(P)$. To be more specific, in the version of holography we are considering the boundary theory lives on $R\times S^2$. And in  more general dynamical situations one associates the entropy with a given location along the real line, $R$, with  this entropy  changing as one moves along $R$, due to  the impinging matter \footnote{One can also consider gravitational waves instead of matter.}.  Similar comments apply for the rotating case where the entropy is now a function of the Pressure, $P$ and angular momentum $J$, eq.(\ref{entdef}). 
	
	It is important to emphasise that the boundary entropy is defined on a co-dimension $1$ slice of the boundary, since it is associated with a particular location along $R$. 
	This is unlike bulk thermodynamic entropy which would be associated with a co-dimension one slice of the bulk. The association of  entropy with co-dimension $1$ surfaces of the boundary is of course what one would expect from the perspective of holography, and we view  our results as  providing  additional evidence for  the hypothesis  that  global dS space admits a hologram  at ${\cal I}^+$.

	The universal form of the first law is attractive, but before going further it is worth reminding the reader of an important caveat.   We deliberately chose a definition for the pressure and angular momentum to make them real.  Actually the boundary stress tensor, $\langle (T^r)^i_j\rangle$,  obtained from the Brown York prescription, is  imaginary, as discussed in \eq{restra}; with $P$ being related to  the integral of $\langle (T^r)^t_t\rangle$ on the boundary $S^2$ by a factor of $i$, \eq{press}. In a similar way the angular momentum is also related to the appropriate components of the boundary stress tensor by a factor of $i$, \eq{shear}. As a result, when we express the first law directly in terms of the components of the boundary stress tensor, ``ugly" factors of $i$ appear in it. We do not have any understanding of these factors from the holographic point of view. It is  clearly   important to understand their origin more deeply. 
	It is also worth noting in this context, that one cannot simply redefine the boundary stress tensor, by multiplying it with  a factor of $i$,  to make it real. 
	This is because, as  follows from the time and spatial reparametrisation invariance of the wave function for any physical state,  the stress tensor is a conserved current.
	And the  Ward identities for translational invariance along the boundary,  which follow from this invariance of the wave function,  completely fixes the   normalisation of the stress tensor, see  eq.(\ref{transward}), and related discussion in section \ref{Sec.Holo}. 
	
	Despite these caveats the universal form of the first law suggests   that   the entropy should have an interpretation in the hologram, possibly as a count of the number of degrees  of freedom. We are of course at an early stage in our understanding of dS holography and it remains to be seen if this is true. One admittedly  simple context in which some progress has been made is that of JT gravity in dS space \cite{sunil_aspJT, KK_probtime, KK_JText, Heldmax_jtds, Harlow_jtds, Cotjen_jtds1, Cotjen_jtds2}. The first law in this theory is discussed in appendix \ref{app.jt1st}. In this case, as discussed in \cite{KK_JText}, a dual holographic description exists and is provided by the SSS matrix model \cite{Saad:2019lba}. The pressure in this case is related, up to a sign, with   the energy, or eigenvalue of the matrix,  in the matrix model. And the entropy of a black hole   as a function of $P$, agrees with the density of states as a function of the energy, in the matrix model. So at least in this case, the entropy in the bulk has an interpretation in terms of  the number of states in the hologram. One caveat to bear in mind is that the SSS matrix model is defined to be in the double scaled limit, with  the rank of the matrix going to infinity. This allows for the density of states to grow without bound as the energy increases. In contrast, in   $4$ dimensional dS space, the entropy has a maximum bound which is obtained  for pure de Sitter space\footnote{JT gravity arises from the near-Nariai limit of the $4$ dimensional Scwarzschild case. Increasing energy in the matrix theory corresponds to going away from the Nariai limit towards empty dS in $4$ dimensions.}. We are currently studying more conventional large $N$ matrix models which might admit dS gravity duals, in the hope that at finite but large $N$ one could  obtain an explanation for the finite dS entropy in terms of the rank of the matrix \cite{KK_JText}. One example is the DSSYK model, see the review \cite{Berkooz:2024lgq} and the references therein; see also \cite{Lin:2022rbf} and \cite{Susskind:2021esx, Susskind:2022bia,Narovlansky:2023lfz, Rahman:2023pgt,Verlinde:2024znh}; and its proposed gravity dual\footnote{We thank O. Parrikar, H. Rajagadia, Pratik Rath and S. Sake for related discussion.} \cite{Blommaert:2023opb}. 
	
	It is worth noting that the analysis we have carried out can be easily generalised to higher dimensions and to include electro-magnetic charge. In all these cases a universal first law would arise in the boundary theory. In the two dimensional context, it can also be easily generalised to  extensions of the JT theory, including models obtained by dimensional reduction from higher dimensions \cite{KK_JText}. 
	
	The discussion of the first law in this paper is in the classical, more precisely, in the semi-classical, limit. Once one  goes beyond this limit, and includes Hawking evaporation and higher order effects, a non-extremal black hole in dS space will eventually evaporate away. The asymptotic boundary will then also receive the Hawking quanta. As discussed in section \ref{sec.evap},  in this case one would expect that  the boundary  entropy increases as we go towards  the  left and right,  starting from the centre region of the boundary, see fig. \ref{fig.evap}. And the  boundary entropy would then  reach its maximum   de Sitter  entropy value, asymptotically,  at the far left and right of the boundary. This asymptotic statement can be taken to be a version of the second law.

	One would  like to extend our  analysis of the bulk  dynamics presented here  by also including the effects of matter entanglement across the cosmological or black hole  horizons to  see if there are  laws of thermodynamics in these cases too, with the generalised entropy having an extra contribution which arises from entanglement across the horizon. We are currently carrying out such an analysis  in the two dimensional context , and hope to report on it soon. Another direction is to consider the effects of higher derivative corrections and analyse whether a first law can be formulated for cosmological horizons involving the Wald entropy \cite{Wald:1993nt}, and a generalisation of the Brown-York stress tensor at ${\cal I}^+$.  
	
	A final comment  is the following. As emphasised above,  our analysis of black hole thermodynamics leads to a definition of boundary  entropy  being associated with a co-dimension $1$  surface of the boundary. This   suggests that if a notion of  entanglement entropy can be associated with the boundary theory, it too should be defined on a co-dimension $1$ surface of the boundary. In fact,  this is line with the explorations of RT surfaces carried out in \cite{Taka_RTds} and \cite{Narayan_RTds}.

	\section*{Acknowledgements}
	%
	We thank Gautam Mandal, Shiraz Minwalla, Onkar Parrikar, Abhijit Gadde, Suvodip Mukherjee and especially Sunil K. Sake for discussions.
	ID, AR, DKS, SPT acknowledge support from Government of India, Department of
	Atomic Energy, under Project Identification No. RTI 4002 and from the Quantum Space-
	Time Endowment of the Infosys Science Foundation. KKN acknowledges the support to CMI from Infosys foundation. 
	{KKN acknowledges the organizers of ``Spring School on Superstring theory and related topics (SMR 4209) held at International Center for Theoretical Physics (ICTP) for hospitality while this work was nearing completion. ID, AR and KKN would like to thank the organizers of ``Asian Winter School 2025" held at IISER Bhopal for their hospitality while the work was in progress and AR is also thankful for giving the opportunity to present parts of this work.}
	AR also acknowledges Nirmalya Kajuri for inviting in the Quantum Cosmos seminar providing the opportunity to present parts of this work in IIT Mandi, and Souparna Nath for discussions.
	ID, AR acknowledge the organizers of ``Indian Strings Meeting 2025" at NISER and IIT Bhubaneswar and ``Progress of Theoretical Bootstrap" at Yukawa Institute for Theoretical Physics; and AR acknowledge ``Student Talks on Trending Topics in Theory 2025" at IISER Bhopal for their hospitality while the work was in progress.
	Most of all, we are grateful to the people of India for generously
	supporting research in String Theory.

	\appendix

	\section{Derivation of Brown York Stress Tensor}
	\label{app.by}
	In this section we derive the Brown York stress tensor starting from the Einstein Hilbert action in presence of a cosmological constant,
	\begin{align}
		16 \pi G S = \int d^4x \sqrt{-g} (R-6) - 2 \int d^3x \sqrt{\abs{h}} K  \label{gravact}
	\end{align}
	We will do it in two ways. First we present the direct method which we will refer to as the Hamiltonian method. We first write the metric in the form
	\begin{align}
		ds^2 = -N^2 dt^2 + h_{ab} dx^a dx^b \label{admmet}
	\end{align}
	where the spacetime is foliated by hypersurface $\Sigma$ which is described by the metric $h_{ab}$. One can then show, see \cite{Poisson_Toolkit},
	\begin{align}
		\dot{h}_{ab} = 2 N K_{ab}~~,~~ h_{ab} \dot{h}^{ab} = 2 N K
	\end{align}
	where $K$ is the trace of extrinsic curvature $K_{ab}$. The action eq.\eqref{gravact} using eq.\eqref{admmet} becomes
	\begin{align}
		16 \pi G S = \int d^4x N \sqrt{\abs{h}} (\mathcal{R} + K_{ab} K^{ab} - K^2 - 6) + 2 \int d^3x \sqrt{\abs{\tilde{h}}} \kappa \label{gravact2}
	\end{align}
	where $\mathcal{R}$ is the Ricci scalar constructed out of $h_{ab}$. We will denote the boundary of the hypersurface $\Sigma$ as $\sigma$ and $\kappa$ is the extrinsic curvature of $\sigma$ embedded in $\Sigma$ and $\tilde{h}$ is the appropriate associated metric.
	
	Now we vary eq.\eqref{gravact2} with respect to $h_{ab}$ which results in the boundary terms after taking into account the equations of motion,
	\begin{align}
		16 \pi G \delta S = \int d^3 x \sqrt{\abs{h}} \delta h_{ab} (K^{ab} - K h^{ab}) \label{gravact3}
	\end{align}
	
	It is instructive to derive the above result in a different method that we will refer to as the Lagrangian method. In this method one lets the boundary metric vary and follow the same path as in the derivation of Einstein field equations from Einstein-Hilbert action keeping in mind that the boundary metric is not fixed. After a somewhat lengthy computation one arrives at the equation
	\begin{align}
		8\pi G	\delta S &= -\frac{1}{2}\int \sqrt{\abs{h}}  n_{\sigma} \left(  g ^{\mu \nu} (\delta \Gamma ^{\sigma} _{\mu \nu}) - g^{\mu \sigma}(\delta \Gamma ^{\nu} _{\nu \mu})\right) - \int \sqrt{\abs{h}} \delta K + \frac{1}{2}  \int \sqrt{\abs{h}} K h_{i j} \delta h^{i j} \nonumber \\
		&= -\frac{1}{2}\int \sqrt{\abs{h}} \left(\mathcal{D}_a E^a+\delta h^{a b} K_{a b} - 2 \delta K \right)- \int \sqrt{\abs{h}} \delta K + \frac{1}{2}  \int \sqrt{\abs{h}} K h_{i j} \delta h^{i j} \nonumber \\
		&= -\frac{1}{2}\int \sqrt{\abs{h}} \mathcal{D}_a E^a - \frac{1}{2} \int \sqrt{\abs{h}} \delta h^{a b} \left(K_{ab} - K h_{ab}\right)
	\end{align}
	The first term is a total derivative, with $\mathcal{D}_a$ being the covariant derivative on the boundary and can be ignored. So we are left with only the final term. Thus,
	\begin{equation}
		\delta S = \frac{1}{16\pi G} \int \sqrt{\abs{h}} \delta h_{a b} \left(K^{ab} - K h^{ab}\right)
	\end{equation}
	which is exactly eq.\eqref{gravact3}.
	
	Thus defining the Brown York stress tensor as
	\begin{equation}
		\langle T^{ab}\rangle = i \frac{2}{\sqrt{\abs{\gamma}}} \frac{\delta S}{\delta \gamma_{ab}}
	\end{equation}
	with
	\begin{equation}
		\gamma_{ab} = \frac{1}{r^2} h_{ab}
	\end{equation}
	we obtain
	\begin{equation}
		\langle T^{ab}\rangle = i r^2 \frac{2}{\sqrt{\abs{\gamma}}} \frac{\delta S}{\delta h_{ab}} = \frac{i}{8\pi G} r^2 \frac{\sqrt{\abs{h}}}{\sqrt{\abs{\gamma}}} (K^{ab}- K h^{ab})|_{r=r_B} \label{Teq}
	\end{equation}
	%
	%
	The above expression in general will be dependent on $r_B$. In the limit $r_B \rightarrow \infty$ the above expression is divergent. To remove the divergences in the stress tensor we can add counter terms to the boundary action, as discussed in section \ref{Sec.Holo}. 
	
	\subsection{Brown York Stress Tensor for Schwarzschild de Sitter Black Hole}
	\label{app.renorm}
	In this section we compute the Brown York stress tensor for a Schwarzschild black hole in dS (SdS). The metric is given by,
	\begin{equation}
		ds^2 = \left(r^2+ \frac{2 G M}{r}-1\right) dt^2 - \frac{dr^2}{\left(r^2+ \frac{2 G M}{r}-1\right)} + r^2 d\Omega_2^2
	\end{equation}
	The boundary is at $r=r_B$. The normal to the boundary then is given by,
	\begin{equation}
		n^{\mu} = \{0, \sqrt{\left(r^2+\frac{2 G M}{r}-1\right)},0,0\}
	\end{equation}
	The extrinsic curvature and its trace can then be computed to be
	\begin{align}
		K^{tt} = \frac{-G M+r^3}{r^2 \left(r^2+ \frac{2 G M}{r}-1\right)^{\frac{3}{2}}}, K^{\theta \theta}=\frac{\left(r^2+ \frac{2 G M}{r}-1\right)^{\frac{1}{2}}}{r^3}, K = \frac{3(r^3+G M)-2r}{r^2 \left(r^2+\frac{2GM}{r}-1\right)^{\frac{1}{2}}}
	\end{align}
	Using the above expressions in eq.\eqref{Teq} we get
	\begin{equation}
		\langle T^{ab}\rangle =i \frac{2}{\sqrt{\gamma}} \frac{\delta S}{\delta \gamma_{ab}} =i r^4 \left(r^2+ \frac{2G M}{r}-1\right)^{\frac{1}{2}}  \frac{1}{8 \pi G}  \left(K^{ab} - K h^{ab}\right)|_{r=r_B}
	\end{equation}
	Thus,
	\begin{equation}
		\langle T^{tt}\rangle = -\frac{2 i r_B^3}{8\pi G}~~,~~ \langle T^{\theta\theta}\rangle =- i\frac{(G M+2r_B^3-r_B) }{8\pi G} \label{Teq1}
	\end{equation}
	Now we add counter terms. Since the boundary is 3 dimensional the counter term action has the form
	\begin{equation}
		S_{ct} = \frac{1}{8\pi G} \int \sqrt{h} (\Lambda_B + {c_2} R[h_{ab}])
	\end{equation}
	from which the stress tensor can be read 
	\begin{equation}
		\langle T^{ab}_{ct}\rangle = \frac{i}{8\pi G} r^4 \left(r^2+ \frac{2GM}{r}-1\right)^{\frac{1}{2}} \left(\Lambda_B h^{ab} - 2c_2 (R^{ab} -\frac{1}{2} R h^{ab})\right)|_{r=r_B}
	\end{equation}
	Evaluating the various components we get
	\begin{equation}
		\langle T^{tt}_{ct}\rangle= - \frac{i M \Lambda_B}{8 \pi} + \frac{i r^3_B \Lambda_B}{8 \pi G}+ \frac{i r_B}{16 \pi G} (\Lambda_B + 4 c_2)~~,~~ \langle T^{\theta \theta}_{ct}\rangle = \frac{i M \Lambda_B}{8 \pi } + \frac{i r^3_B \Lambda_B}{8 \pi G} - \frac{i r_B \Lambda_B}{16 \pi G} \label{Tcteq}
	\end{equation}
	Adding eq.\eqref{Teq1} and eq.\eqref{Tcteq} and demanding that all $r_B$ dependent terms vanish we obtain
	\begin{equation}
		\Lambda_B = 2, c_2 = -{1\over2}
	\end{equation}
	Thus the finite stress tensor components are
	\begin{align}
		\langle (T^r)^{tt}\rangle = \frac{-i M}{4\pi}~~,~~ \langle(T^r)^{\theta \theta}\rangle = \frac{i M}{8 \pi}
	\end{align}
	as mentioned in section \ref{Sec.Holo}. Also, as expected of a stress tensor in a three dimensional boundary CFT it is traceless.
	
	\section{Rotating Black Hole in de Sitter Space}
	\label{app.kdsprop}
	Kerr black hole in de Sitter space (KdS) is described by the metric, written in terms of Boyer-Lindquist (BL) coordinates \cite{entropykds},
	\begin{align}
		\label{kdSmet2}
		ds^2=&-\left({\Delta_r-\Delta_\theta a^2\sin^2\theta\over \sigma}\right)dt^2-{2a\sin^2\theta\over\sigma\zeta}(\Delta_\theta(r^2+a^2)-\Delta_r)dtd\phi\nonumber\\&+{\sin^2\theta\over \sigma\zeta^2}(\Delta_\theta(r^2+a^2)^2-\Delta_ra^2\sin^2\theta)d\phi^2+{\sigma\over \Delta_r}dr^2+{\sigma\over\Delta_\theta}d\theta^2,
	\end{align}
	where the notations are already defined in \eq{not1}.
	\paragraph{Horizons}  Killing horizons of KdS is obtained from the equation 
	\begin{equation}
		\label{horeq}
		\Delta_r=0 \implies r^4+(a^2-1)r^2+2GMr-a^2=0
	\end{equation}
	Among the four roots of \eq{horeq}, one is discarded because it is negative for any parameter value. For general parameter values, the highest root $r_c$ corresponds to cosmological horizon (CH). The explicit expressions of the roots are algebraically complicated and not required explicitly for further discussions, so we do not mention them here. The killing vector
	\begin{equation}
		\label{kil,rot}
		\xi=\del_t+\Omega'_H\del_\phi ~~,~~\Omega'_H={a\zeta\over r_c^2+a^2}
	\end{equation}
	is null at the CH and in fact tangent to the horizon's null generators. Area of the CH is derived as
	\begin{align}
		\label{area}
		A_c=\int d\theta d\phi \sqrt{g_{\theta\theta}[r_c]g_{\phi\phi}[r_c]}=4\pi{r_c^2+a^2\over1+{a^2}}
	\end{align}
	where we used \eq{horeq} and $r_c$ as a root of it.
	
	\paragraph{Surface Gravity}  Surface gravity $\kappa$ is defined as
	\begin{equation}
		\label{sg}
		\kappa^2=-{1\over2}\xi^{\alpha;\beta}\xi_{\alpha;\beta}=-{1\over2}g^{\beta\mu}g_{\nu\alpha}\xi^{\alpha}_{;\mu}\xi^{\nu}_{;\beta}
	\end{equation}
	Using \eq{kil,rot} and the fact that metric components are independent of $t,\phi$, $\kappa$ is found to be,
	\begin{equation}
		\label{sgeval}
		\kappa^2=-{1\over4}g^{\mu\nu}g^{\beta\lambda}(\del_\mu g_{t\beta}.\del_\nu g_{t\lambda}+2\Omega_H\del_\mu g_{t\beta}.\del_\nu g_{\phi\lambda}+\Omega_H^2 \del_\mu g_{\phi\beta}.\del_\nu g_{\phi\lambda})
	\end{equation}
	Using \eq{kdSmet2} and \eq{horeq}, we evaluate the surface gravity $\kappa$ from \eq{sgeval} as
	\begin{equation}
		\label{sg2final}
		\kappa=\frac{a^2 \left(r_c^2+1\right)+3 r_c^4-r_c^2}{2 r_c \left(a^2+r_c^2\right)}
	\end{equation}
	where the particular sign is chosen observing corresponding non-rotating limit $(a\rightarrow0)$. In non-rotating case, the metric is given in \eq{nrmet} from which one can readily derive the surface gravity in terms of radius of cosmological horizon using \eq{khloc} as
	\begin{equation}
		\label{sgsds}
		\kappa_s=\abs{{f'(r_c)\over 2}}={3r_c^2-1\over2r_c}
	\end{equation}
	where $f(r)$ is given in \eq{fdef}. The positivity of the expression of $\kappa$ above is ensured by the fact that $r_c\geq{1\over\sqrt{3}}$ for sub-extremal SdS geometry. We see that \eq{sg2final} agrees with \eq{sgsds} for $a=0$.
	
	
	\subsection{Boundary Charges}
	\label{app.KdScharge}	
	In this section, we compute the boundary pressure and angular momentum from boundary stress tensor in KdS. For this purpose, it is convenient to go to Fefferman-Graham coordinates where the asymptotic form of the metric is given by 
	\begin{equation}
		\label{holomet}
		\lim_{{\bar r}\rightarrow\infty}ds^2={\bar r}^2dt^2-{d\bar{r}^2\over {\bar r}^2}+{\bar r}^2(d{\bar \theta}^2+\sin^2{\bar \theta}~ d\bar{\phi}^2)
	\end{equation}
	However in Boyer-Lindquist coordinates, the metric \eq{kdSmet2} does not become to the asymptotic form \eq{holomet}. Instead we need to do a coordinate transformation of the form given by
	
	\begin{equation}
		\label{coortrans}
		\bar{t}=t~~~~\bar{\phi}=\phi-at~~~~\bar{r}^2={1\over\zeta}(r^2\Delta_{\theta}+a^2\sin^2\theta)~~~~\bar{r}\cos\bar{\theta}=r\cos\theta 
	\end{equation}
	The Killing vector
	\begin{equation}
		\xi=\del_t+\Omega_H' \del_{\phi}
	\end{equation}
	in this new coordinates becomes of the form
	\begin{equation}
		\label{newkil}
		\xi=\del_{\bar t}+\Omega_H\del_{\bar{\phi}}
	\end{equation}
	where $\Omega_H$ is related to $\Omega'_H$, \eq{kil,rot} as
	\begin{equation}
		\Omega_H=\Omega'_H-a
	\end{equation} 
	The inverse transformation of \eq{coortrans} is given in series expansion about $\bar{r}=\infty$ as
	\begin{equation}
		\label{invcotrans}
		r\simeq\bar{r}\sqrt{1+a^2\sin^2\bar{\theta}}~~~~ \cos\theta\simeq {\cos\bar{\theta}\over\sqrt{1+a^2\sin^2\bar{\theta}}}+{a(1+a^2)\cos\bar{\theta}\sin^2\bar{\theta}\over2\bar{r}^2(1+a^2\sin^2\bar{\theta})^{5/2}}
	\end{equation}
	and other two are trivial and exact. It is to note that the transformation closes in $\{t,\phi\}$ and $\{r,\theta\}$. Expressing metric components \eq{kdSmet2} in new coordinates using \eq{invcotrans}, one can show that the metric satisfies the asymptotic form required by \eq{holomet}.\\
	Prescribing \eq{invcotrans}, thereby, as the exact coordinate transformation, we express the bulk metric in coordinates $\{\bar{t},{\bar r},{\bar\theta},{\bar \phi}\}$ and compute ``renormalised" Brown York stress tensor components on the surface $\bar{r}=\bar{r}_b\rightarrow\infty$. For computational simplicity,  we initially expand the bulk metric and consider up to fifth order in $\bar{r}$ about $\bar{r}_b$ which is enough to compute correct stress tensor components. Thereby computing \eq{restra}, we get
	\begin{align}
		\label{bounT1}
		\sqrt{\gamma}\langle (T^r)^{\bar{t}\bar{t}}\rangle=&i\frac{2 \sqrt{2} M \sin\bar{\theta} \left(-a^2\cos (2 \bar{\theta} )+a^2-4\right)}{8\pi\left(-a^2 \cos (2 \bar{\theta} )+a^2+2\right)^{5/2}} \\
		\sqrt{\gamma}\langle (T^r)^{\bar{t}\bar{\phi}}\rangle=&i\frac{12 \sqrt{2} a M \sin\bar{\theta}}{8\pi\left(-a^2 \cos (2 \bar{\theta} )+a^2+2\right)^{5/2}}\\
		\sqrt{\gamma}\langle (T^r)^{\bar{\theta}\bar{\theta}}\rangle=&i\frac{M \csc ^7\bar{\theta} \left(-a^2 \cos ^2\bar{\theta}+a^2+1\right)^{5/2}}{8\pi\left(a^2+\csc ^2\bar{\theta} \right)^4}\\
		\sqrt{\gamma}\langle (T^r)^{\bar{\phi}\bar{\phi}}\rangle=&i\frac{4 \sqrt{2} M \sin \bar{\theta} \left(\csc ^2\bar{\theta}-2 a^2\right)}{8\pi\left(-a^2 \cos (2 \bar{\theta} )+a^2+2\right)^{5/2}}\label{bounT4}
	\end{align}
	where the non-vanishing components of boundary metric $\gamma_{ab}$ is given by
	\begin{equation}
		\label{gamkerr}
		\gamma_{{\bar t}{\bar t}}=1, \gamma_{{\bar \theta}{\bar \theta}}=1, \gamma_{{\bar \phi}{\bar \phi}}=\sin^2{\bar\theta} 
	\end{equation}
	Using \eq{bounT1}--\eqref{bounT4} and \eq{gamkerr}, we compute the boundary pressure $P$ and angular momentum $J$ by \eq{press} and \eq{shear} as
	\begin{equation}
		\label{angcharge}
		P\equiv i  \int_{S} d\theta d\phi \sqrt{\gamma} \langle (T^r)_{\bar t}^{\bar t}\rangle={M\over\zeta^2}~~,~~ J\equiv -i \int_{S} d\theta d\phi \sqrt{\gamma} \langle (T^r)^{\bar t}_{\bar\phi}\rangle = {a M\over \zeta^2}
	\end{equation}
	which agrees with the values mentioned in section \ref{Sec.Holo}. In addition, first law in this case relates the change in area of cosmological horizon to the changes of $P$ and $J$, as discussed in section \ref{sec.1lkerr}.

	\section{Shell Dynamics in Schwarzschild de Sitter Geometry}
	\label{app.shell}
	In this appendix, we study the dynamics of a non-rotating thin spherical shell in the Schwarzschild de Sitter (SdS) background. This will allow us to illustrate some aspects of the first law and also discuss in a concrete context why the fall-off of $\xi^a|_b$, as referred to in section \ref{Sec.hololaw}, goes like $1\over r^3$.
	\subsection{The Dynamics}
	The dynamics of a spherically symmetric shell  has been analysed in \cite{guth} and we will follow this discussion below. Here we are summarise some of the key points. We have in mind, for concreteness, a shell which is first assembled  in the Right Static Patch, fig. \ref{fig.physsds}, and  then evolves, crossing the cosmological horizon $H_2$ and eventually hitting the boundary ${\cal I}^+$. 
	
	Birkoff's theorem implies that the metric inside and outside  the shell  takes the form, in static coordinates, 
	\begin{align}
		&ds_-^2= f_-dt^2-f_-^{-1}dr^2+r^2d\Omega^2~~~~~~~~~~f_-=-1+{2GM_1\over r}+r^2\label{inmetsds}\\
		&ds_+^2= f_+dt^2-f_+^{-1}dr^2+r^2d\Omega^2~~~~~~~~~~f_+=-1+{2GM_2\over r}+r^2\label{outmetsds}
	\end{align}
	where $M_2$ is the BH mass parameter outside the shell and $M_1$ is the mass parameter inside. 
	We expect on physical grounds that $M_2>M_1$; this will be borne out in the calculation below. 
	
	Following \cite{guth} it will be also convenient to use Gaussian Normal coordinates (GNC) to relate the solution inside the shell with that outside. These coordinates are continuous across the shell and in them		
the metric takes the form,
\begin{equation}
	\label{gncmet}
	ds^2=g_{\tau\tau}(\tau,\eta)d\tau^2+d\eta^2+r^2(\tau,\eta)d\Omega^2,
\end{equation}
where $\tau$ is the proper time of the shell and $\eta$ is the proper distance away from the shell along the geodesic normal to the shell. The shell is located at $\eta=0$ and a point outside (or inside) the shell corresponds to taking   $\eta>0$ (or $\eta<0$). The induced metric on the shell is given by
\begin{equation}
	\label{shindm}
	ds^2_{shell}=-d\tau^2+R^2(\tau)d\Omega^2,
\end{equation}
where $r=R(\tau)$ is the radius of the shell at proper time $\tau$. For  \eq{gncmet} and  \eq{shindm} to be true at $\eta=0$, we see that 		
\begin{equation}
	\label{metonsh}
	g_{\tau\tau}(\tau,0)=-1~~,~~r^2(\tau,0)=R^2(\tau)
\end{equation}
Also, comparing eq.(\ref{inmetsds}) and eq.(\ref{outmetsds}) with eq.(\ref{gncmet}) we get 
\begin{align}
	g_{\tau\tau}(\tau,\eta)&=f(r)(dt/d\tau)^2-{1\over f(r)}(dr/d\tau)^2\label{trans00}
	\\
	g_{\tau\eta}=0&=f(r)(dt/d\tau)(dt/d\eta)-{1\over f(r)}(dr/d\tau)(dr/d\eta)\label{trans01}\\
	g_{\eta\eta}=1&=f(r)(dt/d\eta)^2-{1\over f(r)}(dr/d\eta)^2\label{trans11}
\end{align}
where for $\eta<0$, $f(r)=f_-(r)$ and for $\eta>0$, $f(r)=f_+(r)$. The subscript $\pm$ of  $f(r)$ has been suppressed for the moment and will be restored, when   needed, below. 


Using  \eq{metonsh}, we get ${\dot t}$ on the shell $\eta=0$ from \eq{trans00} to be 
\begin{equation}
	\label{dott}
	\dot{t}=\pm{\beta/f(R)}~~,~~\beta\equiv\sqrt{\dot{R}^2-f(R)}
\end{equation}
where `dot' denotes derivative w.r.t. $\tau$. The sign  in \eq{dott}, in front of ${\dot t}$, can be fixed as follows. In our choice of conventions, see fig. \ref{fig.tsds}, the coordinate $t\rightarrow \infty$ at the horizon $H_2$. A shell which crosses $H_2$ and enters the Milne patch $F$, evolves from $t\rightarrow \infty$ to smaller values of $t$. In $F$ the emblackening factor $f>0$, this shows that the sign must be  negative, leading to 
%
\begin{equation}
	\label{dottsign}
	\dot{t}=(-){\beta\over f(R)}
\end{equation}
Note that this relation is also valid in the static patch R where the shell was initially present, since in that case $f<0$ and $t$ increases from finite values to $\infty$. 

The matter stress tensor ${\cal T}^{\mu\nu}$ of the shell is given by, \cite{Poisson_Toolkit}, 
\begin{equation}
	\label{stressenergy}
	{\cal T}^{\mu\nu}=S^{ab}e^\mu_ae^\nu_b\delta(\eta)=\sigma u^\mu u^\nu\delta(\eta),
\end{equation}
where $\sigma$ is the surface tension, $e^\mu_a$ are tangent vectors, with  index $a$ taking values, $(a=\{\tau,\Omega\})$,  and $u^\mu$ is  the velocity vector of the shell.  Note that $\eta=0$ is the location of the shell in GNC, as mentioned  above, where the stress tensor has a delta function singularity. 

Since the Einstein equations are second order in derivatives, in the presence of this stress tensor the metric can be taken to be continuous, as noted earlier, 
and the jump in the extrinsic curvature across the shell can be related to the stress tensor using the Gauss Codazzi relations. These   take the form \cite{Poisson_Toolkit}
\begin{equation}
	\label{2ndjuncul}
	S^a_b={1\over8\pi G}(\Delta K^a_{b}-\delta^a_{b}\Delta K)
\end{equation}
where $\Delta K^a_b=(K^a_b)_+ - (K^a_b)_-$ is the jump in the extrinsic curvature across the surface. 

Computing $K^a_b$ on the surface of the shell $\eta=0$ we get
\begin{equation}
	\label{ExtK}
	(K_{\tau}^{\tau})_{_\pm}={\dot{\beta}_{\pm}\over \dot{R}} \hspace{30pt} (K^{\Omega}_{\Omega})_{_\pm}={\beta_{\pm}\over R}
\end{equation}
where
\begin{equation}
	\label{betasds}
	\beta_+ = \sqrt{1+\dot{R}^2-{2GM_2\over R}-R^2} \hspace{30pt} \beta_- = \sqrt{1+\dot{R}^2-{2GM_1\over R}-R^2}
\end{equation}
It is easy to see that eq.(\ref{2ndjuncul}) is consistent with the relation  $M_2>M_1$, for a positive surface tension, $\sigma>0$. 

Using  \eqref{stressenergy} and \eqref{2ndjuncul}, we can get a constant of motion given by
\begin{equation}
	\label{sol2}
	R(\beta_--\beta_+) =4\pi G\sigma R^2 \equiv G \bar{m}
\end{equation}
The parameter $\bar m$,  can be roughly interpreted to be  the conserved mass of the shell. After  doing some more algebra,  using \eq{sol2} and \eq{betasds}, we then get,  
\begin{equation}
	\label{Msolsds}
	M_2=M_1+\bar{m}\beta_-
	-{G\bar{m}^2\over2R} 
\end{equation}
From eq.(\ref{betasds}) this gives 
\be
\label{eomshella}
{\bar m} \sqrt{1+ {\dot R}^2 - R^2 -{2G M_1\over R}}=M_2-M_1+{G\bar{m}^2\over 2 R}
\ee
This equation determines the radial velocity of the shell ${\dot R}$ as a function of its radial location $R$. In particular we see that the shell approaches ${\cal I}^+$ exponentially rapidly in proper time,  $R\propto e^{\tau}$, due to the exponential expansion of de Sitter space. 
And from eq.(\ref{sol2}) it follows that the surface tension $\sigma$ dilutes away  exponentially $\sigma \propto 1/R^2 \sim e^{-2\tau}$. 
Other features of shell's motion, e.g., for what initial conditions it falls into the black hole of mass $M_1$, or manages to escape to ${\cal I}^+$, can also be determined from eq.(\ref{eomshella}). 

\subsection{Comments on the suppression of $\xi^a|_b$}
\label{app.covxi}

Here we produce some arguments in favour of suppressing the second term in the integral, the RHS of \eq{fm2}, considering thin shell dynamics in background SdS geometry discussed above. To be in accordance with perturbative analysis of first law, we take the energy input $\delta M$ due to shell is small compared to mass of black hole. 
In presence of the shell intersecting the boundary surface $\Sigma$ at $t=t_1$, the induced metric takes the form
\begin{equation}
	\label{shinmet}
	ds_{\Sigma}^2=\brf{r_B^2-1+{2GM\over r_B}+{2G\delta M\over r_B}\Theta(t-t_1)}dt^2 +r_B^2d\Omega^2
\end{equation}
which one can show immediately from \eq{inmetsds} and \eq{outmetsds} considering relabelling of parameters $M_1\rightarrow M+\delta M$ and $M_2 \rightarrow M$.\\
Note that the correction to the induced metric component $h_{tt}$ appears to be $\sim {1\over r_B}$. Therefore, by \eq{relsrb}, $O(\delta M)$ correction to boundary metric component $\gamma_{tt}$ comes to be $\sim {1/r_B^3}$. Now, to compute $\xi^c|_b$ where $\xi$ is a Killing vector given by $\xi=\del_t$, we note that in the absence of time dependence, $\xi^a|_b=0$. In the presence of time dependence, 
\begin{align}
	\label{dxicorr}
	\xi^{a}|_b=\delta(\Gamma_{bc}^a)\xi^c
\end{align}
where $\delta\Gamma^a_{bc}$ is the change in the Christoffel symbol due to the time dependent term and given by
\begin{equation}
	\label{delchr}
	\delta(\Gamma_{bc}^a)={1\over 2}h^{ad}(\del_b \delta h_{dc}+\del_c \delta h_{db}-\del_d \delta h_{bc})\sim {1\over r_B^3}\delta(t-t_1)
\end{equation}
This shows that the quantity $\xi^a|_b$, \eq{dxicorr} falls as ${1\over r^3}$ at large $r$. There is a $\delta$-function in \eq{delchr} which is produced by differentiating the Heaviside function present in \eq{shinmet} but that is an artifact of taking a $\delta$-function thick shell. For more realistic shell, that will get smoothen out resulting no singular answer but $\xi^a|_b$ will fall like $1\over r^3$ at large $r$, as we have argued in section \ref{Sec.hololaw} after \eq{fm2}.

\subsection{A Cross-check}
\label{app.shellfl}
In this section, we compute, to the first order of perturbation, the quantity $\Delta{\cal E}$ as given in \eq{conscur}.
\begin{equation}
	\label{repdelE}
	\Delta{\cal E}=\int_\Sigma \sqrt{h}{\cal T}_{\mu\nu} \xi^\mu n^\nu
\end{equation}
where the matter stress tensor ${\cal T}^{\mu\nu}$ is defined in \eq{stressenergy}, $\xi={\del_t}$ is the Killing vector and $n^\nu$ is the normal to the boundary $r=r_B$
\begin{equation}
	n_\nu=-(f_-(r))^{-{1\over 2}}\delta_\nu^r
\end{equation}
Using these, \eq{repdelE} reduces to 
\begin{equation}
	\label{Isval}
	\Delta{\cal E} = \int_{\Sigma} \sqrt{h} \sigma u^tu^r\xi_tn_r\delta(\eta) = -4\pi \sigma R^2\dot{t} \dot{R}f_-(R)\int_{r=r_B} dt \delta(\eta) = -4\pi \sigma R^2\dot{t} \dot{R}f_-(R){1\over \abs{\del\eta\over\del t}_{r=r_B}}
\end{equation}
The factor ${\del\eta\over\del t}$ on the shell can be evaluated from inverse of the matrix given by
\begin{equation}
	{\cal M}=\left(
	\begin{array}{ccc}
		{\del t\over \del \eta} & {\del t\over \del \tau} \\
		{\del r\over \del \eta} & {\del r\over \del \tau}
	\end{array}
	\right)=\left(
	\begin{array}{ccc}
		{- {\dot{R}\over f(R)}} & {-{\beta\over f(R)}} \\
		{\beta} & {\dot{R}}
	\end{array}
	\right)
\end{equation}
where ${\del t\over \del \tau}$ is taken from \eq{dottsign} and rest of the elements are computed using \eq{trans01} and \eq{trans11}, see \cite{guth} for further discussions.
The particular factor is found to be (1,1) element of its inverse
\begin{equation}
	\abs{{\del\eta\over\del t}}_{r=r_B}=\dot{R} 
\end{equation}
Therefore, \eq{Isval} reduces to
\begin{equation}
	\label{Isval20}
	\Delta{\cal E}= -4\pi \sigma R^2\dot{t}f_-(R)
\end{equation}
Now, we use \eq{dottsign} and \eqref{sol2} to get
\begin{equation}
	\label{Isval2}
	\Delta{\cal E}= 4\pi \sigma \beta_- R^2 = \bar{m}\beta_-
\end{equation}
In the first order of perturbation, i.e, neglecting $O({\bar m}^2)$, \eq{Msolsds} reduces to
\begin{equation}
	\beta_-={M_2-M_1\over {\bar m}}
\end{equation}
Using the above relation, \eq{Isval2} reduces to
\begin{equation}
	\label{IsvaldM}
	\Delta{\cal E}=(-) (M_1-M_2)
\end{equation}
It is to note that $M_1$ (and $M_2$) are the values given by the renormalised Brown York stress tensor inside (and outside) the shell. And, the change in the energy flux comes out to be the difference of the pressure of boundary theory. Particularly, $M_1$ (and $M_2$) is the value of pressure $P$ at a point right (and left) to the shell on the boundary. Hence
\begin{equation}
	\label{etop}
	\Delta{\cal E}=-\Delta P
\end{equation}   
where $\Delta P$ is the difference in $P$ taken in the increasing direction of chronological time.

For the situation where the shell crosses the left cosmological horizon $H_1$, see fig. \ref{fig.physleft}, ${\dot t}$ in the forward Milne patch is positive, since at the $H_1$, $t=-\infty$ and therefore increases in the forward Milne patch along the evolution of the shell, fixing the sign in \eq{dott} as
\begin{equation}
	\label{dott+}
	\dot{t}=(+){\beta\over f(R)}
\end{equation}
Using \eq{dott+} in \eq{Isval20} leads to the result
\begin{equation}
	\Delta{\cal E}=(+) (M_1-M_2)	
\end{equation}
relating the change in the flux to the difference between values of pressure of boundary theory. Here in contrast, $M_1$ (and $M_2$) is the value of pressure $P$ at a point left (and right) to the shell on the boundary. Hence under the same measure of taking difference of $P$ values,
\begin{equation}
	\Delta{\cal E}=+\Delta P
\end{equation} 
For a similar argument, see section \ref{sec.hololaw1}.


\section{First Law in JT Gravity}
\label{app.jt}
In this appendix, we work out in detail how the first law of thermodynamics holds true in JT gravity for the cosmological horizon and how it can be viewed from the boundary theory perspective. 	
\subsection{Basic Setup}

In this section, we review the setup of JT gravity in de Sitter space. The action is given by
\begin{equation}
	\label{JTaction}
	I=I_{B}+I_{GH}+I_{C}
\end{equation}
where
\begin{equation}
	\label{Ibgc}
	I_{B}={1\over16\pi G}\int d^2x\sqrt{-g}\phi(R-2)~~,~~I_{GH}=-{1\over 8\pi G}\int dx\sqrt{\gamma}\phi K~~,~~I_C={1\over 8\pi G}\int dx \sqrt{\gamma}\phi
\end{equation}
In addition, below we consider matter fields which we take to be coupled with gravity alone with no coupling to the dilaton. Variation of the action \eq{JTaction} w.r.t. dilaton $\phi$ thereby gives the corresponding saddle equation as
\begin{equation}
	\label{dilsad}
	R-2=0
\end{equation}
and variation w.r.t. metric $g_{\mu\nu}$ yields the saddle equation as
\begin{equation}
	\label{metsad}
	\nabla_{\mu}\nabla_{\nu}\phi-g_{\mu\nu}\nabla^2\phi-g_{\mu\nu}\phi=-8\pi G{\cal T}_{\mu\nu}
\end{equation}
where ${\cal T}_{\mu\nu}$ is the matter stress tensor.
Representing the metric in conformal gauge
\begin{equation}
	\label{confmet}
	ds^2=e^{2\rho(x_+,x_-)}dx_+dx_-
\end{equation}
\eq{metsad} reduces to
\begin{equation}
	\label{dileom}
	-e^{2\rho}\del_\pm(e^{-2\rho}\del_\pm\phi)=8\pi G{\cal T}_{\pm\pm}~~,~~2\del_+\del_-\phi+e^{2\rho}\phi=16\pi G{\cal T}_{+-}
\end{equation}

\subsubsection{Coordinate Transformations}
\label{sec.basset}
In static coordinates $\{r,t\}$, the vacuum solutions of metric and dilaton are given by
\begin{equation}
	\label{statmet}
	ds^2=-{dr^2\over r^2-m}+(r^2-m)dt^2~~,~~\phi=r
\end{equation}
The Milne patch is defined by $r>\sqrt{m}$ for $m>0$.
Using the following coordinate transformations \footnote{See \cite{sunil_aspJT} for definitions of other coordinate systems in JT gravity and the corresponding transformation formulae.}
\begin{equation}
	\label{Kmilne}
	K_+=\exp\brf{(t+r_*)\sqrt{m}}~,~~K_-=-\exp\brf{(-t+r_*)\sqrt{m}}, \,\, r_*={1\over 2\sqrt{m}} \ln\brf{r-\sqrt{m}\over r+\sqrt{m}}
\end{equation}
metric and dilaton become
\begin{align}
	\label{Kmetdil}
	ds^2={4\over(1+K_+K_-)^2}dK_+dK_-~~,~~\phi=\sqrt{m}{1-K_+K_-\over(1+K_+K_-)}
\end{align}
Comparing the metric, \eq{Kmetdil} with \eq{confmet} in Kruskal coordinates, we get
\begin{equation}
	\label{e2rho}
	e^{2\rho(K_+,K_-)}=4/(1+K_+K_-)^2	
\end{equation}
For ${\cal T}_{++}>0, {\cal T}_{--}>0, {\cal T}_{+-}=0$, solution of \eq{dileom} is given in Kruskal coordinates by
\begin{equation}
	\label{dilgen}
	\phi=-{K_-V(K_+)-K_+U(K_-)\over(1+K_+K_-)}+{V'(K_+)-U'(K_-)\over 2}
\end{equation}
with
\begin{equation}
	\label{triuv}
	V'''(K_+)=-16\pi G {\cal T}_{++}~~,~~ U'''(K_-)=16\pi G{\cal T}_{--}
\end{equation}
Note that the general vacuum solution, i.e. for ${\cal T}_{\mu\nu}=0$, of dilaton in Kruskal coordinates takes the form
\begin{equation}
	\label{vacdilgen}
	\phi={a(1-K_+K_-)+bK_++cK_-\over1+K_+K_-}
\end{equation}
where $a,b,c$ are arbitrary constants.  Eq.\eqref{Kmetdil} comes out for a special choice of the constants viz $a=\sqrt{m},b=0,c=0$.\\~\\
It will turn out to be convenient to work in Fefferman-Graham (FG) coordinates, denoted by $\{{\hat t},{\hat z}\}$, to compute boundary stress-tensor as discussed in \cite{sunil_confjt} for AdS. Metric in this coordinate system is of the form
\begin{equation}
	\label{metbounz}
	ds^2= -{d{\hat z}^2\over {\hat z}^2}+h({\hat z},{\hat t})d{\hat t}^2
\end{equation}
Asymptotically, as ${\hat z}\rightarrow0$, metric up to $O({\hat z}^2)$ behaves as
\begin{align}
	\label{methatz0}
	ds^2= {1\over {\hat z}^2}(-{d{\hat z}^2}+d{\hat t}^2)
\end{align}
and the dilaton behaves as
\begin{equation}
	\label{dilhatz0}
	\phi={1\over {\hat z}}
\end{equation}
Let us assume the coordinate transformation from Kruskal to FG coordinates is given by
\begin{equation}
	\label{ktofg}
	t=H({\hat t})+{\hat z}^2G({\hat t})+\cdots~~;~~z={\hat z}K({\hat t})[1+{\hat z}^2J({\hat t})+\cdots]
\end{equation}
where $\{t,z\}$ are defined as
\begin{equation}
	\label{tz}
	K_+={1-(t+z)\sqrt{m}\over 1+(t+z)\sqrt{m}}~~,~~K_-={1+(t-z)\sqrt{m}\over -1+(t-z)\sqrt{m}}
\end{equation}
Given the condition ${\cal T}_{+-}=0$, e.g. in the case of massless scalar fields, we also assume the general form of dilaton in $\{t,z\}$ as an expansion about $z=0$ \footnote{Allowing quantum backreaction of matter fields in semi-classical limit produces a non-zero ${\cal T}_{+-}$ and introduces a $z$-independent term in \eq{diltz}.}
\begin{equation}
	\label{diltz}
	\phi={f_0(t)\over z}+zf_2(t)+\cdots
\end{equation}
Now, under the conditions given by \eq{dilhatz0} and \eq{metbounz}, we can determine
\begin{equation}
	{\dot H}=K={f_0(t)}~~;~~ G={f_0(t)f_0'(t)\over 2}~~;~~J={f_0'(t)^2\over 4}
\end{equation}
where `dot ' and `prime' denotes derivative w.r.t. $\hat t$ and $t$ respectively. See \cite{sunil_confjt} for similar but more explicit derivation in AdS.
Metric and dilaton are thereby given in FG coordinates as
\begin{equation}
	\label{metfgtop}
	ds^2=-{d{\hat z}^2\over {\hat z}^2}+\brf{{1\over {\hat z}^2}-({f_0'^2\over 2}-f_0f_0'')}d{\hat t}^2
\end{equation}
\begin{equation}
	\label{dilfgtop}
	\phi={1\over{\hat z}}+{{\hat z}}\brf{ {f_0'^2\over 4}+f_0f_2 }
\end{equation}
Note that, using \eq{tz} in the general vacuum solution of $\phi$ \eq{vacdilgen}, 
and then comparing the result with \eq{diltz}, we can find
\begin{equation}
	\label{f0f2}
	f_0(t)=-{(-2a-b+c)+2(b+c)\sqrt{m}t+(2a-b+c)mt^2\over 4\sqrt{m}}~~,~~f_2(t)={(2a-b+c)m\over 4\sqrt{m}}
\end{equation}
which immediately computes
\begin{equation}
	\label{f0f2exp}
	{f_0'^2\over 2}-f_0f_0''=\half(a^2+bc)~~,~~{f_0'^2\over 4}+f_0f_2 = {1\over 4}(a^2+bc)
\end{equation}

\subsubsection{Derivation of Boundary Pressure}
In this section, we compute the boundary pressure
in JT gravity along the lines of appendix A of \cite{KK_probtime}. For this, we first have to compute Brown York stress tensor which can be calculated by working in Fefferman-Graham gauge and computing the variation of action \eq{JTaction} on-shell. Since the resulting analysis is available in appendix A of \cite{KK_probtime} we do not reproduce the derivation here and give only the final result.
The renormalised boundary stress tensor is found to be 
\begin{equation}
	\label{Txxjt}
	(T^r)^{{\hat t}{\hat t}}=i{2\over\sqrt{h}}{\delta I\over\delta h_{{\hat t}{\hat t}}}={i\over 8\pi G h}({\hat z}\del_{\hat z}\phi+\phi)
\end{equation}
We also define the boundary pressure $P$ \footnote{The opposite signature compared to \eq{press} is chosen in the definition so that black hole can have positive value of pressure in dS JT gravity.} by
\begin{equation}
	\label{p}
	P=-i\langle (T^r)^{\hat t}_{\hat t}\rangle=-i{\sqrt{h}\over\sqrt{\gamma}}(T^r)^{\hat t}_{\hat t}={1\over (8\pi G)}{\sqrt{h}\over\sqrt{\gamma}}({\hat z}\del_{\hat z}\phi+\phi)
\end{equation}
where $\gamma$, the boundary metric is defined as
\begin{equation}
	\label{gam}
	\gamma_{{\hat t}{\hat t}}=\lim_{{\hat z}\rightarrow{\hat z}_B}{\hat z}^2 h_{{\hat t}{\hat t}}
\end{equation}
Using \eq{metfgtop} and \eq{dilfgtop}, we can determine the pressure $P$ as
\begin{equation}
	\label{pf0f2}
	P= {1\over 4\pi G}\brf{{f_0'^2\over 4}+f_0f_2}
\end{equation}

\subsubsection{More on the Black Hole in dS JT gravity}
Black hole solution in dS JT gravity is given by the metric \eq{statmet} with $m>0$. The geometry has two horizons -- Cosmological horizon at $r=\sqrt{m}$ and black hole horizon at $r=-\sqrt{m}$. See fig. \ref{fig.bhjt} for the location of horizons in Kruskal coordinates. From \eq{statmet}
the surface gravity is computed as
\begin{equation}
	\label{kappajt}
	\kappa=\sqrt{m}
\end{equation}
at both cosmological and black hole horizons. 
\begin{figure}[h]
	\centering

	\tikzset{every picture/.style={line width=0.75pt}} 
	
	\begin{tikzpicture}[x=0.75pt,y=0.75pt,yscale=-1,xscale=1]
		
		\draw  [fill={rgb, 255:red, 248; green, 231; blue, 28 }  ,fill opacity=1 ] (286.73,65.21) -- (427.52,65.21) -- (427.52,206) -- (286.73,206) -- cycle ;
		\draw  [fill={rgb, 255:red, 248; green, 231; blue, 28 }  ,fill opacity=1 ] (286.88,206) -- (286.88,65.21) -- (215.64,135.6) -- cycle ;
		\draw  [fill={rgb, 255:red, 248; green, 231; blue, 28 }  ,fill opacity=1 ] (427.52,206) -- (427.52,65.21) -- (498.76,135.6) -- cycle ;
		\draw   (286.73,65.21) -- (427.52,65.21) -- (427.52,206) -- (286.73,206) -- cycle ;
		\draw    (286.73,65.21) -- (427.52,206) ;
		\draw    (427.52,65.21) -- (286.73,206) ;
		\draw    (286.73,65.21) .. controls (285.06,66.88) and (283.4,66.88) .. (281.73,65.21) .. controls (280.06,63.54) and (278.4,63.54) .. (276.73,65.21) .. controls (275.06,66.88) and (273.4,66.88) .. (271.73,65.21) .. controls (270.06,63.54) and (268.4,63.54) .. (266.73,65.21) .. controls (265.06,66.88) and (263.4,66.88) .. (261.73,65.21) .. controls (260.06,63.54) and (258.4,63.54) .. (256.73,65.21) .. controls (255.06,66.88) and (253.4,66.88) .. (251.73,65.21) .. controls (250.06,63.54) and (248.4,63.54) .. (246.73,65.21) .. controls (245.06,66.88) and (243.4,66.88) .. (241.73,65.21) .. controls (240.06,63.54) and (238.4,63.54) .. (236.73,65.21) .. controls (235.06,66.88) and (233.4,66.88) .. (231.73,65.21) .. controls (230.06,63.54) and (228.4,63.54) .. (226.73,65.21) .. controls (225.06,66.88) and (223.4,66.88) .. (221.73,65.21) .. controls (220.06,63.54) and (218.4,63.54) .. (216.73,65.21) .. controls (215.06,66.88) and (213.4,66.88) .. (211.73,65.21) .. controls (210.06,63.54) and (208.4,63.54) .. (206.73,65.21) .. controls (205.06,66.88) and (203.4,66.88) .. (201.73,65.21) .. controls (200.06,63.54) and (198.4,63.54) .. (196.73,65.21) .. controls (195.06,66.88) and (193.4,66.88) .. (191.73,65.21) .. controls (190.06,63.54) and (188.4,63.54) .. (186.73,65.21) .. controls (185.06,66.88) and (183.4,66.88) .. (181.73,65.21) .. controls (180.06,63.54) and (178.4,63.54) .. (176.73,65.21) .. controls (175.06,66.88) and (173.4,66.88) .. (171.73,65.21) .. controls (170.06,63.54) and (168.4,63.54) .. (166.73,65.21) .. controls (165.06,66.88) and (163.4,66.88) .. (161.73,65.21) .. controls (160.06,63.54) and (158.4,63.54) .. (156.73,65.21) .. controls (155.06,66.88) and (153.4,66.88) .. (151.73,65.21) .. controls (150.06,63.54) and (148.4,63.54) .. (146.73,65.21) -- (145.94,65.21) -- (145.94,65.21) ;
		\draw    (145.94,65.21) -- (145.94,206) ;
		\draw    (286.73,206) .. controls (285.06,207.67) and (283.4,207.67) .. (281.73,206) .. controls (280.06,204.33) and (278.4,204.33) .. (276.73,206) .. controls (275.06,207.67) and (273.4,207.67) .. (271.73,206) .. controls (270.06,204.33) and (268.4,204.33) .. (266.73,206) .. controls (265.06,207.67) and (263.4,207.67) .. (261.73,206) .. controls (260.06,204.33) and (258.4,204.33) .. (256.73,206) .. controls (255.06,207.67) and (253.4,207.67) .. (251.73,206) .. controls (250.06,204.33) and (248.4,204.33) .. (246.73,206) .. controls (245.06,207.67) and (243.4,207.67) .. (241.73,206) .. controls (240.06,204.33) and (238.4,204.33) .. (236.73,206) .. controls (235.06,207.67) and (233.4,207.67) .. (231.73,206) .. controls (230.06,204.33) and (228.4,204.33) .. (226.73,206) .. controls (225.06,207.67) and (223.4,207.67) .. (221.73,206) .. controls (220.06,204.33) and (218.4,204.33) .. (216.73,206) .. controls (215.06,207.67) and (213.4,207.67) .. (211.73,206) .. controls (210.06,204.33) and (208.4,204.33) .. (206.73,206) .. controls (205.06,207.67) and (203.4,207.67) .. (201.73,206) .. controls (200.06,204.33) and (198.4,204.33) .. (196.73,206) .. controls (195.06,207.67) and (193.4,207.67) .. (191.73,206) .. controls (190.06,204.33) and (188.4,204.33) .. (186.73,206) .. controls (185.06,207.67) and (183.4,207.67) .. (181.73,206) .. controls (180.06,204.33) and (178.4,204.33) .. (176.73,206) .. controls (175.06,207.67) and (173.4,207.67) .. (171.73,206) .. controls (170.06,204.33) and (168.4,204.33) .. (166.73,206) .. controls (165.06,207.67) and (163.4,207.67) .. (161.73,206) .. controls (160.06,204.33) and (158.4,204.33) .. (156.73,206) .. controls (155.06,207.67) and (153.4,207.67) .. (151.73,206) .. controls (150.06,204.33) and (148.4,204.33) .. (146.73,206) -- (145.94,206) -- (145.94,206) ;
		\draw    (145.94,65.21) -- (286.73,206) ;
		\draw    (285.88,65.21) -- (145.09,206) ;
		
		\draw    (73.85,128.82) -- (100.99,155.96) ;
		\draw [shift={(102.4,157.37)}, rotate = 225] [color={rgb, 255:red, 0; green, 0; blue, 0 }  ][line width=0.75]    (10.93,-3.29) .. controls (6.95,-1.4) and (3.31,-0.3) .. (0,0) .. controls (3.31,0.3) and (6.95,1.4) .. (10.93,3.29)   ;
		\draw    (73,135.6) -- (99.29,109.31) ;
		\draw [shift={(100.71,107.9)}, rotate = 135] [color={rgb, 255:red, 0; green, 0; blue, 0 }  ][line width=0.75]    (10.93,-3.29) .. controls (6.95,-1.4) and (3.31,-0.3) .. (0,0) .. controls (3.31,0.3) and (6.95,1.4) .. (10.93,3.29)   ;
		\draw   (427.52,65.21) -- (286.73,65.21) -- (286.73,206) -- (427.52,206) -- cycle ;
		\draw    (427.52,65.21) .. controls (429.19,63.54) and (430.85,63.54) .. (432.52,65.21) .. controls (434.19,66.88) and (435.85,66.88) .. (437.52,65.21) .. controls (439.19,63.54) and (440.85,63.54) .. (442.52,65.21) .. controls (444.19,66.88) and (445.85,66.88) .. (447.52,65.21) .. controls (449.19,63.54) and (450.85,63.54) .. (452.52,65.21) .. controls (454.19,66.88) and (455.85,66.88) .. (457.52,65.21) .. controls (459.19,63.54) and (460.85,63.54) .. (462.52,65.21) .. controls (464.19,66.88) and (465.85,66.88) .. (467.52,65.21) .. controls (469.19,63.54) and (470.85,63.54) .. (472.52,65.21) .. controls (474.19,66.88) and (475.85,66.88) .. (477.52,65.21) .. controls (479.19,63.54) and (480.85,63.54) .. (482.52,65.21) .. controls (484.19,66.88) and (485.85,66.88) .. (487.52,65.21) .. controls (489.19,63.54) and (490.85,63.54) .. (492.52,65.21) .. controls (494.19,66.88) and (495.85,66.88) .. (497.52,65.21) .. controls (499.19,63.54) and (500.85,63.54) .. (502.52,65.21) .. controls (504.19,66.88) and (505.85,66.88) .. (507.52,65.21) .. controls (509.19,63.54) and (510.85,63.54) .. (512.52,65.21) .. controls (514.19,66.88) and (515.85,66.88) .. (517.52,65.21) .. controls (519.19,63.54) and (520.85,63.54) .. (522.52,65.21) .. controls (524.19,66.88) and (525.85,66.88) .. (527.52,65.21) .. controls (529.19,63.54) and (530.85,63.54) .. (532.52,65.21) .. controls (534.19,66.88) and (535.85,66.88) .. (537.52,65.21) .. controls (539.19,63.54) and (540.85,63.54) .. (542.52,65.21) .. controls (544.19,66.88) and (545.85,66.88) .. (547.52,65.21) .. controls (549.19,63.54) and (550.85,63.54) .. (552.52,65.21) .. controls (554.19,66.88) and (555.85,66.88) .. (557.52,65.21) .. controls (559.19,63.54) and (560.85,63.54) .. (562.52,65.21) .. controls (564.19,66.88) and (565.85,66.88) .. (567.52,65.21) -- (568.31,65.21) -- (568.31,65.21) ;
		\draw    (568.31,65.21) -- (568.31,206) ;
		\draw    (427.52,206) .. controls (429.19,204.33) and (430.85,204.33) .. (432.52,206) .. controls (434.19,207.67) and (435.85,207.67) .. (437.52,206) .. controls (439.19,204.33) and (440.85,204.33) .. (442.52,206) .. controls (444.19,207.67) and (445.85,207.67) .. (447.52,206) .. controls (449.19,204.33) and (450.85,204.33) .. (452.52,206) .. controls (454.19,207.67) and (455.85,207.67) .. (457.52,206) .. controls (459.19,204.33) and (460.85,204.33) .. (462.52,206) .. controls (464.19,207.67) and (465.85,207.67) .. (467.52,206) .. controls (469.19,204.33) and (470.85,204.33) .. (472.52,206) .. controls (474.19,207.67) and (475.85,207.67) .. (477.52,206) .. controls (479.19,204.33) and (480.85,204.33) .. (482.52,206) .. controls (484.19,207.67) and (485.85,207.67) .. (487.52,206) .. controls (489.19,204.33) and (490.85,204.33) .. (492.52,206) .. controls (494.19,207.67) and (495.85,207.67) .. (497.52,206) .. controls (499.19,204.33) and (500.85,204.33) .. (502.52,206) .. controls (504.19,207.67) and (505.85,207.67) .. (507.52,206) .. controls (509.19,204.33) and (510.85,204.33) .. (512.52,206) .. controls (514.19,207.67) and (515.85,207.67) .. (517.52,206) .. controls (519.19,204.33) and (520.85,204.33) .. (522.52,206) .. controls (524.19,207.67) and (525.85,207.67) .. (527.52,206) .. controls (529.19,204.33) and (530.85,204.33) .. (532.52,206) .. controls (534.19,207.67) and (535.85,207.67) .. (537.52,206) .. controls (539.19,204.33) and (540.85,204.33) .. (542.52,206) .. controls (544.19,207.67) and (545.85,207.67) .. (547.52,206) .. controls (549.19,204.33) and (550.85,204.33) .. (552.52,206) .. controls (554.19,207.67) and (555.85,207.67) .. (557.52,206) .. controls (559.19,204.33) and (560.85,204.33) .. (562.52,206) .. controls (564.19,207.67) and (565.85,207.67) .. (567.52,206) -- (568.31,206) -- (568.31,206) ;
		\draw    (568.31,65.21) -- (427.52,206) ;
		\draw    (428.37,65.21) -- (569.16,206) ;

		\draw (100.32,92.91) node [anchor=north west][inner sep=0.75pt]  [font=\small] [align=left] {$\displaystyle K_{+}$};
		\draw (101.24,156.52) node [anchor=north west][inner sep=0.75pt]  [font=\small] [align=left] {$\displaystyle K_{-}$};
		\draw (293.6,91.21) node [anchor=north east][inner sep=0.75pt]  [font=\small,xscale=-1] [align=left] {$ $};
		\draw (321.06,80.62) node [anchor=north west][inner sep=0.75pt]  [font=\scriptsize,rotate=-45] [align=left] {$\displaystyle {\textstyle K_{+} =0}$};
		\draw (365.49,109.07) node [anchor=north west][inner sep=0.75pt]  [font=\scriptsize,rotate=-315] [align=left] {$\displaystyle {\textstyle K_{-} =0}$};
		\draw (445.49,169.07) node [anchor=north west][inner sep=0.75pt]  [font=\scriptsize,rotate=-315] [align=left] {$\displaystyle {\textstyle K_{-} =\infty }$};
		\draw (240.27,139.08) node [anchor=north west][inner sep=0.75pt]  [font=\scriptsize,rotate=-45] [align=left] {$\displaystyle {\textstyle K_{+} =-\infty }$};
		\draw (443.27,83.08) node [anchor=north west][inner sep=0.75pt]  [font=\scriptsize,rotate=-45] [align=left] {$\displaystyle {\textstyle K_{+} =\infty }$};
		\draw (236.49,120.07) node [anchor=north west][inner sep=0.75pt]  [font=\scriptsize,rotate=-315] [align=left] {$\displaystyle {\textstyle K_{-} =-\infty }$};

	\end{tikzpicture}
	\caption{Penrose diagram of Black hole sector in dS JT gravity. The region covered by Kruskal coordinates in depicted in yellow.}
	\label{fig.bhjt}
\end{figure}
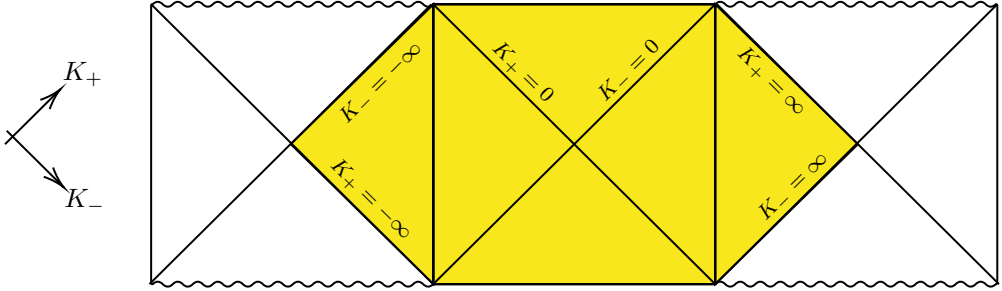
The value of dilaton \eq{statmet} is computed on the cosmological horizon, i.e., at $r=\sqrt{m}$ as
\begin{equation}
	\phi=\sqrt{m}\equiv \phi_h
\end{equation}
Varying w.r.t. $m$ and using \eq{kappajt}, we get
\begin{equation}
	\label{varym}
	\kappa\delta\phi_h={1\over 2}\delta m
\end{equation}
Now, given the expression of dilaton in Kruskal coordinates \eq{Kmetdil}, identifying the constants $a,b,c$ in \eq{vacdilgen}, one can compute boundary pressure $P$ from \eq{pf0f2} using \eq{f0f2exp} as
\begin{equation}
	\label{pbh}
	P={m\over 16\pi G}
\end{equation}
Hence \eq{varym} reduces to
\begin{equation}
	\label{1lawjt}
	\kappa\delta\phi_h=8\pi G \delta P
\end{equation}   
\subsection{Derivation of First Law considering Physical Process}
\label{app.jt1st}

Having derived the first law of thermodynamics in higher dimensional dS background and viewed it from the perspective of hologram on ${\cal I}^+$, it is interesting to ask whether this perspective holds even in JT gravity in de Sitter space. The crucial difference in JT gravity is that for the matter not coupled with dilaton, stress tensor does not back-react on the geometry, rather changes the vacuum dilaton solution, see \eq{dilsad} and \eqref{metsad}. Nonetheless, in this section, we show that first law of thermodynamics holds in this setting in terms of boundary pressure and change in dilaton value which mimics the area of the horizon. The derivation takes into consideration the matter flux crossing $H_2$, fig. \ref{fig.jtphcom+}. Similar calculation for matter crossing $H_1$, e.g. see fig. \ref{fig.physleft}, reproduces the first law as well.
\subsubsection{Change in the Value of Dilaton at the Cosmological Horizon}
In this section, we derive the change in the value of dilaton at the right cosmological horizon $H_2$ in terms of the flux crossing through $H_2$. The location of $H_2$ can be derived by solving 
\be
\label{H2equ}
\del_+\phi=0
\ee
where $\phi$ is the dilaton.
We assume the flux having a compact support within $s_1<K_+<s_2$.\\ In the region $K_+<s_1$, dilaton takes the value as given in \eq{Kmetdil}. Solving \eq{H2equ}, we find $H_2$ located at $K_-=0$ and the dilaton has the value $\phi=\sqrt{m}$ on $H_2$.
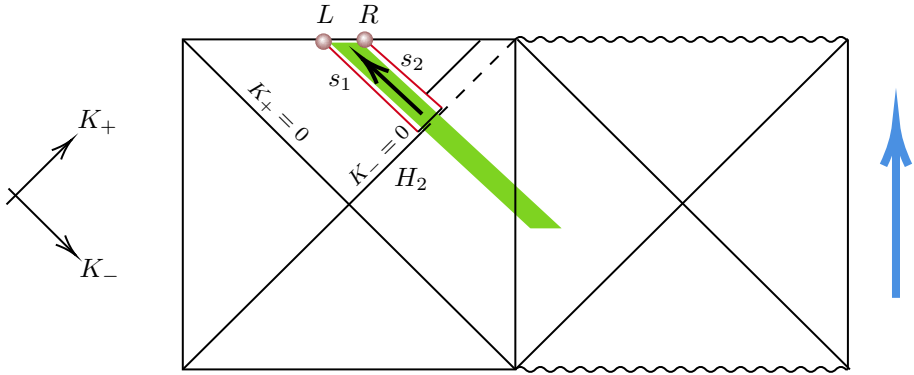
\begin{figure}[h]
	\centering	
	
	
	\tikzset {_4lw790swn/.code = {\pgfsetadditionalshadetransform{ \pgftransformshift{\pgfpoint{89.1 bp } { -128.7 bp }  }  \pgftransformscale{1.32 }  }}}
	\pgfdeclareradialshading{_8eqx4ubad}{\pgfpoint{-72bp}{104bp}}{rgb(0bp)=(1,1,1);
		rgb(0bp)=(1,1,1);
		rgb(25bp)=(0.48,0.15,0.15);
		rgb(400bp)=(0.48,0.15,0.15)}
	
	
	\tikzset {_7j9nygm3w/.code = {\pgfsetadditionalshadetransform{ \pgftransformshift{\pgfpoint{89.1 bp } { -128.7 bp }  }  \pgftransformscale{1.32 }  }}}
	\pgfdeclareradialshading{_eb7f6bhud}{\pgfpoint{-72bp}{104bp}}{rgb(0bp)=(1,1,1);
		rgb(0bp)=(1,1,1);
		rgb(25bp)=(0.48,0.15,0.15);
		rgb(400bp)=(0.48,0.15,0.15)}
	\tikzset{every picture/.style={line width=0.75pt}} 
	
	\begin{tikzpicture}[x=0.75pt,y=0.75pt,yscale=-1,xscale=1]
		
		\draw [color={rgb, 255:red, 208; green, 2; blue, 27 }  ,draw opacity=1 ]   (229.27,72.13) -- (275.43,116.54) ;
		\draw [color={rgb, 255:red, 208; green, 2; blue, 27 }  ,draw opacity=1 ]   (248.86,70.14) -- (287.43,104.54) ;
		\draw  [color={rgb, 255:red, 255; green, 255; blue, 255 }  ,draw opacity=1 ][fill={rgb, 255:red, 126; green, 211; blue, 33 }  ,fill opacity=1 ] (247.17,71.13) -- (230.03,71.12) -- (331.27,165.55) -- (348.41,165.56) -- cycle ;
		\draw   (324,70) -- (158,70) -- (158,236) -- (324,236) -- cycle ;
		\draw    (277.46,116.54) -- (158,236) ;
		\draw    (158,70) -- (324,236) ;
		\draw [color={rgb, 255:red, 74; green, 144; blue, 226 }  ,draw opacity=1 ][line width=3]    (514,200) -- (514,117.6) ;
		\draw [shift={(514,112.6)}, rotate = 90] [color={rgb, 255:red, 74; green, 144; blue, 226 }  ,draw opacity=1 ][line width=3]    (20.77,-6.25) .. controls (13.2,-2.65) and (6.28,-0.57) .. (0,0) .. controls (6.28,0.57) and (13.2,2.66) .. (20.77,6.25)   ;
		\draw    (324,70) .. controls (325.67,68.33) and (327.33,68.33) .. (329,70) .. controls (330.67,71.67) and (332.33,71.67) .. (334,70) .. controls (335.67,68.33) and (337.33,68.33) .. (339,70) .. controls (340.67,71.67) and (342.33,71.67) .. (344,70) .. controls (345.67,68.33) and (347.33,68.33) .. (349,70) .. controls (350.67,71.67) and (352.33,71.67) .. (354,70) .. controls (355.67,68.33) and (357.33,68.33) .. (359,70) .. controls (360.67,71.67) and (362.33,71.67) .. (364,70) .. controls (365.67,68.33) and (367.33,68.33) .. (369,70) .. controls (370.67,71.67) and (372.33,71.67) .. (374,70) .. controls (375.67,68.33) and (377.33,68.33) .. (379,70) .. controls (380.67,71.67) and (382.33,71.67) .. (384,70) .. controls (385.67,68.33) and (387.33,68.33) .. (389,70) .. controls (390.67,71.67) and (392.33,71.67) .. (394,70) .. controls (395.67,68.33) and (397.33,68.33) .. (399,70) .. controls (400.67,71.67) and (402.33,71.67) .. (404,70) .. controls (405.67,68.33) and (407.33,68.33) .. (409,70) .. controls (410.67,71.67) and (412.33,71.67) .. (414,70) .. controls (415.67,68.33) and (417.33,68.33) .. (419,70) .. controls (420.67,71.67) and (422.33,71.67) .. (424,70) .. controls (425.67,68.33) and (427.33,68.33) .. (429,70) .. controls (430.67,71.67) and (432.33,71.67) .. (434,70) .. controls (435.67,68.33) and (437.33,68.33) .. (439,70) .. controls (440.67,71.67) and (442.33,71.67) .. (444,70) .. controls (445.67,68.33) and (447.33,68.33) .. (449,70) .. controls (450.67,71.67) and (452.33,71.67) .. (454,70) .. controls (455.67,68.33) and (457.33,68.33) .. (459,70) .. controls (460.67,71.67) and (462.33,71.67) .. (464,70) .. controls (465.67,68.33) and (467.33,68.33) .. (469,70) .. controls (470.67,71.67) and (472.33,71.67) .. (474,70) .. controls (475.67,68.33) and (477.33,68.33) .. (479,70) .. controls (480.67,71.67) and (482.33,71.67) .. (484,70) .. controls (485.67,68.33) and (487.33,68.33) .. (489,70) -- (490,70) -- (490,70) ;
		\draw    (490,70) -- (490,236) ;
		\draw    (324,236) .. controls (325.67,234.33) and (327.33,234.33) .. (329,236) .. controls (330.67,237.67) and (332.33,237.67) .. (334,236) .. controls (335.67,234.33) and (337.33,234.33) .. (339,236) .. controls (340.67,237.67) and (342.33,237.67) .. (344,236) .. controls (345.67,234.33) and (347.33,234.33) .. (349,236) .. controls (350.67,237.67) and (352.33,237.67) .. (354,236) .. controls (355.67,234.33) and (357.33,234.33) .. (359,236) .. controls (360.67,237.67) and (362.33,237.67) .. (364,236) .. controls (365.67,234.33) and (367.33,234.33) .. (369,236) .. controls (370.67,237.67) and (372.33,237.67) .. (374,236) .. controls (375.67,234.33) and (377.33,234.33) .. (379,236) .. controls (380.67,237.67) and (382.33,237.67) .. (384,236) .. controls (385.67,234.33) and (387.33,234.33) .. (389,236) .. controls (390.67,237.67) and (392.33,237.67) .. (394,236) .. controls (395.67,234.33) and (397.33,234.33) .. (399,236) .. controls (400.67,237.67) and (402.33,237.67) .. (404,236) .. controls (405.67,234.33) and (407.33,234.33) .. (409,236) .. controls (410.67,237.67) and (412.33,237.67) .. (414,236) .. controls (415.67,234.33) and (417.33,234.33) .. (419,236) .. controls (420.67,237.67) and (422.33,237.67) .. (424,236) .. controls (425.67,234.33) and (427.33,234.33) .. (429,236) .. controls (430.67,237.67) and (432.33,237.67) .. (434,236) .. controls (435.67,234.33) and (437.33,234.33) .. (439,236) .. controls (440.67,237.67) and (442.33,237.67) .. (444,236) .. controls (445.67,234.33) and (447.33,234.33) .. (449,236) .. controls (450.67,237.67) and (452.33,237.67) .. (454,236) .. controls (455.67,234.33) and (457.33,234.33) .. (459,236) .. controls (460.67,237.67) and (462.33,237.67) .. (464,236) .. controls (465.67,234.33) and (467.33,234.33) .. (469,236) .. controls (470.67,237.67) and (472.33,237.67) .. (474,236) .. controls (475.67,234.33) and (477.33,234.33) .. (479,236) .. controls (480.67,237.67) and (482.33,237.67) .. (484,236) .. controls (485.67,234.33) and (487.33,234.33) .. (489,236) -- (490,236) -- (490,236) ;
		\draw    (490,70) -- (324,236) ;
		\draw    (325,70) -- (491,236) ;
		
		\draw  [draw opacity=0][shading=_8eqx4ubad,_4lw790swn] (244.71,70.14) .. controls (244.71,67.85) and (246.57,66) .. (248.86,66) .. controls (251.15,66) and (253,67.85) .. (253,70.14) .. controls (253,72.43) and (251.15,74.29) .. (248.86,74.29) .. controls (246.57,74.29) and (244.71,72.43) .. (244.71,70.14) -- cycle ;
		\draw  [draw opacity=0][shading=_eb7f6bhud,_7j9nygm3w] (224.13,71.13) .. controls (224.13,68.84) and (225.99,66.99) .. (228.27,66.99) .. controls (230.56,66.99) and (232.42,68.84) .. (232.42,71.13) .. controls (232.42,73.42) and (230.56,75.27) .. (228.27,75.27) .. controls (225.99,75.27) and (224.13,73.42) .. (224.13,71.13) -- cycle ;
		\draw [line width=1.5]    (277.43,107.54) -- (249.62,81.48) ;
		\draw [shift={(247.43,79.43)}, rotate = 43.14] [color={rgb, 255:red, 0; green, 0; blue, 0 }  ][line width=1.5]    (14.21,-4.28) .. controls (9.04,-1.82) and (4.3,-0.39) .. (0,0) .. controls (4.3,0.39) and (9.04,1.82) .. (14.21,4.28)   ;
		\draw    (275.43,116.54) -- (287.43,104.54) ;
		\draw    (279.71,97.29) -- (306.43,70.57) ;
		\draw  [dash pattern={on 4.5pt off 4.5pt}]  (277.46,116.54) -- (324,70) ;
		\draw    (71,145) -- (103.25,177.25) ;
		\draw [shift={(104.67,178.67)}, rotate = 225] [color={rgb, 255:red, 0; green, 0; blue, 0 }  ][line width=0.75]    (10.93,-3.29) .. controls (6.95,-1.4) and (3.31,-0.3) .. (0,0) .. controls (3.31,0.3) and (6.95,1.4) .. (10.93,3.29)   ;
		\draw    (70,153) -- (101.25,121.75) ;
		\draw [shift={(102.67,120.33)}, rotate = 135] [color={rgb, 255:red, 0; green, 0; blue, 0 }  ][line width=0.75]    (10.93,-3.29) .. controls (6.95,-1.4) and (3.31,-0.3) .. (0,0) .. controls (3.31,0.3) and (6.95,1.4) .. (10.93,3.29)   ;
		
		\draw (167,102) node [anchor=north east][inner sep=0.75pt]  [font=\small,xscale=-1] [align=left] {$ $};
		\draw (262,133) node [anchor=north west][inner sep=0.75pt]  [font=\small] [align=left] {$\displaystyle H_{2}$};
		\draw (223,51) node [anchor=north west][inner sep=0.75pt]  [font=\small] [align=left] {$\displaystyle L$};
		\draw (244,51.14) node [anchor=north west][inner sep=0.75pt]  [font=\small] [align=left] {$\displaystyle R$};
		\draw (229,86) node [anchor=north west][inner sep=0.75pt]  [font=\small] [align=left] {$\displaystyle s_{1}$};
		\draw (265,76) node [anchor=north west][inner sep=0.75pt]  [font=\small] [align=left] {$\displaystyle s_{2}$};
		\draw (104,104) node [anchor=north west][inner sep=0.75pt]  [font=\small] [align=left] {$\displaystyle K_{+}$};
		\draw (105,179) node [anchor=north west][inner sep=0.75pt]  [font=\small] [align=left] {$\displaystyle K_{-}$};
		\draw (237.27,139.06) node [anchor=north west][inner sep=0.75pt]  [font=\scriptsize,rotate=-315.04] [align=left] {$\displaystyle {\textstyle K_{-} =0}$};
		\draw (195.06,87.62) node [anchor=north west][inner sep=0.75pt]  [font=\scriptsize,rotate=-45] [align=left] {$\displaystyle {\textstyle K_{+} =0}$};

	\end{tikzpicture}
	\caption{Example of Physical process in de Sitter JT gravity : Matter crossing through right cosmological horizon $H_2$.}
	\label{fig.jtphcom+}
\end{figure}

After the flux crosses the horizon, i.e., region $K_+>s_2$, the horizon $H_2$ shifts to a new location and dilaton takes different value on the new horizon. To compute the corresponding change in the value of dilaton in terms of the flux value, 
we first define the flux crossing through horizon as
\begin{equation}
	\label{Ih}
	\Delta {\cal E}=\int_{H_2} d\lambda ~{\cal T}_{\mu\nu}\xi^\mu k^\nu
\end{equation}
where the integral is defined on the right cosmological horizon. $\xi^\mu$ is the Killing vector ${\del\over\del t}$ related to $k^\mu={\del\over\del \lambda}$, the parallel transported generator of horizon ${H_2}$, as
\begin{equation}
	\xi^\mu=\kappa\lambda k^\mu
\end{equation}
where $\kappa$ is the surface gravity given by \eq{kappajt}
and $\lambda\in [0,\infty]$ is the affine parameter along $H_2$. Using these, \eq{Ih} reduces to
\begin{equation}
	\label{Ihjt}
	\Delta {\cal E}=\kappa\int_{H_2} d\lambda ~{\cal T}_{++}\lambda\left({dK_+\over d\lambda}\right)^2=\kappa\int_{H_2} dK_+ ~{\cal T}_{++}\lambda\left({dK_+\over d\lambda}\right)
\end{equation}
Note again that we only consider ${\cal T}_{++}$ to be non-vanishing and contributing to the change in dilaton profile. Using \eq{dileom}, we get
\begin{equation}
	\label{Ihent}
	8\pi G\Delta {\cal E}=-\kappa\int_{H_2} dK_+ ~e^{2\rho}\del_+(e^{-2\rho}\del_+\phi)\lambda\left({dK_+\over d\lambda}\right)
\end{equation}
Noting that the coordinate $K_+$ running along $H_2$ satisfies null geodesic equation 
\begin{equation}
	{d^2K_+\over d\lambda^2}+\Gamma_{++}^+ \brf{{dK_+\over d\lambda}}^2=0
\end{equation}
we can evaluate ${dK_+\over d\lambda}$ to $0$-th order of perturbation to be
\begin{equation}
	\label{dk+dlmval}
	{dK_+\over d\lambda}=1
\end{equation}
and henceforth the integral \eq{Ihent} gives
\begin{equation}
	\label{Ihentval}
	8\pi G\Delta {\cal E}=-\kappa\int_{H_2} dK_+ ~e^{2\rho}\del_+(e^{-2\rho}\del_+\phi)\lambda
\end{equation}
Using \eq{e2rho} and the fact that $\del_+\phi$ vanishes at 0-th order on $H_2$, we get
\begin{equation}
	\label{bnterm}
	8\pi G\Delta {\cal E}=-\kappa\brt{\lambda\del_+\phi}_{0}^{\infty}+\kappa\int_{H_2} dK_+\del_+\phi=-\kappa\brt{\lambda\del_+\phi}_{0}^{\infty}+\kappa\Delta\phi
\end{equation}
where $\Delta\phi=\phi(\infty)-\phi(0)$ is the difference between final and initial values of dilaton. Due to the matter flux having compact support, we can define location of $H_2$ by $\del_+\phi=0$ after the flux having crossed $H_2$. Therefore, the first term in RHS of second equality in \eq{bnterm} vanishes and we get
\begin{equation}
	\label{areatoEjt}
	8\pi G\Delta {\cal E}=\kappa\Delta\phi
\end{equation}

\subsubsection{Change in Boundary Pressure}
Next, by conservation of matter current $j^\mu$ given by
\begin{equation}
	j^\mu={\cal T}^{\mu\nu}\xi_\mu
\end{equation}
the flux integral at horizon $H_2$, \eq{Ih}, can be related to flux through the boundary as
\begin{equation}
	\label{delEjtb}
	\int_{\Sigma} d\hat{t}\sqrt{h}{\cal T}_{\mu\nu}\xi^\mu n^\nu = \Delta{\cal E}
\end{equation}
where the boundary $\Sigma$ is the surface where $\phi$, the dilaton, is constant
\begin{equation}
	\label{phib}
	\phi=\phi_B \gg 1
\end{equation}
It is convenient to work here in FG coordinates $\{{\hat t},{\hat z}\}$ as introduced in section \ref{sec.basset}. In these coordinates, \eq{phib} corresponds to the value of $\hat z$ at ${\Sigma}$ by \eq{dilhatz0} as
\begin{equation}
	\hat{z}={1\over \phi_B}\equiv{\hat z}_B\rightarrow0
\end{equation}
Thereby the boundary metric is defined as in \eq{gam}.
Also, $\xi=\del_{\hat t}$ is the Killing vector in the unperturbed solution and the outward normal to the boundary $\Sigma$ is given by
\begin{equation}
	\label{n}
	n^\mu=(-){\hat z} \delta^\mu_{\hat z}
\end{equation}
Note that the $-ve$ sign above appears because of ${\hat z}$ decreasing along the outward direction. 
Therefore in the first order of matter perturbation, we can write \eq{delEjtb} using \eq{n} and eq.\eqref{metsad}
\begin{align}
	8 \pi G \Delta{\cal E}&=\int d\hat{t}\sqrt{h} \,\, 8 \pi G {\cal T}_{{\hat t}{\hat z}}(-{\hat z}) = \int d\hat{t}\sqrt{h}(\nabla_{\hat t}\nabla_{\hat z}\phi){\hat z} \nonumber \\
	&=\int d\hat{t}\sqrt{h}(\del_{\hat t}\del_{\hat z}\phi-\Gamma^{\hat t}_{{\hat t}{\hat z}}\del_{\hat t}\phi){\hat z} =\int d\hat{t}\sqrt{h}\del_{\hat t}({\hat z}\del_{\hat z}\phi+\phi)+O({\hat z}^2)
\end{align}
In the fourth equality we have used the fact that in the absence of matter perturbation, dilaton is independent on $\hat t$. One can verify this by putting \eq{f0f2exp} in \eq{dilfgtop}. We also computed the Christoffel symbol using the metric \eq{metfgtop}.
Since $\del_{\hat t}\phi$ starts contributing at first order of matter perturbation and the contribution comes in $O({\hat z})$, one can check that the correction in fourth equality goes as $O({\hat z}^2)$. 
Now since the full derivative term in the first integral of fourth equality starts contributing at first order of matter perturbation we can take $\sqrt{h}$ inside the $\hat t$ derivative. Using $\gamma_{{\hat t}{\hat t}}=1$ by \eq{gam}, we get
\begin{equation}
	\label{ptoEjt}
	8\pi G\Delta{\cal E}=\int d\hat{t}\del_{\hat t}[{\sqrt{h}\over \sqrt{\gamma}}({\hat z}\del_{\hat z}\phi+\phi)]= 8\pi G\Delta P
\end{equation}
where $\Delta P$ is the difference between final and initial value of $P$ defined in \eq{p}.

Comparing \eq{ptoEjt} with \eq{areatoEjt}, we can reproduce \eq{1lawjt}. 

\subsection{Shockwave Analysis}
As a concrete example, we consider a general set-up with two shockwaves which, having crossed $H_1,H_2$ intersecting at $K_-=s_-, K_+=s_+$ respectively, impinge on the boundary as shown in fig. \ref{fig.jtshLR}. 
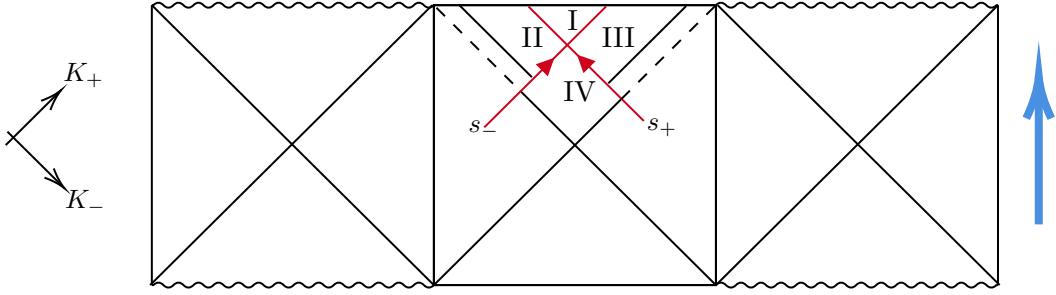
\begin{figure}[h]
	\centering

	\tikzset{every picture/.style={line width=0.75pt}} 
	
	\begin{tikzpicture}[x=0.75pt,y=0.75pt,yscale=-1,xscale=1]
		
		\draw   (246.73,105.21) -- (387.52,105.21) -- (387.52,246) -- (246.73,246) -- cycle ;
		\draw    (289.7,148.18) -- (387.52,246) ;
		\draw    (340.31,152.42) -- (246.73,246) ;
		\draw    (246.73,105.21) .. controls (245.06,106.88) and (243.4,106.88) .. (241.73,105.21) .. controls (240.06,103.54) and (238.4,103.54) .. (236.73,105.21) .. controls (235.06,106.88) and (233.4,106.88) .. (231.73,105.21) .. controls (230.06,103.54) and (228.4,103.54) .. (226.73,105.21) .. controls (225.06,106.88) and (223.4,106.88) .. (221.73,105.21) .. controls (220.06,103.54) and (218.4,103.54) .. (216.73,105.21) .. controls (215.06,106.88) and (213.4,106.88) .. (211.73,105.21) .. controls (210.06,103.54) and (208.4,103.54) .. (206.73,105.21) .. controls (205.06,106.88) and (203.4,106.88) .. (201.73,105.21) .. controls (200.06,103.54) and (198.4,103.54) .. (196.73,105.21) .. controls (195.06,106.88) and (193.4,106.88) .. (191.73,105.21) .. controls (190.06,103.54) and (188.4,103.54) .. (186.73,105.21) .. controls (185.06,106.88) and (183.4,106.88) .. (181.73,105.21) .. controls (180.06,103.54) and (178.4,103.54) .. (176.73,105.21) .. controls (175.06,106.88) and (173.4,106.88) .. (171.73,105.21) .. controls (170.06,103.54) and (168.4,103.54) .. (166.73,105.21) .. controls (165.06,106.88) and (163.4,106.88) .. (161.73,105.21) .. controls (160.06,103.54) and (158.4,103.54) .. (156.73,105.21) .. controls (155.06,106.88) and (153.4,106.88) .. (151.73,105.21) .. controls (150.06,103.54) and (148.4,103.54) .. (146.73,105.21) .. controls (145.06,106.88) and (143.4,106.88) .. (141.73,105.21) .. controls (140.06,103.54) and (138.4,103.54) .. (136.73,105.21) .. controls (135.06,106.88) and (133.4,106.88) .. (131.73,105.21) .. controls (130.06,103.54) and (128.4,103.54) .. (126.73,105.21) .. controls (125.06,106.88) and (123.4,106.88) .. (121.73,105.21) .. controls (120.06,103.54) and (118.4,103.54) .. (116.73,105.21) .. controls (115.06,106.88) and (113.4,106.88) .. (111.73,105.21) .. controls (110.06,103.54) and (108.4,103.54) .. (106.73,105.21) -- (105.94,105.21) -- (105.94,105.21) ;
		\draw    (105.94,105.21) -- (105.94,246) ;
		\draw    (246.73,246) .. controls (245.06,247.67) and (243.4,247.67) .. (241.73,246) .. controls (240.06,244.33) and (238.4,244.33) .. (236.73,246) .. controls (235.06,247.67) and (233.4,247.67) .. (231.73,246) .. controls (230.06,244.33) and (228.4,244.33) .. (226.73,246) .. controls (225.06,247.67) and (223.4,247.67) .. (221.73,246) .. controls (220.06,244.33) and (218.4,244.33) .. (216.73,246) .. controls (215.06,247.67) and (213.4,247.67) .. (211.73,246) .. controls (210.06,244.33) and (208.4,244.33) .. (206.73,246) .. controls (205.06,247.67) and (203.4,247.67) .. (201.73,246) .. controls (200.06,244.33) and (198.4,244.33) .. (196.73,246) .. controls (195.06,247.67) and (193.4,247.67) .. (191.73,246) .. controls (190.06,244.33) and (188.4,244.33) .. (186.73,246) .. controls (185.06,247.67) and (183.4,247.67) .. (181.73,246) .. controls (180.06,244.33) and (178.4,244.33) .. (176.73,246) .. controls (175.06,247.67) and (173.4,247.67) .. (171.73,246) .. controls (170.06,244.33) and (168.4,244.33) .. (166.73,246) .. controls (165.06,247.67) and (163.4,247.67) .. (161.73,246) .. controls (160.06,244.33) and (158.4,244.33) .. (156.73,246) .. controls (155.06,247.67) and (153.4,247.67) .. (151.73,246) .. controls (150.06,244.33) and (148.4,244.33) .. (146.73,246) .. controls (145.06,247.67) and (143.4,247.67) .. (141.73,246) .. controls (140.06,244.33) and (138.4,244.33) .. (136.73,246) .. controls (135.06,247.67) and (133.4,247.67) .. (131.73,246) .. controls (130.06,244.33) and (128.4,244.33) .. (126.73,246) .. controls (125.06,247.67) and (123.4,247.67) .. (121.73,246) .. controls (120.06,244.33) and (118.4,244.33) .. (116.73,246) .. controls (115.06,247.67) and (113.4,247.67) .. (111.73,246) .. controls (110.06,244.33) and (108.4,244.33) .. (106.73,246) -- (105.94,246) -- (105.94,246) ;
		\draw    (105.94,105.21) -- (246.73,246) ;
		\draw    (245.88,105.21) -- (105.09,246) ;
		
		\draw    (295.64,141.4) -- (259.09,104.85) ;
		\draw  [dash pattern={on 4.5pt off 4.5pt}]  (286.21,144.69) -- (246.73,105.21) ;
		\draw    (33.85,168.82) -- (60.99,195.96) ;
		\draw [shift={(62.4,197.37)}, rotate = 225] [color={rgb, 255:red, 0; green, 0; blue, 0 }  ][line width=0.75]    (10.93,-3.29) .. controls (6.95,-1.4) and (3.31,-0.3) .. (0,0) .. controls (3.31,0.3) and (6.95,1.4) .. (10.93,3.29)   ;
		\draw    (33,175.6) -- (59.29,149.31) ;
		\draw [shift={(60.71,147.9)}, rotate = 135] [color={rgb, 255:red, 0; green, 0; blue, 0 }  ][line width=0.75]    (10.93,-3.29) .. controls (6.95,-1.4) and (3.31,-0.3) .. (0,0) .. controls (3.31,0.3) and (6.95,1.4) .. (10.93,3.29)   ;
		\draw [color={rgb, 255:red, 208; green, 2; blue, 27 }  ,draw opacity=1 ]   (332.67,105.87) -- (271.89,166.66) ;
		\draw [shift={(306.88,131.67)}, rotate = 135] [fill={rgb, 255:red, 208; green, 2; blue, 27 }  ,fill opacity=1 ][line width=0.08]  [draw opacity=0] (8.93,-4.29) -- (0,0) -- (8.93,4.29) -- cycle    ;
		\draw   (387.52,105.21) -- (246.73,105.21) -- (246.73,246) -- (387.52,246) -- cycle ;
		\draw [color={rgb, 255:red, 74; green, 144; blue, 226 }  ,draw opacity=1 ][line width=3]    (548.67,215.47) -- (548.67,146.34) ;
		\draw [shift={(548.67,141.34)}, rotate = 90] [color={rgb, 255:red, 74; green, 144; blue, 226 }  ,draw opacity=1 ][line width=3]    (20.77,-6.25) .. controls (13.2,-2.65) and (6.28,-0.57) .. (0,0) .. controls (6.28,0.57) and (13.2,2.66) .. (20.77,6.25)   ;
		\draw    (387.52,105.21) .. controls (389.19,103.54) and (390.85,103.54) .. (392.52,105.21) .. controls (394.19,106.88) and (395.85,106.88) .. (397.52,105.21) .. controls (399.19,103.54) and (400.85,103.54) .. (402.52,105.21) .. controls (404.19,106.88) and (405.85,106.88) .. (407.52,105.21) .. controls (409.19,103.54) and (410.85,103.54) .. (412.52,105.21) .. controls (414.19,106.88) and (415.85,106.88) .. (417.52,105.21) .. controls (419.19,103.54) and (420.85,103.54) .. (422.52,105.21) .. controls (424.19,106.88) and (425.85,106.88) .. (427.52,105.21) .. controls (429.19,103.54) and (430.85,103.54) .. (432.52,105.21) .. controls (434.19,106.88) and (435.85,106.88) .. (437.52,105.21) .. controls (439.19,103.54) and (440.85,103.54) .. (442.52,105.21) .. controls (444.19,106.88) and (445.85,106.88) .. (447.52,105.21) .. controls (449.19,103.54) and (450.85,103.54) .. (452.52,105.21) .. controls (454.19,106.88) and (455.85,106.88) .. (457.52,105.21) .. controls (459.19,103.54) and (460.85,103.54) .. (462.52,105.21) .. controls (464.19,106.88) and (465.85,106.88) .. (467.52,105.21) .. controls (469.19,103.54) and (470.85,103.54) .. (472.52,105.21) .. controls (474.19,106.88) and (475.85,106.88) .. (477.52,105.21) .. controls (479.19,103.54) and (480.85,103.54) .. (482.52,105.21) .. controls (484.19,106.88) and (485.85,106.88) .. (487.52,105.21) .. controls (489.19,103.54) and (490.85,103.54) .. (492.52,105.21) .. controls (494.19,106.88) and (495.85,106.88) .. (497.52,105.21) .. controls (499.19,103.54) and (500.85,103.54) .. (502.52,105.21) .. controls (504.19,106.88) and (505.85,106.88) .. (507.52,105.21) .. controls (509.19,103.54) and (510.85,103.54) .. (512.52,105.21) .. controls (514.19,106.88) and (515.85,106.88) .. (517.52,105.21) .. controls (519.19,103.54) and (520.85,103.54) .. (522.52,105.21) .. controls (524.19,106.88) and (525.85,106.88) .. (527.52,105.21) -- (528.31,105.21) -- (528.31,105.21) ;
		\draw    (528.31,105.21) -- (528.31,246) ;
		\draw    (387.52,246) .. controls (389.19,244.33) and (390.85,244.33) .. (392.52,246) .. controls (394.19,247.67) and (395.85,247.67) .. (397.52,246) .. controls (399.19,244.33) and (400.85,244.33) .. (402.52,246) .. controls (404.19,247.67) and (405.85,247.67) .. (407.52,246) .. controls (409.19,244.33) and (410.85,244.33) .. (412.52,246) .. controls (414.19,247.67) and (415.85,247.67) .. (417.52,246) .. controls (419.19,244.33) and (420.85,244.33) .. (422.52,246) .. controls (424.19,247.67) and (425.85,247.67) .. (427.52,246) .. controls (429.19,244.33) and (430.85,244.33) .. (432.52,246) .. controls (434.19,247.67) and (435.85,247.67) .. (437.52,246) .. controls (439.19,244.33) and (440.85,244.33) .. (442.52,246) .. controls (444.19,247.67) and (445.85,247.67) .. (447.52,246) .. controls (449.19,244.33) and (450.85,244.33) .. (452.52,246) .. controls (454.19,247.67) and (455.85,247.67) .. (457.52,246) .. controls (459.19,244.33) and (460.85,244.33) .. (462.52,246) .. controls (464.19,247.67) and (465.85,247.67) .. (467.52,246) .. controls (469.19,244.33) and (470.85,244.33) .. (472.52,246) .. controls (474.19,247.67) and (475.85,247.67) .. (477.52,246) .. controls (479.19,244.33) and (480.85,244.33) .. (482.52,246) .. controls (484.19,247.67) and (485.85,247.67) .. (487.52,246) .. controls (489.19,244.33) and (490.85,244.33) .. (492.52,246) .. controls (494.19,247.67) and (495.85,247.67) .. (497.52,246) .. controls (499.19,244.33) and (500.85,244.33) .. (502.52,246) .. controls (504.19,247.67) and (505.85,247.67) .. (507.52,246) .. controls (509.19,244.33) and (510.85,244.33) .. (512.52,246) .. controls (514.19,247.67) and (515.85,247.67) .. (517.52,246) .. controls (519.19,244.33) and (520.85,244.33) .. (522.52,246) .. controls (524.19,247.67) and (525.85,247.67) .. (527.52,246) -- (528.31,246) -- (528.31,246) ;
		\draw    (528.31,105.21) -- (387.52,246) ;
		\draw    (388.37,105.21) -- (529.16,246) ;
		
		\draw    (333.81,144.51) -- (372.62,105.69) ;
		\draw [color={rgb, 255:red, 208; green, 2; blue, 27 }  ,draw opacity=1 ]   (293.94,105.87) -- (351.52,163.45) ;
		\draw [shift={(318.13,130.06)}, rotate = 45] [fill={rgb, 255:red, 208; green, 2; blue, 27 }  ,fill opacity=1 ][line width=0.08]  [draw opacity=0] (8.93,-4.29) -- (0,0) -- (8.93,4.29) -- cycle    ;
		\draw  [dash pattern={on 4.5pt off 4.5pt}]  (340.31,152.42) -- (387.52,105.21) ;
		
		\draw (370.65,131.21) node [anchor=north west][inner sep=0.75pt]  [font=\small] [align=left] {$ $};
		\draw (60.32,132.91) node [anchor=north west][inner sep=0.75pt]  [font=\small] [align=left] {$\displaystyle K_{+}$};
		\draw (61.24,196.52) node [anchor=north west][inner sep=0.75pt]  [font=\small] [align=left] {$\displaystyle K_{-}$};
		\draw (253.6,131.21) node [anchor=north east][inner sep=0.75pt]  [font=\small,xscale=-1] [align=left] {$ $};
		\draw (262.55,161.74) node [anchor=north west][inner sep=0.75pt]  [font=\small] [align=left] {$\displaystyle s_{-}$};
		\draw (351.53,160.9) node [anchor=north west][inner sep=0.75pt]  [font=\small] [align=left] {$\displaystyle s_{+}$};
		\draw (312,106) node [anchor=north west][inner sep=0.75pt]   [align=left] {I};
		\draw (289,115) node [anchor=north west][inner sep=0.75pt]   [align=left] {II};
		\draw (329,115) node [anchor=north west][inner sep=0.75pt]   [align=left] {III};
		\draw (310,142) node [anchor=north west][inner sep=0.75pt]   [align=left] {IV};

	\end{tikzpicture}

	\caption{Shockwave crossing through both left and right cosmological horizons in de Sitter JT gravity.}
	\label{fig.jtshLR}	
\end{figure}
~\\	
Stress-tensor is given in this case as
\begin{equation}
	\label{shockT}
	{\cal T}_{--}={\mu_-\over 8\pi G}\delta(K_--s_-)~~,~~{\cal T}_{++}={\mu_+\over 8\pi G}\delta(K_+-s_+)~~,~~{\cal T}_{+-}=0
\end{equation}
We also assume that the shockwaves cross each other inside the upper Milne patch, see fig. \ref{fig.jtshLR}, without any interaction. One can therefore partition the patch into four mutually exclusive regions I, II, III, IV defined as follows
\begin{align}
	&{\rm {\bf Region~I~:~}} K_+>s_+ {~\rm and~} K_-<s_- ~~,~~ {\rm {\bf Region~II~:~}} K_+<s_+ {~\rm and~} K_-<s_-\nonumber\\
	&{\rm {\bf Region~III~:~}} K_+>s_+ {~\rm and~} K_->s_- ~~,~~ {\rm {\bf Region~IV~:~}} K_+<s_+ {~\rm and~} K_->s_- \label{regions}
\end{align}
In these regions, solutions of dilaton takes the form given in \eq{vacdilgen}. The dilaton profile over the entire patch is given by \eq{dilgen}. The functions $U,V$ in that equation satisfy \eq{triuv} which can be written using \eq{shockT} as
\begin{equation}
	\label{triuvsh}
	V'''(K_+)=-2\mu_+\delta(K_+-s_+)~~,~~U'''(K_-)=2\mu_-\delta(K_--s_-)
\end{equation}

\paragraph{Region IV :}
In this region, solution of dilaton is given by \eq{Kmetdil}. 
One can check that the following expressions of $U,V$
\begin{equation}
	\label{iniuv}
	U(K_-)=0~~,~~V(K_+)=2\sqrt{m}K_+
\end{equation}
produces the expression of dilaton in \eq{Kmetdil}.

\paragraph{For all region :}
Now, we solve \eq{triuvsh} with \eq{iniuv} as initial conditions. The solutions are given by :
\begin{align}
	V(K_+)=2\sqrt{m}K_+-\mu_+(K_+-s_+)^2\Theta(K_+-s_+)
\end{align}
and,
\begin{align}
	U(K_-)=-\mu_-(s_--K_-)^2\Theta(s_--K_-)
\end{align}
Using the above relations in \eq{dilgen}, we get
\begin{equation}
	\label{dilall}
	\phi={1\over (1+K_+K_-)}\brt{\sqrt{m}(1-K_+K_-)+\mu_+\Theta(K_+-s_+)f_+(K_+,K_-)-\mu_-\Theta(s_--K_-)f_-(K_+,K_-)}
\end{equation}
where
\begin{equation}
	f_+(K_+,K_-)\equiv s_+(1-K_+K_-)-K_++s_+^2K_-~~,~~f_-(K_+,K_-)\equiv s_-(1-K_+K_-)-K_-+s_-^2K_+
\end{equation}
Now, we can easily find the dilaton and the pressure in regions II, III, I as follows

\paragraph{Region II :} As specified in \eqref{regions}, \eq{dilall} reduces to \eq{vacdilgen}
where
\begin{equation}
	a=\sqrt{m}-\mu_-s_-~~,~~b=-\mu_-s_-^2~~,~~c=\mu_-
\end{equation}
which gives $P$ \eq{pf0f2}, using \eq{f0f2exp}, to the first order of perturbation as
\begin{equation}
	\label{apl}
	P={m\over 16\pi G}-{1\over 8\pi G}\sqrt{m}\mu_-s_-
\end{equation}
The location of $H_1$ is given by the value of $K_+$ satisfying $\del_-\phi=0$, 
and thereby the value of dilaton on $H_1$ as
\begin{equation}
	\label{phihl}
	\phi_h=\sqrt{m}-\mu_-s_-
\end{equation}

\paragraph{Region III :} As specified in \eqref{regions}, \eq{dilall} reduces to \eq{vacdilgen}
where
\begin{equation}
	a=\sqrt{m}+\mu_+s_+~~,~~b=-\mu_+~~,~~c=\mu_+s_+^2
\end{equation}
which gives $P$ \eq{pf0f2}, using \eq{f0f2exp}, to the first order of perturbation as
\begin{equation}
	\label{ap}
	P={m\over 16\pi G}+{1\over 8\pi G}\sqrt{m}\mu_+s_+
\end{equation}
The location of $H_2$ is given by the value of $K_-$ satisfying $\del_+\phi=0$,  
and thereby the value of dilaton on $H_2$ as
\begin{equation}
	\label{phihr}
	\phi_h=\sqrt{m}+\mu_+s_+
\end{equation} 

To note, in region II and III, one can relate entropy $S={\phi_h\over 4G}$ and $P$ before shockwave, at first order of perturbation, as
\begin{equation}
	\label{s(p)+}
	S(P)=\sqrt{\pi \over G}\sqrt{P}
\end{equation}

\paragraph{Region I :} As specified in \eqref{regions}, \eq{dilall} reduces to \eq{vacdilgen}
where
\begin{equation}
	a=\sqrt{m}+\mu_+s_+-\mu_-s_-~~,~~b=-\mu_+-\mu_-s_-^2~~,~~c=\mu_-+\mu_+s_+^2
\end{equation}
which gives $P$ \eq{pf0f2}, using \eq{f0f2exp}, to the first order of perturbation as
\begin{equation}
	\label{ap+-}
	P={m\over 16\pi G}+{\sqrt{m}\over 8\pi G}(\mu_+s_+-\mu_-s_-)
\end{equation}

We now thereby attribute an entropy to the boundary which is given by \eq{s(p)+}.  
Noting the value of pressure given in \eq{ap+-} and the entropy $S$ of horizons in either region II or III given by the values of $\phi_h$ in \eq{phihl} and \eq{phihr} respectively, one can immediately see that the pressure in the intermediate region $I$ cannot be related by \eq{s(p)+} to entropy of any of the $H_1$ or $H_2$ as discussed in section \ref{sec.hololaw1}.

		\newpage
		\bibliographystyle{JHEP}
		\bibliography{dSthermo}	
	\end{document}